\documentclass[11pt, oneside]{article}   	
\usepackage{slashed}
\usepackage{jheppub}
\usepackage{amsmath}
\usepackage{amssymb}
\usepackage{amsfonts}
\usepackage{amsthm}
\usepackage{amsbsy}
\usepackage{array}
\usepackage{mathtools}
\usepackage[usenames,dvipsnames]{xcolor}
\usepackage{subcaption}
\usepackage{dsdshorthand}
\usepackage{graphicx}
\usepackage[vcentermath]{youngtab}
\usepackage{hyperref}
\usepackage{cleveref}
\usepackage{cancel}
\usepackage[normalem]{ulem}

\usepackage{tikz}
\usetikzlibrary{snakes}
\usetikzlibrary{decorations}
\usetikzlibrary{trees}
\usetikzlibrary{decorations.pathmorphing}
\usetikzlibrary{decorations.markings}
\usetikzlibrary{external}
\usetikzlibrary{intersections}
\usetikzlibrary{shapes,arrows}
\usetikzlibrary{arrows.meta}
\usetikzlibrary{calc}
\usetikzlibrary{shapes.misc}
\usetikzlibrary{decorations.text}
\usetikzlibrary{backgrounds}
\usetikzlibrary{fadings}
\usetikzlibrary{tikzmark,calc,arrows,shapes,decorations.pathreplacing}
\usetikzlibrary{patterns}
\tikzset{
        cross/.style={cross out, draw=black, minimum size=2*(#1-\pgflinewidth), inner sep=0pt, outer sep=0pt},
	branchCut/.style={postaction={decorate},
		snake=zigzag,
		decoration = {snake=zigzag,segment length = 2mm, amplitude = 2mm}	
    }}

\newcommand{\bea}{\setlength\arraycolsep{2pt} \begin{eqnarray}}
\newcommand{\eea}{\end{eqnarray}}

\def\fft#1#2{{\frac{#1}{#2}}}

\newcommand{\baa}{\begin{align}}
\newcommand{\eaa}{\end{align}}

\def\Mpl{M_{\rm pl}}
\def\M{\mathcal{M}}
\def\cutoff{\Lambda_\text{\cancel{QFT}}}
\def\hs{\Lambda_{\rm o}}
\def\mubar{\bar{\mu}}

\makeatletter
\def\@fpheader{\ }
\makeatother

\title{Gravity and a universal cutoff for field theory}
\author{Simon Caron-Huot$^1$, Yue-Zhou Li$^2$}
\affiliation{
${}^1$Department of Physics, McGill University, 3600 Rue University, Montr\'eal, H3A 2T8, QC Canada \\
${}^2$Department of Physics, Princeton University, Princeton, NJ 08544, USA
}
\emailAdd{schuot@physics.mcgill.ca}
\emailAdd{liyuezhou@princeton.edu}
\date{}
\abstract{
We analyze the one-loop effects of massive fields on 2-to-2 scattering processes involving gravitons.
It has been suggested that in the presence of gravity, any local effective field theory description must break down at the ``species scale''. We first observe that unitarity and analyticity of the amplitude indeed imply a species-type bound $G\Lambda^{d-2}N\leq O(1)$, where $N$ counts parametrically light species and $\Lambda$ is an energy scale above which 
new unknown ingredients must modify the graviton amplitude.
To clarify what happens at this scale, we contrast the partial wave decomposition of calculated amplitudes with that of some ultraviolet scenarios: string theory and strongly interacting Planck-scale physics.
Observing that the latter exhibit a markedly stronger high-spin content, we define nonperturbatively the \emph{high-spin onset} scale $\hs$, which coincides with the string scale and higher-dimensional Planck scale in respective examples. We argue that, generally, no local field description can exist at distances shorter than $1/\hs$.
}

\begin{document}

\maketitle
\pagenumbering{roman}
\setcounter{page}{2}
\newpage
\pagenumbering{arabic}
\setcounter{page}{1}

\section{Introduction}

The nature of gravity at very short distances is an outstanding mystery, with spacetime fluctuating so strongly at the Planck length that it ceases to make sense.
Given that current and foreseeable experiments probe much longer length scales, 
it might well be that all observable consequences of these effects can be captured by a local effective field theory (EFT):
\be
  S= S_{\rm EH} + S_{\rm matter} + \mbox{higher-derivative corrections}\,. \label{EFT}
\ee
The idea is that the Einstein-Hilbert action minimally coupled to matter
is supplemented by derivative corrections which parametrize our ignorance of physics
at distances shorter than some scale $\Lambda^{-1}$.
To our knowledge, all currently observed gravitational phenomena fit this framework \cite{Donoghue:1994dn,Donoghue:2022eay}.
In the absence of a unique predictive theory of quantum gravity, it makes sense to
theoretically chart out the space of EFTs which admit a meaningful UV completion.
The Swampland program, inspired by string theory, has led to a number of intriguing
conjectures about such EFTs  \cite{Ooguri:2006in,Brennan:2017rbf,Palti:2019pca,vanBeest:2021lhn,Arkani-Hamed:2006emk,Harlow:2022ich}.
A basic question to be discussed in this paper is: down to which length scales can one use EFT logic?

A longstanding observation is that in the presence of a large number $N(\Lambda)$ of matter fields below the scale $\Lambda$,
the ``quantum gravity cutoff'' $\Lambda$ should be parametrically lower than the Planck mass \cite{Dvali:2000xg,Veneziano:2001ah,Dvali:2007hz}:
\be
\Lambda^{d-2}N(\Lambda) \lesssim \mathcal{O}(1) M_{\rm pl}^{d-2}\,.\label{eq: rough bound}
\ee
The simplest way to understand this ``species scale'' is that it is where loop corrections to the graviton propagator
overwhelm its tree-level expression (see e.g. \cite{Dvali:2007wp}, section 3.2).
One could try to reach higher energies by simply resuming self-energies, however this does not appear to yield a sensible graviton propagator (as is further discussed below), unless an infinite number of derivative corrections simultaneously become important; hence, when \eqref{eq: rough bound} is violated, the framework \eqref{EFT} loses any predictive power and we say that it ``breaks down''.
Violation of \eqref{eq: rough bound} would also be in tension with the Bekenstein-Hawking entropy formula, by enabling, roughly, too many distinct ways of producing a black hole of radius $\Lambda^{-1}$ \cite{Brustein:2009ex}.
Scenarios with $N$ very large were initially considered as a way to ameliorate the hierarchy problem between the weak scale and the gravity scale.  Since the species scale is closely linked with quantum gravity, it has recently drawn significant attention, see, e.g., \cite{vandeHeisteeg:2022btw,Cribiori:2022nke,vandeHeisteeg:2023dlw,Cribiori:2023ffn,Basile:2024dqq}.

In calculable string theory examples, there are at least two distinct ways to realize a low cutoff $\Lambda\ll M_{\rm pl}$.
The first is through the decompactification of $k$ extra dimensions whose volume ${\rm Vol}_{k}$ tends to infinity. 
In this limit, $N(\Lambda)\sim \Lambda^{k}{\rm Vol}_{k}$ from counting Kaluza-Klein modes and the cutoff \eqref{eq: rough bound} coincides
parametrically with the Planck mass of the higher-dimension theory:
$\Lambda^{d+k-2}\sim M_{\rm pl}^{d-2}/{\rm Vol}_{k}\equiv M_{{\rm pl},d+k}^{d+k-2}$.
The second way is that a fundamental string can become tensionless, meaning that an infinite tower of high-spin states on the graviton Regge trajectory become light. The local EFT framework embodied by \eqref{EFT} then breaks down at the string scale.  (This is unrelated to the question of whether a more specialized framework such as string field theory holds.)
In fact, it has been conjectured that these are the only options, known as the emergent string conjecture \cite{Lee:2019wij,Castellano:2022bvr,Blumenhagen:2023yws} (see also \cite{Basile:2023blg,Bedroya:2024ubj} in the context of black holes).

The cutoff $\Lambda$ enters a number of Swampland conjectures, such as the distance conjecture \cite{Ooguri:2006in,Etheredge:2022opl}
and the recently observed ``pattern'' \cite{Castellano:2023stg,Castellano:2023jjt} which relates the way that the UV cutoff varies as a function of scalar moduli, $\vec{\nabla}_\phi \log\Lambda$,
with the mass of other massive towers: $(\vec{\nabla}_\phi \log \Lambda){\cdot}(\vec{\nabla}_\phi \log M)=\frac{8\pi G}{d-2}$.

In order to better understand statements and conjectures involving $\Lambda$, it would be helpful to have a more precise characterization of what happens at this scale without specifying the UV theory.
A single review \cite{Harlow:2022ich} mentions ``the fundamental quantum gravity cutoff'',
the ``cutoff on low energy effective field theory in any form'',
that ``effective field theory breaks down irrevocably at some scale'',
and ``the species bound scale'' $\Lambda_{\rm sp}$.
While the phrase ``species scale'' is becoming more widely used in the literature as a proxy for the UV cutoff, 
we find this terminology unsatisfying because it mixes concepts that are a priori distinct.
Furthermore, a precise definition of ``$N$'' remains challenging in our opinion.\footnote{
For modes much below the cutoff, what is the correct relative weight to assign to different types of fields---scalars, fermions, gauge fields etc.---and why?
For modes near a weakly coupled cutoff, ie. string theory with $M_s\ll M_{\rm pl}$, 
different natural counts change $\Lambda_{\rm sp}$ by $O(\log(M_{\rm pl}/M_s))$ \cite{Dvali:2010vm,Blumenhagen:2023yws}---which should we use?
And when interactions are strong, is there any well-defined procedure to count modes near the cutoff?}
From now on, we will refer thus to the universal cutoff of \eqref{EFT} as $\cutoff$.

The central notion which seems consensual is that at lengths shorter then a certain scale $\cutoff^{-1}$, the concept of local fields becomes inapplicable.
We stress that this scale is unique and unlike any other EFT cutoff we have encountered before in physics.
We are accustomed to the idea that when an EFT breaks down, it simply gets replaced by a better one: 
QED with a point-like proton gets replaced by QED+QCD above $\Lambda_{\rm QCD}$, where quarks and gluons become dynamical;
Fermi's four-fermion description of the weak force upgrades to the electroweak theory above the mass scale of the $W$, $Z$, etc.
There is no local field theory, effective or not, above $\cutoff$. 

The main goal of this paper is to clarify what happens at the scale $\cutoff$ and to explain how precise bounds of the form \eqref{eq: rough bound} naturally arise from suitable definitions.\footnote{We stress that the cutoff represents a minimal length, not a maximal energy:
apple-Earth scattering can be well described by EFT despite having a large center-of-mass energy in Planck units, if short distances are not probed.  Also, we consider only simple measurements that are not exponentially complicated in $1/G$.}
We will consider thought experiments involving the scattering of gravitons and calculate the one-loop contributions from various types of matter,
including scalars, fermions, vectors, and additional spin-$3/2$ and spin-$2$ fields, as well as two- and three-form fields in higher dimensions.
We will analyze these amplitudes from the prism of \emph{dispersive sum rules},
which the amplitudes must satisfy if standard S-matrix axioms embodying unitarity and relativistic causality continue to apply above the scale $\cutoff$.

\begin{figure}[t]
\centering\includegraphics[width=0.9\textwidth]{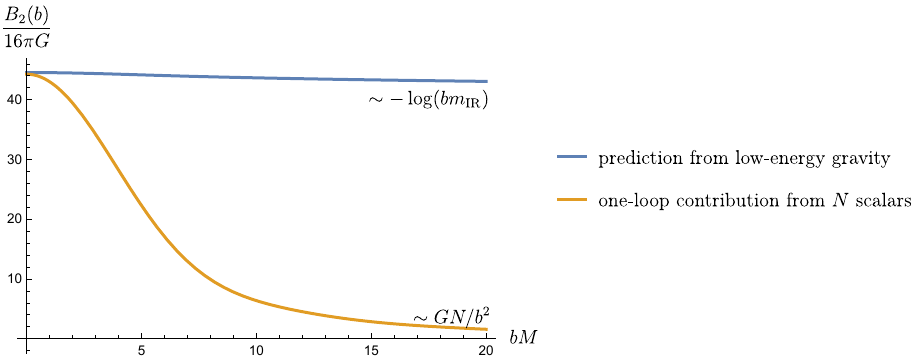}
\caption{Illustration of sum rules for Newton's constant: for any $b$,
calculable positive contributions from $N$ scalars (orange line) cannot exceed the total required by low-energy gravity
(blue line). Other light matter fields have a qualitatively similar shape while heavier fields decay more rapidly with $b$;
all are nonnegative.
}\label{fig: compare b scalar}
\end{figure}

The basic idea (further reviewed in section \ref{sec: sum rules})
is that under these assumptions, graviton scattering satisfy Kramers-Kronig-type sum rules which relate Newton's constant at low energies to nonnegative scattering probabilities at high energies:
\be \label{schematic sum rule}
  \frac{8\pi G}{p_\perp^2} = \int \fft{ds}{\pi} (2s-p^2)  {\rm Im}\, f(s,-p_\perp^2) \equiv B_2(p_\perp),
\ee
which makes sense after integrating both sides against a suitable ``focusing'' wavepacket $\psi(p_\perp)$ (called smearing function in \cite{Caron-Huot:2021rmr}). This generalizes the positivity argument of \cite{Adams:2006sv} to a gravitational context and also connects with the causality argument of \cite{Camanho:2014apa}. In practice, we separate the integral into low energies $s<M^2$, where it can be calculated using EFT, and high energies $s>M^2$.  Importantly, one can choose $\psi(p_\perp)$ such that the contribution from any computable low-energy loop below $M$ is positive, \emph{and} any unknown high-energy physics above the scale $M$ also contributes positively. In fact, one can design an infinite set of such wavefunctions, which are loosely labelled by impact parameter $b\gtrsim M^{-1}$, leading to a family of sum rules whose typical shape is sketched in figure \ref{fig: compare b scalar}.\footnote{
For illustration, we used $B_2(b) = \int_0^{M} dp_\perp \int \fft{ds}{\pi} (2s-p^2) (1-p)^2 p_\perp J_{0}(b p_\perp)   {\rm Im}\, f(s,-p_\perp^2)$ in this plot.
Technically, for $b$ fixed, only high-energy states with $s\gg M^2$ are guaranteed to contribute positively to this specific sum rule, which allow slight negative contributions from states with $s\sim M^2$.
Better sum rules that are rigorously sign-definite are described in section \ref{sec: sum rules}.
}

For any separation scale $M$, we learn two things from such plots:  i) Newton's constant sets an upper bound on computable loop contributions to graviton scattering from modes lighter than $M$, ie. a bound on $N(M)$
ii) there must exist states heavier than $M$ whose contributions is qualitatively stronger at $b M\gg 1$ than any computable matter loop.
Given the usual relation between angular momentum of two-particle states and impact parameter ($J= \frac{bm}{2}$),
we refer to this second phenomenon as the \emph{onset} of high-spin states.

We propose that the quantum gravity cutoff is simply the scale of high-spin onset: $$\cutoff\equiv \hs.$$
High-spin onset is a counterpart to the ``low-spin dominance'' phenomenon observed in \cite{Bern:2021ppb} for computable matter loops.
The idea is that the physics which UV completes graviton scattering must be quantifiably
less low-spin-dominated than any physics we can describe using local field theory below the cutoff. The onset of high-spin states at the scale $\hs$ thus signals the breakdown of any EFT description, and triggers a nontrivial UV completion of gravity.

More precisely, we propose to define $\hs$ as the lowest scale at which certain ratios of integrated spectral densities exceed certain order-one constants:
\be
\fft{\int_0^{\hs} s^{-\fft{d}{2}}ds\,{\rm Im}\, a_{J}(s)}{\int_0^{\hs} s^{-\fft{d}{2}}ds\,{\rm Im}\, a_{J-2}(s)} \gtrsim C_J
\label{def cutoff}
\ee
which we will compute in a number of spacetime dimensions and SO($d-1$) representations.
Generally, from this definition and dispersive sum rules, we expect that $\hs\leq M_{\rm pl}\times O(1)$.
We will find that the high-spin scale coincides parametrically with the higher-dimensional Planck scale in decompactification limits,
while in weakly coupled string theory it coincides with the mass of spin ${\geq}4$ string states that couple strongly to two gravitons.

A key feature of the criterion \eqref{def cutoff} is that it makes sense even when physics at the cutoff is not weakly coupled.
As we will discuss, in the context of AdS/CFT, the analogous definition of $\hs$ (using three-point couplings between two stress tensors and heavy spinning states) gives a precise meaning to the notion of ``higher-spin gap'' at finite $N_c$.  It was conjectured in \cite{Heemskerk:2009pn} that CFTs exhibiting such a gap had a local bulk dual to sub-AdS distances.

While the technology to calculate one-loop graviton amplitudes is not new, to our knowledge this is the first time that these
amplitudes and their partial waves are systematically analyzed in $d$ dimensions. Anticipating possible other applications, we record self-energy corrections, vertex corrections and on-shell four-graviton amplitudes in an ancillary file.

Our graviton-graviton scattering setup deserves a few comments.
While the operation of a graviton collider (or even of a gravity wave collider) is hard to envision with current technology and budgets,
the essential point of thought experiments involving graviton scattering is that there is no \emph{in principle} obstruction to preparing a pair of gravitons with center-of-mass energy above $\cutoff$ in an otherwise locally flat region of spacetime, in which our ``collider'' resides. By unitarity, we mean that the probability that they scatter nontrivially should remain nonnegative at any relevant energy.
By causality, we mean that near-forward signals do not arrive superluminally compared with the flat metric in which the collider resides,\footnote{In $d=4$, infrared divergences make this notion logarithmically sensitive to the size of the collider.} even with time resolution better than $\cutoff^{-1}$.%
\footnote{
For relativistic S-matrices, we are unaware of any \emph{in principle} limit on the accuracy of arrival time measurements. Naively, one could imagine using highly boosted detectors to take advantage of time dilation factors and thus achieve super-Planckian accuracy on the retarded time at which gravitational radiation arrives.}
Together, these two assumptions ensure analyticity and dispersive sum rules of the type \eqref{schematic sum rule}.

While the just stated assumptions are standard for the S-matrices of local quantum field theories,
it would be fair to question them in the presence of gravity. For example, the convergence of dispersive sum rules has been questioned in \cite{Giddings:2009gj}
in relation to  ``non-locality'' above Planck energies.
Our best answer is that in AdS/CFT,
equivalent sum rules can be derived using the exact causality properties of the boundary CFT \cite{Caron-Huot:2021enk}.\footnote{AdS space also serves as a well-defined infrared regulator, such that the familiar infrared divergence of the four-dimensional Shapiro time delay appear in the flat space limit as logarithms of the AdS${}_4$ radius $R_{\rm AdS}\to\infty$.}
It appears that quantum gravity is perfectly compatible with standard S-matrix dispersion relations.

The paper is organized as follows. In section \ref{sec: general}, we present the prerequisites for our discussions of the high-spin onset, including the kinematic structure of graviton scattering, low-energy effective actions, and the partial wave decompositions. In section \ref{sec: amplitudes}, we compute all one-loop contributions to the graviton amplitudes at low energy in general dimensions, from fields of spin $j\leq 2$. This also includes the resummed graviton propagator and effective vertices as byproducts. In section \ref{sec: sum rules}, we first argue that the resummed propagators do not make sense at scales parametrically above the species scale, although it remains unclear whether this fact can be used to obtain sharp species-type bound.
Then, we review the gravitational sum rules and use the fixed impact parameter ones to demonstrate the phenomena of high-spin onset. We also apply the sum rules to put rigorous bounds on the number of light species, followed by comments on the low-energy Wilson coefficients.

In section \ref{sec: QFT breaking}, we define the integrated spectral densities, as motivated by gravitational sum rules. We compute such quantities in different spacetime dimensions and SO$(d-1)$ irreducible representations for a variety of models, including matter loops, string theory, and strongly interacting gravitational physics at the Planck scale. We demonstrate that the high-spin spectral densities of these UV completion scenarios of gravity are qualitatively stronger than any of the content calculated in field theory. This completes our proposal of defining the scale of high-spin onset $\hs$
as the universal field theory cutoff,
which we also discuss in the AdS/CFT context.
We summarize in section \ref{sec: conclusion},
which includes a number of conjectures regarding $\hs$.

In appendix \ref{app: ghost}, we review technical aspects of ghost actions for massless Rarita-Schwinger and two-form fields. In appendix \ref{app: branch}, we record discontinuities of one-loop master integrals. In appendix \ref{app: string amp}, we review tree-level amplitudes in various string theories.

\section{Generalities and setup}
\label{sec: general}

In this section, we prepare necessary ingredients that enter our loop amplitudes and analysis of $\cutoff$ and $\hs$.

\subsection{Polarization structures for graviton scattering}

We consider graviton $2$-to-$2$ scattering. The amplitude for this process
depends on the polarization and momenta of the gravitons.
In $d=4$,  polarizations can be classified as two helicity states and
the independent amplitudes up to Bose symmetry are \cite{Elvang:2013cua,Bern:2021ppb}
\begin{align}
{\cal M}(1^+ 2^- 3^- 4^+) &=   \langle 23\rangle^4 [14]^4\, f(s,u)\,,\\
 \mathcal{M}(1^{+}2^{+}3^{+}4^{-}) &= ([12][13]\langle 14\rangle)^4 \,g(s,u)\,,\\
 \mathcal{M}(1^{+}2^{+}3^{+}4^{+}) &= \fft{[12]^2 [34]^2}{\langle 12\rangle^2 \langle 34\rangle^2}\,
h(s,u)\,,
\label{eq:Mdef}
\end{align}
where the Mandelstam variables are defined by
\begin{equation}
        s = - (p_1+ p_2)^2\,, \quad t = - (p_2+ p_3)^2\,, \quad u = - (p_1+ p_3)^2\,.
\end{equation}
We take the conventions that all momenta are out-going. Energy-momentum conservation gives $s+t+u=0$.
The spinor brackets satisfy $|\<12\>|^2=|[12]|^2=|s|$ and permutations thereof.

In general dimensions, similarly, graviton amplitudes can be factorized into the product of Lorentz-invariant polynomials of momenta and polarizations times pure functions of Mandelstam variables
\be
 \cM = \sum_{(i)} {\rm P}^{(i)}(\{p_j,\e_j\})\times \mathcal{M}^{(i)}(s,t)\,, \label{generic M}
\ee
in which the polarizations vectors are transverse traceless and possess the gauge redundancies
\be
p_j^2=  \e_j^2=p_j{\cdot}\e_j =0, \qquad \e^\mu_j \simeq \e^\mu_j + \# p^\mu_j\,.
\ee
In generic dimensions $d\geq 8$, there are $29$ independent structures of ${\rm P}^{(i)}$. An appropriate choice of them, called as the local module, ensures that no spurious poles appear in the pure functions $\mathcal{M}^{(i)}$ \cite{Chowdhury:2019kaq,Caron-Huot:2022jli}
\def\spc{,\hspace{2mm}}
\be \label{graviton basis}
{\rm singlets\!:}\ & \mathcal{G}\M^{(1)}(s,u)\spc S^2 \M^{(10)}(s,u),\nn\\
{\rm triplets\!:}\ &H_{14}^2H_{23}^2 \M^{(2)}(s,u), H_{12}H_{13}H_{24}H_{34}\M^{(3)}(s,u), H_{14}H_{23}(X_{1243}{-}X_{1234}{-}X_{1324})\M^{(4)}(s,u),\nn\\
& X_{1243}^2 \M^{(6)}(s,u)\spc X_{1234}X_{1324} \M^{(7)}(s,u), H_{14}H_{23} S\M^{(8)}(s,u)\spc X_{1243}S\M^{(9)}(s,u),\nonumber\\
{\rm sextuplet\!:}\  &H_{12}H_{34}X_{1243} \M^{(5)}(s,u)\,.
\ee
Two singlets, seven cyclic triplets and one sextuplet describe their transformations under permutations.
The building blocks, each gauge and Lorentz invariant, are defined as
\begin{align}
\label{generators dumb}
& H_{12} = F_{1\nu}^{\mu} F_{2\mu}^\nu\,,\quad H_{123} = F_{1\nu}^{\mu} F_{2\sigma}^\nu F_{3\mu}^\sigma \,,\quad
H_{1234} = F_{1\nu}^{\mu} F_{2\sigma}^\nu F_{3\rho}^\sigma F_{4\mu}^\rho\,,\quad V_1 = p_{4\mu}F_{1\nu}^{\mu} p_2^\nu\,,\nn\\
& X_{1234} = H_{1234}-\tfrac14H_{12}H_{34}-\tfrac14H_{13}H_{24}-\tfrac14H_{14}H_{23}\,,
\nn\\
 & S = V_1 H_{234} + V_2 H_{341} + V_3 H_{412} + V_4 H_{123}\,,\quad \mathcal{G}={\rm det}\, v_i\cdot v_j\,,
\end{align}
where $F_{i\nu}^\mu=p_i^\mu\epsilon_{i\nu}-p_{i\nu}\epsilon_i^\mu$ and $v=(p_1,p_2,p_3,\epsilon_1,\epsilon_2,\epsilon_3,\epsilon_4)$. Since the $\mathcal{M}^{(i)}$ do not have spurious poles in this basis, the interpretation of amplitudes and analysis of dispersion relation become more straightforward \cite{Caron-Huot:2022jli}; for example,
gauge invariant contact interactions between four gravitons are in one-to-one correspondence with arbitrary polynomial amplitudes
$\mathcal{M}^{(i)}(s,t)$ subject only to the correct permutation symmetries.

\subsection{Stress-tensor correlators and effect of matter on gravity amplitudes}
\label{subsec: ingredients}

The gravity theory we consider is Einstein's gravity perturbed by its coupling
to a large number $N\gg 1$ of matter fields, plus possible higher-derivative terms involving the metric:
\be \label{Sfull}
 S&= S_{\rm EH}[g] +  S_{\rm higher-derivatives} [g] + S_{\rm matter}, \quad S_{\rm EH}= \frac{1}{16\pi G} \int d^d x \sqrt{-g}R \,.
\ee

As discussed in introduction, at the ``species scale'' scale $M$ where $G N M^{\fft{D-2}{2}}\sim 1$,
loop corrections to graviton-graviton scattering become comparable to tree-level contributions.
A reasonable effective field theorist might not immediately conclude that perturbation theory breaks down at that scale,
however, since only a very restricted set of diagrams need to be resummed:
those that are enhanced by the maximal power of $N$ for each loop.

We will thus start by entertaining a scenario in which such a resummation were to make sense
up to a scale parametrically above $M$.  This will indeed be the case at a purely power-counting level.
However, the resummed theory will turn out to be pathological in various ways.
Thus, after we realize the foolishness of this scenario,
we will return to doing ordinary perturbation theory at energies $s,t\ll M^2$.

The leading effects of $N$ matter fields on graviton scattering are captured by
a non-local effective action $S_{\rm eff}[g]$ obtained by integrating out the matter fields.
We refer to this framework as ``matter weakly coupled to gravity''.
For bookkeeping purposes, it will be convenient to lump matter loops and higher-derivative terms of the metric
into the same effective action:
\be\begin{aligned} \label{Seff}
 e^{iS_{\rm eff}[g]}\equiv e^{i S_{\rm higher-derivatives}} \int {\cal D}\phi_{\rm matter} e^{i S_{\rm matter}}\,.
\end{aligned}\ee
To be fully precise, we include also in $S_{\rm eff}$ counter-terms proportional to the cosmological constant
and Einstein-Hilbert action, as well as quadratic terms:
\be
S_{\rm higher-derivatives} = \int d^dx \left(
\frac{\delta \Lambda+ \delta_R R}{16\pi G}+ c_{R^2} R^2 +\fft{4(d-3)}{d-2} \delta c \big(C^2-E^{(4)}\big) + c_{\rm GB} E^{(4)}\right) \label{Sgrav ders}
\ee
with $C$ the Weyl tensor and $E^{(4)}$ the four-dimensional Euler density  (Gauss-Bonnet term)
\be
E^{(4)}=R_{\mu\nu\rho\sigma}R^{\mu\nu\rho\sigma}-4 R_{\mu\nu}R^{\mu\nu}+R^2\,.
\ee
This parametrization will be especially convenient for conformal matter,
where $\delta c\big|_{\rm div}\sim c$ and $c_{\rm GB}\big|_{\rm div}\sim (a-c)$ with $a$ and $c$ the two central charges in four dimensional conformal field theories \cite{Duff:1977ay,Duff:1993wm}. 

In $d=4$ we stop at quadratic terms in \eqref{Sgrav ders} since their effect has the same parametric dependence on momenta as matter loops, ie. $\Pi(p) \sim p^4$ for self energies. In $d>4$ we include a finite number of additional terms in agreement with the degree of ultraviolet divergences of matter loops (ie. we keep up to ${\rm Riem}^3$ in $d=6$).

The effective action $S_{\rm eff}[g]$ is simply the generating function of renormalized
stress-tensor correlators of a non-gravitational QFT,
with the higher-derivative terms \eqref{Sgrav ders} representing different choices of renormalization schemes.

We will initially work in the following power-counting scheme: we take $c_{R^2}\sim 1/(8\pi G M^2)$ so that the tree-level effect of these terms are of the same order at \emph{and above} the species scale as the one-loop effects from $N$ matter fields.
We initially also assume that higher-derivative terms are further suppressed by a higher scale (ie. $c_{R^3}\sim M^{-2}M'^{-2}$ with 
$M\ll M'\lesssim M_{\rm pl}$ in $d=4$).

The stated power-counting assumptions are self-consistent and they make it reasonable, \emph{a priori}, to calculate amplitudes up to energies $s\lesssim M'^2$ by treating
the effective action \eqref{Seff} at tree-level,
with the 1-loop amplitudes entering it computed ignoring higher-derivative corrections. (We will ultimately see, however, that such a treatment does not make much sense at scales parametrically above $M$ due to pathologies in the effective propagator.)

The first correlator is the one-point function.
As usual when expanding around Minkowski space, we fine-tune the bare cosmological constant $\delta \Lambda$
to exactly cancel the matter loop contribution so that the renormalized vacuum energy density vanishes:
\be
\< 0| T^{\mu\nu}|0\> \equiv 2\frac{\delta S_{\rm eff}}{\delta g_{\mu\nu}}\Big|_{g_{\mu\nu}=\eta_{\mu\nu}} = 0.
\ee
With this choice (and only with this choice!), the two-point function of the effective stress tensor gives a transverse self-energy.
It can be decomposed into scalar and spin-2 parts under SO($d-1$):
\begin{align}
\Sigma^{\mu\nu,\rho\sigma}(p)&\equiv -i \<0|T^{\mu\nu}(p) T^{\rho\sigma}(x=0)|0\>
\nonumber\\
&= \frac{\Pi^{\mu\nu}(p)\Pi^{\rho\sigma}(p)}{d-1} \Sigma^{(0)}(p)+ \left(\frac{\Pi^{\mu\rho}(p)\Pi^{\nu\sigma}(p)+(\rho{\leftrightarrow}\sigma)}{2}
-\frac{\Pi^{\mu\nu}(p)\Pi^{\rho\sigma}(p)}{d-1}\right)\Sigma^{(2)}(p)\,,\label{eq: TT}
\end{align}
where we introduced the transverse projector
\be
\Pi^{\mu\nu}(p)\equiv \eta^{\mu\nu}-p^\mu p^\nu/p^2\,.
\ee
Note that in many formulas here we work in units where $8\pi G=1$, which can be easily restored when necessary.
We represent the self-energy using the following diagram:
\begin{equation}\centering
\raisebox{-0.45\height}{\begin{tikzpicture}
    \pgfdeclarepatternformonly{my custom lines}{\pgfqpoint{-1pt}{-1pt}}{\pgfqpoint{10pt}{10pt}}{\pgfqpoint{9pt}{9pt}}%
    {
        \pgfsetlinewidth{0.4pt}
        \pgfpathmoveto{\pgfqpoint{0pt}{0pt}}
        \pgfpathlineto{\pgfqpoint{9pt}{9pt}}
        \pgfusepath{stroke}
    }

    \node[draw, thick, circle, shading=radial, inner color=gray!30, outer color=white, minimum size=2cm, inner sep=0pt] (vertex) at (0,0) {};

    \draw[decorate, decoration={snake, amplitude=.4mm, segment length=2mm, post length=1mm}] (-2.5,0) -- (vertex.west); 
    \draw[decorate, decoration={snake, amplitude=.4mm, segment length=2mm, post length=1mm}] (vertex.east) -- (2.5,0); 
\end{tikzpicture}}
\end{equation}
As usual in QFTs, we can resum the self-energy to define the full dressed propagator.
This is simply the tree-level two-point function in the theory with action $S_{\rm EH}+S_{\rm eff}$.
In a schematic notation that omits only Lorentz indices, we have (see also the first line of Fig \ref{fig: Gull and Vfull})
\be
G_{\rm full} &= \frac{1}{G_{\rm GR}^{-1}-\Sigma} = G_{\rm GR} + G_{\rm GR}\Sigma G_{\rm GR}+ G_{\rm GR}\Sigma G_{\rm GR}\Sigma G_{\rm GR} + \ldots\,.
\ee

The next correlator is the three-point one, which gives an effective vertex. We will only need it when two legs are on-shell and the third off-shell
\be
 V_{3,\rm full}^{\mu\nu}(p_1,p_2) =  
V_{3,{\rm GR}}^{\mu\nu}(p_1,p_2)+  \<0| (\epsilon_1{\cdot} T(p_1))\, (\epsilon_2{\cdot}T(p_2))\,
 T^{\mu\nu}(x{=}0)|0\>\,,\label{eq: effective vertices definition}
\ee
where $\epsilon_i{\cdot}T$ stands for $\epsilon_{i\mu}\epsilon_{i\nu}T^{\mu\nu}$.
Finally, the last ingredient we will need is the four-point function with all legs on-shell:
\be
 V_{4,\rm full}(p_1,p_2,p_3) =  
V_{4,{\rm GR}}(p_1,p_2,p_3)+ \<0| (\epsilon_{1}{\cdot}T(p_1))\,(\epsilon_{2}{\cdot}T(p_2))\,
 (\epsilon_{3}{\cdot}T(p_3))\,(\epsilon_{4}{\cdot}T(x{=}0))|0\>\,.
\ee
Our diagrammatic notation for the GR and matter contributions is shown in Figure \ref{fig: Gull and Vfull}.

\begin{figure}[tbp]
\centering
\begin{tikzpicture}
 \draw[decorate, decoration={snake, amplitude=.4mm, segment length=2mm, post length=1mm}] (0,0) -- (0.75,0);
 \node[draw, thick, circle,  fill=black, minimum size=0.1cm, inner sep=0pt] (vertex) at (0.75,0) {};
  \draw[decorate, decoration={snake, amplitude=.4mm, segment length=2mm, post length=1mm}] (0.75,0) -- (1.5,0)
  node at (1.9,0) {$=$};
    \draw[decorate, decoration={snake, amplitude=.4mm, segment length=2mm, post length=1mm}] (2.3,0) -- (3.8,0)
    node at (4.2,0) {$+$};
    \draw[decorate, decoration={snake, amplitude=.4mm, segment length=2mm, post length=1mm}] (4.6,0) -- (5.35,0);
     \node[draw, thick, circle, shading=radial, inner color=gray!30, outer color=white, minimum size=0.5cm, inner sep=0pt] (vertex) at (5.5,0) {};
     \draw[decorate, decoration={snake, amplitude=.4mm, segment length=2mm, post length=1mm}] (5.75,0) -- (6.5,0)
      node at (6.9,0) {$+$};
      \draw[decorate, decoration={snake, amplitude=.4mm, segment length=2mm, post length=1mm}] (7.3,0) -- (8.15,0);
       \node[draw, thick, circle, shading=radial, inner color=gray!30, outer color=white, minimum size=0.5cm, inner sep=0pt] (vertex) at (8.3,0) {};
        \draw[decorate, decoration={snake, amplitude=.4mm, segment length=2mm, post length=1mm}] (8.55,0) -- (9.3,0);
        \node[draw, thick, circle, shading=radial, inner color=gray!30, outer color=white, minimum size=0.5cm, inner sep=0pt] (vertex) at (9.45,0) {};
          \draw[decorate, decoration={snake, amplitude=.4mm, segment length=2mm, post length=1mm}] (9.7,0) -- (10.45,0)
           node at (11.25,0) {$+\, \cdots$};
%
\begin{scope}[xshift=-1cm]
\node[draw, thick, circle, fill=black, minimum size=0.1cm, inner sep=0pt] (vertex) at (0.75, -2) {}; 
\draw[decorate, decoration={snake, amplitude=.4mm, segment length=2mm, post length=1mm}] (0, -1.25) -- (0.75, -2); 
\draw[decorate, decoration={snake, amplitude=.4mm, segment length=2mm, post length=1mm}] (0, -2.75) -- (0.75, -2); 
\draw[decorate, decoration={snake, amplitude=.4mm, segment length=2mm, post length=1mm}] (1.5, -2) -- (0.75, -2)
node at (1.9,-2) {$=$}; 
\draw[decorate, decoration={snake, amplitude=.4mm, segment length=2mm, post length=1mm}] (2.3, -1.25) -- (3.05, -2); 
\draw[decorate, decoration={snake, amplitude=.4mm, segment length=2mm, post length=1mm}] (2.3, -2.75) -- (3.05, -2); 
\draw[decorate, decoration={snake, amplitude=.4mm, segment length=2mm, post length=1mm}] (3.8, -2) -- (3.05, -2)
node at (4.2,-2) {$+$};
\node[draw, thick, circle, shading=radial, inner color=gray!30, outer color=white, minimum size=0.5cm, inner sep=0pt] (vertex) at (5.5, -2) {}; 
\draw[decorate, decoration={snake, amplitude=.4mm, segment length=2mm, post length=1mm}] (4.6, -1.25) -- (5.35, -1.8); 
\draw[decorate, decoration={snake, amplitude=.4mm, segment length=2mm, post length=1mm}] (4.6, -2.75) -- (5.35, -2.2); 
\draw[decorate, decoration={snake, amplitude=.4mm, segment length=2mm, post length=1mm}] (6.51, -2) -- (5.75, -2);
\end{scope}
%
\begin{scope}[xshift=7cm,yshift=2cm]
\node[draw, thick, circle, fill=black, minimum size=0.1cm, inner sep=0pt] (vertex) at (0.75, -4) {};
\draw[decorate, decoration={snake, amplitude=.4mm, segment length=2mm, post length=1mm}] (0, -3.25) -- (0.75, -4); 
\draw[decorate, decoration={snake, amplitude=.4mm, segment length=2mm, post length=1mm}] (0, -4.75) -- (0.75, -4);
\draw[decorate, decoration={snake, amplitude=.4mm, segment length=2mm, post length=1mm}] (1.5, -3.25) -- (0.75, -4); 
\draw[decorate, decoration={snake, amplitude=.4mm, segment length=2mm, post length=1mm}] (1.5, -4.75) -- (0.75, -4)
node at (1.9,-4) {$=$};
\draw[decorate, decoration={snake, amplitude=.4mm, segment length=2mm, post length=1mm}] (2.3, -3.25) -- (3.05, -4); 
\draw[decorate, decoration={snake, amplitude=.4mm, segment length=2mm, post length=1mm}] (2.3, -4.75) -- (3.05, -4);
\draw[decorate, decoration={snake, amplitude=.4mm, segment length=2mm, post length=1mm}] (3.8, -3.25) -- (3.05, -4); 
\draw[decorate, decoration={snake, amplitude=.4mm, segment length=2mm, post length=1mm}] (3.8, -4.75) -- (3.05, -4)
node at (4.2,-4) {$+$};
\node[draw, thick, circle, shading=radial, inner color=gray!30, outer color=white, minimum size=0.5cm, inner sep=0pt] (vertex) at (5.5, -4) {};
\draw[decorate, decoration={snake, amplitude=.4mm, segment length=2mm, post length=1mm}] (4.6, -3.25) -- (5.35, -3.8); 
\draw[decorate, decoration={snake, amplitude=.4mm, segment length=2mm, post length=1mm}] (4.6, -4.75) -- (5.35, -4.2);
\draw[decorate, decoration={snake, amplitude=.4mm, segment length=2mm, post length=1mm}] (6.5, -3.25) -- (5.65, -3.8); 
\draw[decorate, decoration={snake, amplitude=.4mm, segment length=2mm, post length=1mm}] (6.51, -4.75) -- (5.65, -4.2);
\end{scope}
\end{tikzpicture}
\caption{Effective Feynman rules obtained from the action $S_{\rm EH}+S_{\rm eff}$, indicated with dots.
We have separated contributions from pure GR and matter loops, the latter containing counter-terms as well as
higher-derivative corrections to GR.}
\label{fig: Gull and Vfull}
\end{figure}
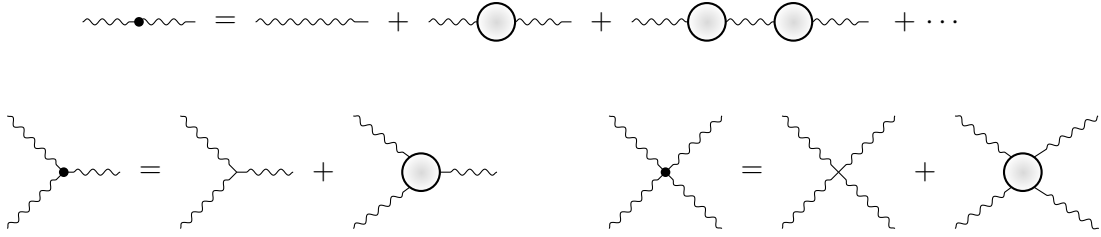

In summary, in terms of ordinary loop diagrams, the four-graviton scattering amplitudes in the ``matter weakly coupled to gravity'' framework takes the general form:
\be \label{M 012 form}
 {\cal M}_4 = {\cal M}_{4,\rm GR} +{\cal M}_{4,\rm 1-loop}+{\cal M}_{4,\geq 2\,\rm loops},
\ee
where the first two terms follow the naive loop expansion (treating higher-derivative corrections to GR as the same size as loop effects) and the last term necessarily involves the resummed propagator and effective vertex \eqref{eq: effective vertices definition}. 

The calculation of these ingredients from various types of matter fields will be detailed in the following subsections.
As mentioned, the resummed propagator will turn out to be pathological above the species scale, so unfortunately we will not find any situation in which the ${\cal M}_{4,\geq 2\, \rm loops}$ term is both sensible and important.

\subsection{Actions for various matter types}\label{ssec: actions}

In this subsection, we spell out the action for the matter fields of various spins that we consider, which all have spin $j\leq 2$:
a scalar, Dirac fermion, 1-form, Rarita-Schwinger spin-3/2, massive graviton, as well as 2-form and 3-form fields.
Since we wanted to consistently obtain the resummed propagator and vertices as well as one-loop amplitudes, we did not use on-shell techniques, but rather used direct Feynman diagrams.

The total action is
\be
S_{\rm matter}=\sum_j n_j S_{j}\,.
\ee
The most familiar actions are those of spin $\leq 1$, which are well behaved in the massless limit
\be
& S_{0}= -\fft{1}{2}\int d^dx\sqrt{g} ((\nabla\phi)^2 + (m^2+ \xi   R) \phi^2)\,,\nn\\
& S_{1/2}= {\rm i} \int d^dx\sqrt{g} \, \bar{\psi}(\slashed{D}+m) \psi\,,\nn\\
& S_{1}=-\fft{1}{2} \int d^dx\sqrt{g}(\tfrac{1}{2}F_{\mu\nu}F^{\mu\nu} + m^2 A_\mu A^\mu)\,.
\ee
Physically, we imagine that for fields of each spin we have a distribution of masses $m$ (ie. organized in a Kaluza-Klein tower) but we suppress this from our notation.  Note that we don't add non-minimal couplings except for $\xi R$ in the scalar case. In principle one could add terms such as $R_{\mu\nu}A^\mu A^\nu$ for a spin-$1$ massive particle, however, we find that such terms give rise to amplitudes that are power divergent in the massless limit $1/m^\#$,
and grow correspondingly faster with energy.  This seems to violate the power-counting scheme articulated in subsection \ref{subsec: ingredients} and for this reason we did not consider such terms further.

For higher spin particles $j=3/2, 2$, it is rather challenging to directly write down appropriate minimal-coupling actions. A shortcut is to consider the Scherk-Schwarz dimension reduction \cite{Scherk:1978ta,Scherk:1979zr} of $d+1$ dimensional massless Rarita-Schwinger and Einstein-Hilbert actions. The minimal couplings of massless spin-$3/2$ and gravitons are uniquely well-defined, and the Scherk-Schwarz dimension reduction serves as a ``geometric'' Higgs mechanism that breaks the gauge symmetry
and thus generates the masses \cite{Scherk:1978ta,Scherk:1979zr}.

For the massless spin-$3/2$ in higher dimension we have
\be
S_{3/2}^{(d+1)}={\rm i} \int d^{d+1}x\sqrt{g}\, \bar{\Psi}_\mu \big(\gamma^{\mu\nu\rho}-\fft{d-1}{4}\gamma^\mu\gamma^\nu\gamma^\rho\big)D_\nu \Psi^\rho + S_{3/2,{\rm FP}}\,.
\ee
The first term is the standard Rarita-Schwinger action with an appropriate gauge-fixing term. On the other hand, $S_{3/2,{\rm FP}}$ refers to the Faddeev-Popov (FP) ghost action, and it behaves like $-3$ times fermion contributions, as shown in appendix \ref{app: ghost}. We then perform the Scherk-Schwarz dimension reduction for $\Psi_\mu$ to generate a tower of massive spin-$3/2$ fields minimally coupled to gravitons in $d$ dimensions.

Massive spin-2 particles have an added subtlety related to the gauge choice.  In order to minimize pain and ensure that the effective action is transverse, we use the background field method in $d+1$ dimensions, distinguishing the field running inside the loop and that outside.  This means that we consider linearized perturbation of Einstein gravity in $d+1$ dimensions around the ``background field''
\be
g_{\mu\nu}=g_{\mu\nu}^{d}+\sqrt{32\pi G} \, h^s_{\mu\nu}\,.
\ee
Then $h^s_{\mu\nu}$ denotes the ``virtual'' graviton, which will be dimensionally reduced to become the massive spin-$2$ particles. The background $g_{\mu\nu}^d = \eta_{\mu\nu} + \sqrt{32\pi G}h_{\mu\nu}$ describes the real gravitons, and it remains as massless gravitons after the dimensional reduction.
Upon adding the background field de-Donder gauge-fixing term, we find
\be
S_2^{d+1}&=-\fft{1}{2}\int d^{d+1}x\sqrt{g}\Big(\nabla_\mu h^s_{\nu\rho}\nabla^\mu h^{s\nu\rho}-\fft{1}{2}\nabla_\mu h^s \nabla^\mu h^s-\big(\fft{1}{2}(h^s)^2-h^s_{\mu\nu}h^{s\mu\nu} \big)R\nn\\
&+2\big(h^s h^{s\mu\nu}+h^{s\nu}_\rho h^{s\rho\mu}\big)R_{\mu\nu}+2h^{s\mu\nu}h^{s\rho\sigma}R_{\mu\rho\nu\sigma}\Big)+S_{2,{\rm FP}}\,,
\ee
where all the covariant derivatives and the curvatures are defined with respect to the background field $g^{d}_{\mu\nu}$. The FP ghost action is also straightforward to obtain
\be
S_{2,{\rm FP}}=-\int d^{d+1}x\sqrt{g}\,\nabla_\mu \bar{c}_\nu \nabla_\rho c_\sigma \big(g^{\mu\rho}g^{\nu\sigma}+g^{\mu\sigma}g^{\nu\rho}-g^{\mu\nu}g^{\rho\sigma}\big)\,.
\ee

It is worth emphasizing that Scherk-Schwarz reduction plays nicely with dimensional regularization and that it is never necessary to work out any reduced action in order to compute amplitudes.
In practice,  we simply perform all numerator algebra using the $(d+1)$-dimensional Lorentz invariant Feynman rules of the massless theory (\emph{not} discarding scale-less integrals), then plug in the $(d+1)$-dimensional loop momentum
$(\ell^\mu,\sqrt{\ell_\perp^2+m^2})$ where $\ell_\perp$ is the conventional $(-2\epsilon)$-dimensional part of the loop momentum in $d{-}2\epsilon$ dimensions.

In higher dimensions $d> 4$, it is also necessary to discuss two-form and three-form fields, which are fully antisymmetric representations with Young-Tableaux $(1,1)$ and $(1,1,1)$.  Note that with the conventional definition of ``spin'' as the length of the first row, these are all spin-1 fields. These are low-spin matter fields that can exist in the low-energy EFT; for example, they play important roles in supergravity.  In $d=4$, they reduce to combinations of scalars and vectors via duality.
To include them, we start with the $(d+1)$ action and perform dimension reduction to generate the masses. The general action for a massless $p$-form $B_p$ is
\be
S_{p}^{d+1}= -\fft{(p+1)^2}{2(p+1)!}\int d^{d+1}x\sqrt{g}\, \nabla_{[\mu} B_{\nu_1\cdots \nu_p]}\nabla^{[\mu} B^{\nu_1\cdots \nu_p]} +S_{gf}+ S_{\rm ghosts}\,.
\ee
We use the covariant derivative to write the action, emphasizing that the $p$-form is coupled to background gravitons.
Again there is a technical subtlety.  A massless $p$-form for $p>1$ is known as a reducible gauge theory because the gauge-fixing condition itself can be further constrained by the background field \cite{Siegel:1980jj,Batalin:1983ggl}. Therefore, the usual FP ghost does not fully capture the gauge redundancy and one needs ghosts for ghosts. We review the counting of ghosts for the $2$-form by closely follwoing \cite{Siegel:1980jj} in appendix \ref{app: ghost}, where we show that the ghost of $2$-form contributes like three scalars.

For $3$-form fields, the ghosts become non-minimally coupled to gravity \cite{Batalin:1983ggl,Kimura:1980aw}, which would substantially increase the difficulty of the analysis. Therefore, we by-passed a direct calculation and instead inferred the $3$-form contribution to one-loop by imposing the supergravity Ward identities and subtracting other matter contributions (see \eqref{susy decomp} below).

When recording one-loop amplitudes below and in ancillary files, an effective data compression device is to use a supersymmetric decomposition \cite{Dunbar:1994bn,Bern:1995db,Bern:1993tz}, since the contributions of whole supermultiplets tends to be significantly simpler than that of individual fields.  In addition, we used duality to simplify the $p$-form results.
In general dimensions we use:
\be
\mathcal{M}_{\mathcal{N}=0}&= {\cal M}_{0}
\nn\\ \mathcal{M}_{\mathcal{N}=1}&=4 {\cal M}_{0} +\fft{4}{d_F} {\cal M}_{\fft{1}{2}}\,,
\nn\\ \mathcal{M}_{\mathcal{N}=4}&=(9-d) {\cal M}_{0} +\fft{8}{d_F} {\cal M}_{\fft{1}{2}} + {\cal M}_{1}\,,\nn\\
 \mathcal{M}_{\mathcal{N}=6}&=4\big(20+\fft{1}{4}d(d-17)\big) {\cal M}_{0} +\fft{4(18-d)}{d_F} {\cal M}_{\fft{1}{2}} -2(d-9) {\cal M}_{1} + \fft{4}{d_F} {\cal M}_{\fft{3}{2}}+2{\cal M}_{(1,1)}\,,\nn\\
\mathcal{M}_{\mathcal{N}=8}&=-\frac{1}{6} (d-10) \big((d-20) d+105\big){\cal M}_{0} +\fft{16}{d_F}(10-d) {\cal M}_{\fft{1}{2}} +\frac{1}{2} (d-11) (d-10) {\cal M}_{1} \nn\\
& \phantom{=}+\fft{16}{d_F}{\cal M}_{\fft{3}{2}}+{\cal M}_{2} + (10-d){\cal M}_{(1,1)}+ {\cal M}_{(1,1,1)}\,,
\nn\\
{\cal M}_{\bar{(1,1)}}&={\cal M}_{(1,1)}+\fft{1}{2}(d-1)(d-4) {\cal M}_{0} - (d-3){\cal M}_{1}\,,\nn\\
 {\cal M}_{\bar{(1,1,1)}}&={\cal M}_{(1,1,1)}+(5-d){\cal M}_{(1,1)} -\frac{1}{6} (d-6) (d-5) (d-1) {\cal M}_{0}+ \frac{1}{2} (d-6) (d-3) {\cal M}_{1}\,,\label{susy decomp}
\ee
where $\mathcal{M}$ here can refer to either the one-loop amplitudes, self-energy or three-point effective vertices;
$d_F={\rm Tr}[1]$ counts the dimension of the spinor representation, for example $d_F=4$ for a Dirac fermion in four dimensions.
We will confirm soon that $\mathcal{N}=8$ $1$-loop amplitudes are indeed simplified to scalar box diagrams \cite{Green:1982sw,Bern:1998ug},
and $\mathcal{N}=6$ amplitudes contain only box and triangle integrals in any dimension. 

The duality-subtracted 2-form ${\cal M}_{\bar{(1,1)}}$ vanishes \emph{in any spacetime dimension} whenever all external momenta and polarizations of the four-graviton process lie within a $4$-dimensional subspace; ${\cal M}_{\bar{(1,1,1)}}$ vanishes when all data lie within a $6$-dimensional subspace. (In general, the coefficients in these relations are obtained by dimensionally reducing from $4$ and $6$ dimensions, respectively.) In particular, this implies that self-energy $\Sigma_{\bar{(1,1)}}$ and  $\Sigma_{\bar{(1,1,1)}}$ are identically zero, as well as the effective vertex $V_{\bar{(1,1,1)}}=0$. This also implies that the one-loop amplitude ${\cal M}_{\bar{(1,1,1)}}\propto {\cal G}$ defined in \eqref{generators dumb}, a fact which eased its extraction using supersymmetry of $\mathcal{M}_{\mathcal{N}=8}$. In addition, we find that ${\cal M}_{\bar{(1,1,1)}}$ contains only box and triangle integrals.

\subsection{Partial wave decomposition}
\label{subsec: partial wave}

A powerful technique to understand the deep physics of loop amplitudes and the nature of their possible UV completion is to analyze them in the partial wave basis. In particular, in section \ref{sec: QFT breaking}, we will use partial wave coefficients to propose a sharp definition of the UV cutoff $\cutoff$.

Properly normalized partial waves for graviton scattering have been computed in \cite{Caron-Huot:2022jli} and we used the ancillary files provided there; we refer to that paper for further details.
Generally the partial wave expansion for $2\rightarrow 2$ graviton scattering takes the form
\be
\mathcal{M}= s^{\fft{4-d}{2}}\sum_\rho n_\rho^{(d)} \sum_{ij} (a_\rho(s))_{ji} \pi_\rho^{ij}(x)\,,\quad n_\rho^{(d)}=\fft{2^{d}(2\pi)^{d-2}{\rm dim}\,\rho}{{\rm vol}\, S^{d-2}}\,.
\ee
where $\rho$ labels all finite-dimensional irreducible representations of the little group SO$(d-1)$ that preserves $p_1 + p_2$. Physically, $\rho$ represents the intermediate states in the center-of-mass frame. Therefore, properly normalized partial waves $\pi_\rho$ can be constructed by gluing the vertices of two gravitons and one heavy particle \cite{Caron-Huot:2022jli}, where different vertices are labeled by $(i,j)$. In this formula, $x=1+2t/s$ is the cosine of the scattering angle, and we normalize the partial waves by
\be
\sum_{\epsilon_4=\epsilon_1^\ast,\epsilon_3=\epsilon_2^\ast}\pi_\rho^{ij}(1) = \delta^{ij}\,.
\ee
In four dimensions, the partial waves of graviton scattering for any helicity configurations are known as the Wigner-d function
\begin{equation} \label{Wigner d}
\pi_J^{h,h'}(x)= d_{h,h'}^J(x) = {\cal N}_{h,h'}^J \left(\frac{1+x}{2}\right)^{\frac{h+ h'}{2}}  \left(\frac{1-x}{2}\right)^{\frac{h-h'}{2}} {}_2F_1\left(h - J, J + h +1; h - h' + 1;\tfrac{1-x}{2}\right)\,,
\end{equation}
where
\begin{equation}
 {\cal N}_{h,h'}^J =  \frac{1}{\Gamma(h-h'+1)} \sqrt{\frac{\Gamma(J + h + 1) \Gamma(J - h' + 1)}{\Gamma(J - h + 1)\Gamma(J + h' + 1) }}\,,\quad d_{h,h'}^J(1) = \delta_{h,h'}\,.
\end{equation}
In higher dimensions, there are more irreducible representations of the little group SO$(d-1)$. For example, there are 20 graviton-graviton-massive vertices for $d \geq 8$. All these vertices are constructed in \cite{Caron-Huot:2022jli}, and were glued to give the partial waves, recorded in the ancillary file therein\footnote{Note that the partial waves recorded in the ancillary file of \cite{Caron-Huot:2022jli} differ by an overall factor ${\rm dim}(J)/{\rm dim}\,\rho$ to ensure \eqref{eq: orth partial wave}.}. For alternative but equivalent constructions, see \cite{Buric:2023ykg}.

We have verified that the partial waves satisfy the orthogonality relation
\be
 \int_{-1}^1 dx (1-x^2)^{\fft{d-1}{2}} {\rm Tr}\left[\pi_\rho^{ij}(x) \pi_{\rho^\prime}^{ij}(x)\right] = \fft{\sqrt{\pi}\, \Gamma\big(\fft{d}{2}-1\big)}{\Gamma\big(\fft{d-1}{2}\big) {\rm dim}\,\rho} \delta_{\rho\rho^\prime}\,,\label{eq: orth partial wave}
\ee
where ``Tr'' stands for a sum over all polarizations.
We have used this in practice to extract the partial wave coefficients $a_\rho$\footnote{
 This method requires calculating the pairing between the 29 tensor structures listed in \eqref{generators dumb}.
 This pairing matrix is available from the authors upon request.}
\be
a_\rho^{ij} = \fft{2^{3-2d} s^{\fft{d-4}{2}} \pi^{\fft{2-d}{2}}}{\Gamma\big(\fft{d-2}{2}\big)}  \int_{-1}^1 dx (1-x^2)^{\fft{d-4}{2}} {\rm Tr}\left[\pi_{\rho}^{ij}(x) \mathcal{M}\right]\,.
\ee

\section{Graviton amplitudes with matter loops}
\label{sec: amplitudes}

\subsection{Master integrals}

In this section, we review the master integrals that will appear in our final amplitudes. The master integrals constitute a minimal set of independent integrals to which all other integrals can be reduced. Although we are considering the nonperturbative resummation due to the enhancement from a large-$N$ number of particles, all the loop ingredients are limited to just one loop. Therefore, the master integrals are enumerated by the tadpole, bubbles, triangles and boxes topologies.
Permuting external momenta produces more sub-topologies. Generally, we can define the inverse of the Feynman propagator by
\be
D(\ell,p)=(\ell-p)^2+m^2\,.
\ee
The one-loop family is
\be
G_{a_0 a_1 a_2 a_3}(q_1,q_2,q_3)=\int \fft{d^d\ell}{(2\pi)^d} \fft{1}{D(\ell,0)^{a_0}D(\ell,q_1)^{a_1}D(\ell,q_2)^{a_2}D(\ell,q_3)^{a_3}}\,,
\ee
where we use $q_i$ to distinguish with the external momenta $p_i$. If $a_i=0$, we simply drop it out in the arguments. 
Eventually, the integrals only depend on the external momenta, especially as functions of Mandelstam variables. We then define
\be
& G_{1000}=I_{\rm tad}(m)\,,\quad G_{1100}(p_1+p_2)=I_{\rm bub}(m,s)\,,\quad G_{1110}(p_1,-p_2)=G_{1110}(p_3,-p_4)=I_{\rm tri}(m,s)\,,\nn\\
& G_{1111}(p_1,p_1+p_2,-p_4)=I_{\rm box}(m,s,t)\,,
\ee
and permutations therein. It is important to note that the box integral is defined in the Euclidean region $s<0, t<0$, and we should analytically continue it to the physical region.

In $d=4$, all master integrals are solved by the massive differential equation (see \cite{Caron-Huot:2014lda} and references therein) in the dimensional regularization $d\rightarrow 4-2\epsilon$ and $\bar{{\rm MS}}$ scheme $\mubar^2=\mu^2 e^{\gamma_E}/(4\pi)$. We simply collect the results here up to $O(\epsilon)$ terms:
\be
\mu^{2\epsilon} I_{\rm tad}(m) &= -\fft{1}{16\pi^2} m^2\Big(\fft{1}{\epsilon}+1+\log\big(\fft{\mubar^2}{m^2}\big)\Big)\,,\nn\\
I_{\rm tri}(m,s)&=-\fft{1}{32\pi^2  s} \log^2 \Big(\fft{\beta(s,m)-1}{\beta(s,m)+1}\Big)\,, \nn\\
\mu^{2\epsilon} I_{\rm bub}(m,s)&=\fft{1}{16\pi^2}\Big(\fft{1}{\epsilon}+2 + \log\Big(\fft{\mubar^2}{m^2}\Big) + \log \Big(\fft{\beta(s,m)-1}{\beta(s,m)+1}\Big)\beta(s,m) \Big)\,,\nn\\
 I_{\rm box}(m,s,t)&=-\frac{1}{16\pi^2 s t \, \beta(s,t,m) } \Big(\pi^2-4 \log ^2\Big(\frac{\beta(s,t,m)+\beta(s,m)}{\beta(s,t,m)+\beta(t,m)}\Big)\nn\\ 
&\quad-2 \log \Big(\frac{\beta(s,t,m)- \beta(s,m)}{\beta(s,t,m)+\beta(s,m)}\Big) \log \Big(\frac{\beta(s,t,m)- \beta(t,m)}{\beta(s,t,m)+\beta(t,m)}\Big)\nn\\
&\quad+2\Big\{ \log ^2\Big(\frac{\beta(s,m)+1}{\beta(s,t,m)+\beta(s,m)}\Big)+2 \text{Li}_2\Big(\frac{\beta(s,m)-\beta(s,t,m)}{\beta(s,m)+1}\Big)\nn\\
&\quad\quad-2 \text{Li}_2\Big(\frac{\beta(s,m)-1}{\beta(s,m)+\beta(s,t,m)}\Big)
 + \big(s\leftrightarrow t\big)\Big\} \Big)\,,
\ee
where
\be
\beta(s,m)=\Big(1-\fft{4m^2}{s}\Big)^{\fft{1}{2}}\,,\quad \beta(s,t,m)=\Big(1-\fft{4m^2}{s}-\fft{4m^2}{t}\Big)^{\fft{1}{2}}\,.
\ee
For even higher dimensions, we then use the dimensional shift formula to link the master integral to four dimensional solutions
\begin{equation}\begin{aligned}
I_{\rm tad}^{(d)}(m) &= -\fft{m^2}{2\pi (d-2)} I_{\rm tad}^{(d-2)}(m)\,,
\\
I_{\rm bub}^{(d)}(m,s)&=\fft{1}{4\pi (d-3)} \Big(I_{\rm tad}^{(d-2)}(m)-\fft{4m^2-s}{2} I_{\rm bub}^{(d-2)}(m,s)\Big)\,,\\
I_{\rm tri}^{(d)}(m,s)&=\fft{1}{4\pi(d-4)}\Big(I_{\rm bub}^{(d-2)}(m,s)-2m^2 I_{\rm tri}^{(d-2)}(m,s)\Big)\,,\\
I_{\rm box}^{(d)}(m,s,t)&=-\fft{1}{4\pi (d-5) u}\Big(s I_{\rm tri}^{(d-2)}(m,s)+t  I_{\rm tri}^{(d-2)}(m,t) + \fft{s t+4m^2 u}{2} I_{\rm box}^{(d-2)}(m,s,t) \Big)\,.
\end{aligned}\end{equation}
We do not have access to the integral formulas in generic odd dimensions. Nevertheless, for our purposes in section \ref{sec: QFT breaking}, only the imaginary part of the master integrals is necessary, which can be obtained analytically in arbitrary dimensions, as recorded in appendix \ref{app: branch}.

\subsection{Gravitational self-energy}

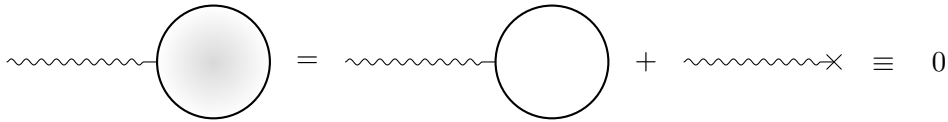
\begin{figure}[htbp]
\centering
\begin{tikzpicture}
  \draw[decorate, decoration={snake, amplitude=.4mm, segment length=2mm, post length=1mm}]
    (-4.5,0) -- (-2.5,0) 
    node at (-0.5,0) {$=$};
     \node[draw, thick, circle, shading=radial, inner color=gray!30, outer color=white, minimum size=1.5cm, inner sep=0pt] (vertex) at (-1.75,0) {};
  \draw[decorate, thick, decoration={amplitude=.4mm, segment length=2mm, post length=1mm}]
    (2.75,0) circle (0.75 cm);

  \draw[decorate, decoration={snake, amplitude=.4mm, segment length=2mm, post length=1mm}]
    (0,0) -- (2,0)
    node at (4,0) {$+$};
     \draw[decorate, decoration={snake, amplitude=.4mm, segment length=2mm, post length=1mm}]
    (4.5,0) -- (6.5,0);
   \draw (6.5-0.1, 0-0.1) -- (6.5+0.1, 0+0.1);
\draw (6.5-0.1, 0+0.1) -- (6.5+0.1, 0-0.1)
node at (7.5,0) {$\equiv\quad 0$};
\end{tikzpicture}
\caption{The tadpole one-point stress-tensor correlators and the counterterm from the cosmological constant are considered together. The net result is identically zero, ensuring that the graviton one-loop vanishes in the vacuum.}
\label{fig: tadpole}
\end{figure}

Now we start to compute the gravitational one-loop self-energy. As noted in section \ref{sec: general}, there are non-vanishing tadpole one-point stress-tensor correlators for massive particles, as shown in Figure \ref{fig: tadpole}. These tadpoles should be absorbed by the counterterm $\delta \Lambda$. A simple exercise yields $\delta\Lambda=(8\pi G)^2 \sum_{j}n_j \delta\Lambda_j$ with
\be
\delta\Lambda_j =&\Big\{-\fft{m_{0}^2}{d}I_{\rm tad}(m_0),\fft{m_{1/2}^2}{d}I_{\rm tad}(m_{1/2}),-\fft{(d-1)m_1^2}{d}I_{\rm tad}(m_1),\fft{(d-2)m_{3/2}^2}{d}I_{\rm tad}(m_{3/2}),\nn\\
& -\fft{(d^2-d-2)m_{2}^2}{2d}I_{\rm tad}(m_2),-\fft{(d-2)(d-1)m_{(1,1)}^2}{2d} I_{\rm tad}(m_{(1,1)}),\nn\\
&-\fft{(d-3)(d-2)(d-1)m_{(1,1,1)}^2}{6d} I_{\rm tad}(m_{(1,1,1)})\Big\}\,.
\ee

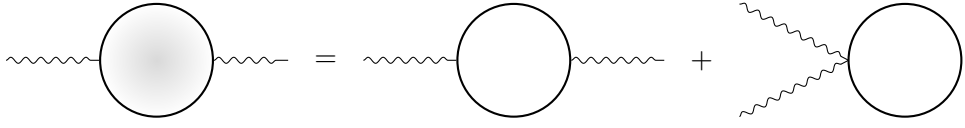
\begin{figure}[tbp]
\centering
\begin{tikzpicture}
  \node[draw, thick, circle, shading=radial, inner color=gray!30, outer color=white, minimum size=1.5cm, inner sep=0pt] (vertex) at (-4.75,0) {};
   \draw[decorate, decoration={snake, amplitude=.4mm, segment length=2mm, post length=1mm}] (-6.75,0) -- (vertex.west) ;
 \draw[decorate, decoration={snake, amplitude=.4mm, segment length=2mm, post length=1mm}] (vertex.east) -- (-3,0)  node at (-2.5,0) {$=$};
    \node[draw, thick, circle, minimum size=1.5cm, inner sep=0pt] (vertex) at (0,0) {};

    \draw[decorate, decoration={snake, amplitude=.4mm, segment length=2mm, post length=1mm}] (-2,0) -- (vertex.west); 
    \draw[decorate, decoration={snake, amplitude=.4mm, segment length=2mm, post length=1mm}] (vertex.east) -- (2,0)
    node at (2.5,0) {$+$};
      \draw[decorate, decoration={snake, amplitude=.4mm, segment length=2mm, post length=1mm}] (3,0.75) -- (4.45,0);
       \draw[decorate, decoration={snake, amplitude=.4mm, segment length=2mm, post length=1mm}] (3,-0.75) -- (4.45,0);
       \node[draw, thick, circle, minimum size=1.5cm, inner sep=0pt] (vertex) at (5.2,0) {};
      
       \end{tikzpicture}
\caption{The Feynman diagrams for graviton self-energy involve both bubble and tadpole topologies, because there are both cubic and quartic vertices involving gravitons and matter. The solid loops represent matters.}
\label{fig: self-energy Feynman}
\end{figure}

There are two Feynman diagrams contributing to the self-energy because we have both cubic and quartic vertices, as shown in Fig \ref{fig: self-energy Feynman}. The resulting integrands are intricate, as they involve tensor products of loop momentum with the polarizations. We then perform the Passarino-Veltman (PV) reduction \cite{Passarino:1978jh} and apply Integration by Parts (IBP) \cite{Chetyrkin:1981qh} to reduce the integrals to master integrals with rational coefficients.\footnote{Although public codes exist to generate Feynman rules with gravity and manipulate them \cite{Latosh:2024lhl}, we used our own implementation.} 
To simplify the analysis, it is beneficial to choose a renormalization condition so that the self-energy does not disrupt low-energy measurement of $G$. This requires that the low-energy expansion of the self-energy starts at an order higher than $p^2$. We then find $\delta_R=(8\pi G)^2 \sum_j n_j (\delta_{R})_j$, where
\be
(\delta_R)_j=& \Big\{\fft{1}{6}(6\xi -1) I_{\rm tad}(m_0),-\fft{1}{12}I_{\rm tad}(m_{1/2}),\fft{1}{6}(7-d) I_{\rm tad}(m_{1}),-\fft{1}{12}(d-2)I_{\rm tad}(m_{3/2}),\nn\\
& \fft{1}{12}(5d^2+7d+26)I_{\rm tad}(m_2),-\fft{1}{12}(d^2-15d+38)I_{\rm tad}(m_{(1,1)}),\nn\\
& -\fft{1}{36}(d-3)(d^2-21d+74)I_{\rm tad}(m_{(1,1,1)})\Big\}\,.
\ee

Consistently, the results organize into the transverse form \eqref{eq: TT}. For example, for scalar loops, we find
\be
 \Sigma_0^{(0)}(p^2)&= -\frac{(d-2) \left(24 d p^2 \xi -24 p^2 \xi -5 d p^2+12 m_0^2+8 p^2\right)}{12 (d-1)} I_{\text{tad}}\left(m_0\right)\nn\\
&\phantom{=} -\frac{ \left(4 d p^2 \xi -4 p^2 \xi -d p^2+4 m_0^2+2 p^2\right)^2}{8 (d-1)}I_{\text{bub}}\left(m_0,-p^2\right)\,,\nn\\
 \Sigma_0^{(2)}(p^2)&=-\frac{(d-2) \left(d p^2+12 m_0^2+2 p^2\right)}{6 (d-1) (d+1)}I_{\text{tad}} \left(m_0\right)-\frac{\left(4 m_0^2+p^2\right)^2}{4 (d-1) (d+1)} I_{\text{bub}}\left(m_0,-p^2\right)\,.
\ee
This, as well as the corrections from fermions, vectors, and the graviton itself, has been computed a long time ago \cite{Capper:1973pv,Capper:1973mv,Capper:1973bk,Capper:1974ed,DeMeyer:1974ed,tHooft:1974toh}; see also \cite{Burns:2014bva} for a recent revisit. However, to the best of our knowledge, we did not find the individual self-energy from the gravitino loop. We record the self-energy from other fields in the ancillary file. 
As will be further discussed below, however, the contributions from gravitino and graviton loops to the self-energy are not gauge invariant and have unclear physical significance.

\subsection{Transverse effective vertices}

This organization in section \ref{subsec: ingredients} is standard but is not quite optimal.
The matter contribution to $V_{3,\rm full}$ is not transverse with respect to $\epsilon_1$ nor $\epsilon_2$,
since Ward identities relate its divergence to the self-energy $\Sigma$. 
Similar Ward identities apply to $V_{3,\rm GR}$, which is not transverse when one of its legs is off-shell.
However, it is possible to combine these into a ``nice'' vertex which is transverse
with respect to both $\epsilon_1$ and $\epsilon_2$:
\be
 V_{3,\rm T}=V_{3,\rm full} - V_{3,\rm GR}(1-G_{\rm GR}\Sigma)\,.\label{eq: Vtras}
\ee
We can illustrate this definition as:
\begin{equation}
\centering
\raisebox{-0.45\height}{
\begin{tikzpicture}
\node[draw, thick, circle, fill=black, minimum size=0.1cm, inner sep=0pt] (vertex) at (0.75, 0) {}; 
\draw[decorate, decoration={snake, amplitude=.4mm, segment length=2mm, post length=1mm}] (0, 0.75) -- (0.75, 0); 
\draw[decorate, decoration={snake, amplitude=.4mm, segment length=2mm, post length=1mm}] (0, -0.75) -- (0.75, 0)
node at (0.9,0.3) {$T$} ; 
\draw[decorate, decoration={snake, amplitude=.4mm, segment length=2mm, post length=1mm}] (1.5, 0) -- (0.75, 0)
node at (1.9,0) {$=$}; 
 \node[draw, thick, circle, shading=radial, inner color=gray!30, outer color=white, minimum size=0.4cm, inner sep=0pt] (vertex) at (3.05,0) {};
\draw[decorate, decoration={snake, amplitude=.4mm, segment length=2mm, post length=1mm}] (2.3, 0.75) -- (3.05-0.14, 0.14); 
\draw[decorate, decoration={snake, amplitude=.4mm, segment length=2mm, post length=1mm}] (2.3, -0.75) -- (3.05-0.14, -0.14); 
\draw[decorate, decoration={snake, amplitude=.4mm, segment length=2mm, post length=1mm}] (3.8, 0) -- (3.05+0.2, 0)
node at (4.2,0) {$+$};
\draw[decorate, decoration={snake, amplitude=.4mm, segment length=2mm, post length=1mm}] (4.6, 0.75) -- (5.35, 0); 
\draw[decorate, decoration={snake, amplitude=.4mm, segment length=2mm, post length=1mm}] (4.6, -0.75) -- (5.35, 0); 
\draw[decorate, decoration={snake, amplitude=.4mm, segment length=2mm, post length=1mm}] (6.1, 0) -- (5.35, 0);
\node[draw, thick, circle, shading=radial, inner color=gray!30, outer color=white, minimum size=0.4cm, inner sep=0pt] (vertex) at (6.1, 0) {}; 
\draw[decorate, decoration={snake, amplitude=.4mm, segment length=2mm, post length=1mm}] (6.85, 0) -- (6.3, 0);
\end{tikzpicture}}
\end{equation}

Notice that the parenthesis in \eqref{eq: Vtras} is the inverse of the \emph{full} propagator.
This decomposition enables us to isolate terms that require the resummed propagator, and those that involve the ordinary GR propagator:
\be\begin{aligned} \label{eq: VGV_full}
V_{3,\rm full} G_{\rm full} V_{3,\rm full} &= V_{3,\rm GR}G_{\rm GR}V_{3,\rm GR}
 \\ &\phantom{=}-V_{3,\rm GR}G_{\rm GR}\Sigma G_{\rm GR}V_{3,\rm GR} + V_{3,\rm GR}G_{\rm GR} V_{3,\rm T}+V_{3,\rm T}G_{\rm GR} V_{3,\rm GR}
 \\ &\phantom{=} +V_{3,\rm T} G_{\rm full} V_{3,\rm T}\,.
\end{aligned}\ee
Each line has a simple interpretation. The first line is the pure GR result (without any higher-derivative correction),
the second line collects all one-loop corrections to it, while the last line collects all terms with two or more loops, exactly the general form anticipated in \eqref{M 012 form}. 
Note that the term with the explicit self-energy has a minus sign, to avoid double-counting the self-energies includes in $V_{3,T}$. The total can be depicted as:
\begin{equation}\centering\raisebox{-0.45\height}{
\begin{tikzpicture}
\node[draw, thick, circle, fill=black, minimum size=0.1cm, inner sep=0pt] (vertex) at (0.75, 0) {}; 
\node[draw, thick, circle, fill=black, minimum size=0.1cm, inner sep=0pt] (vertex) at (1.5, 0) {}; 
\node[draw, thick, circle, fill=black, minimum size=0.1cm, inner sep=0pt] (vertex) at (1.125, 0) {}; 
\draw[decorate, decoration={snake, amplitude=.4mm, segment length=2mm, post length=1mm}] (0, 0.75) -- (0.75, 0); 
\draw[decorate, decoration={snake, amplitude=.4mm, segment length=2mm, post length=1mm}] (0, -0.75) -- (0.75, 0) ; 
\draw[decorate, decoration={snake, amplitude=.4mm, segment length=2mm, post length=1mm}] (1.5, 0) -- (0.75, 0);
\draw[decorate, decoration={snake, amplitude=.4mm, segment length=2mm, post length=1mm}] (2.25, 0.75) -- (1.5, 0); 
\draw[decorate, decoration={snake, amplitude=.4mm, segment length=2mm, post length=1mm}] (2.25, -0.75) -- (1.5, 0)
node at (2.45,0) {$=$}; 
\draw[decorate, decoration={snake, amplitude=.4mm, segment length=2mm, post length=1mm}] (2.65, 0.75) -- (3.4, 0); 
\draw[decorate, decoration={snake, amplitude=.4mm, segment length=2mm, post length=1mm}] (2.65, -0.75) -- (3.4, 0) ; 
\draw[decorate, decoration={snake, amplitude=.4mm, segment length=2mm, post length=1mm}] (4.15, 0) -- (3.4, 0);
\draw[decorate, decoration={snake, amplitude=.4mm, segment length=2mm, post length=1mm}] (4.9, 0.75) -- (4.15, 0); 
\draw[decorate, decoration={snake, amplitude=.4mm, segment length=2mm, post length=1mm}] (4.9, -0.75) -- (4.15, 0)
node at (5.1,0) {$-$}; 
 \node[draw, thick, circle, shading=radial, inner color=gray!30, outer color=white, minimum size=0.25cm, inner sep=0pt] (vertex) at (6.425,0) {};
\draw[decorate, decoration={snake, amplitude=.4mm, segment length=2mm, post length=1mm}] (5.3, 0.75) -- (6.05, 0); 
\draw[decorate, decoration={snake, amplitude=.4mm, segment length=2mm, post length=1mm}] (5.3, -0.75) -- (6.05, 0) ; 
\draw[decorate, decoration={snake, amplitude=.4mm, segment length=2mm, post length=1mm}] (6.3, 0) -- (6.05, 0);
\draw[decorate, decoration={snake, amplitude=.4mm, segment length=2mm, post length=1mm}] (6.8, 0) -- (6.55, 0);
\draw[decorate, decoration={snake, amplitude=.4mm, segment length=2mm, post length=1mm}] (7.55, 0.75) -- (6.8, 0); 
\draw[decorate, decoration={snake, amplitude=.4mm, segment length=2mm, post length=1mm}] (7.55, -0.75) -- (6.8, 0)
node at (7.75,0) {$+$};
 \node[draw, thick, circle, fill=black, minimum size=0.1cm, inner sep=0pt] (vertex) at (8.7,0) {};
\draw[decorate, decoration={snake, amplitude=.4mm, segment length=2mm, post length=1mm}] (7.95, 0.75) -- (8.7, 0)
node at (8.85,0.3) {$T$} ;
\draw[decorate, decoration={snake, amplitude=.4mm, segment length=2mm, post length=1mm}] (7.95, -0.75) -- (8.7, 0);
\draw[decorate, decoration={snake, amplitude=.4mm, segment length=2mm, post length=1mm}] (9.45, 0) -- (8.7, 0);
\draw[decorate, decoration={snake, amplitude=.4mm, segment length=2mm, post length=1mm}] (10.2, 0.75) -- (9.45, 0);
\draw[decorate, decoration={snake, amplitude=.4mm, segment length=2mm, post length=1mm}] (10.2, -0.75) -- (9.45, 0)
node at (10.4,0) {$+$};
\node[draw, thick, circle, fill=black, minimum size=0.1cm, inner sep=0pt] (vertex) at (12.1,0) {};
\draw[decorate, decoration={snake, amplitude=.4mm, segment length=2mm, post length=1mm}] (10.6, 0.75) -- (11.35, 0);
\draw[decorate, decoration={snake, amplitude=.4mm, segment length=2mm, post length=1mm}] (10.6, -0.75) -- (11.35, 0);
\draw[decorate, decoration={snake, amplitude=.4mm, segment length=2mm, post length=1mm}] (12.1, 0) -- (11.35, 0);
\draw[decorate, decoration={snake, amplitude=.4mm, segment length=2mm, post length=1mm}] (12.85, 0.75) -- (12.1, 0);
\draw[decorate, decoration={snake, amplitude=.4mm, segment length=2mm, post length=1mm}] (12.85, -0.75) -- (12.1, 0)
node at (11.95,0.3) {$T$} 
node at (13.05,0) {$+$};
\node[draw, thick, circle, fill=black, minimum size=0.1cm, inner sep=0pt] (vertex) at (0.75+13.25, -2+2) {}; 
\node[draw, thick, circle, fill=black, minimum size=0.1cm, inner sep=0pt] (vertex) at (1.5+13.25, -2+2) {}; 
\node[draw, thick, circle, fill=black, minimum size=0.1cm, inner sep=0pt] (vertex) at (1.125+13.25, -2+2) {}; 
\draw[decorate, decoration={snake, amplitude=.4mm, segment length=2mm, post length=1mm}] (13.25, -1.25+2) -- (0.75+13.25, -2+2); 
\draw[decorate, decoration={snake, amplitude=.4mm, segment length=2mm, post length=1mm}] (13.25, -2.75+2) -- (0.75+13.25, -2+2) ; 
\draw[decorate, decoration={snake, amplitude=.4mm, segment length=2mm, post length=1mm}] (1.5+13.25, -2+2) -- (0.75+13.25, -2+2);
\draw[decorate, decoration={snake, amplitude=.4mm, segment length=2mm, post length=1mm}] (2.25+13.25, -1.25+2) -- (1.5+13.25, -2+2); 
\draw[decorate, decoration={snake, amplitude=.4mm, segment length=2mm, post length=1mm}] (2.25+13.25, -2.75+2) -- (1.5+13.25, -2+2)
node at (0.7+13.25+0.15,-2+0.3+2) {$T$} 
node at (1.5+13.25,-2+0.3+2) {$T$} ;
\end{tikzpicture}}
\end{equation}

The other diagram contributing to the four-graviton amplitude is simply the four-point effective vertex, which involves no resummation.

The transverse vertices include Feynman diagrams for one-loop contributions to the three-point vertices, with triangle, bubble and the tadpole topologies:
\begin{equation}
\centering
\begin{tikzpicture}
 \node[draw, thick, circle, shading=radial, inner color=gray!30, outer color=white, minimum size=1.5cm, inner sep=0pt] (vertex) at (-2.7,0) {};
 \draw[decorate, decoration={snake, amplitude=.4mm, segment length=2mm, post length=1mm}] (-0.75-0.3-2.7,1.5) -- (0-0.3-2.7,0.7); 
 \draw[decorate, decoration={snake, amplitude=.4mm, segment length=2mm, post length=1mm}] (-0.75-0.3-2.7,-1.5) -- (0-0.3-2.7,-0.7); 
 \draw[decorate, decoration={snake, amplitude=.4mm, segment length=2mm, post length=1mm}] (-1.95,0) -- (-0.85,0)
 node at (-0.55,0) {$=$};
 
\draw[decorate, thick] (0,0.75) -- (0,-0.75);
\draw[decorate, thick] (0,0.75) -- (1.3,0);
\draw[decorate, thick] (0,-0.75) -- (1.3,0);
 \draw[decorate, decoration={snake, amplitude=.4mm, segment length=2mm, post length=1mm}] (-0.75,1.5) -- (0,0.75); 
 \draw[decorate, decoration={snake, amplitude=.4mm, segment length=2mm, post length=1mm}] (-0.75,-1.5) -- (0,-0.75); 
 \draw[decorate, decoration={snake, amplitude=.4mm, segment length=2mm, post length=1mm}] (1.3,0) -- (2.4,0)
 node at (2.7,0) {$+$};
  \draw[decorate, decoration={snake, amplitude=.4mm, segment length=2mm, post length=1mm}] (3,1) -- (3+1,0); 
 \draw[decorate, decoration={snake, amplitude=.4mm, segment length=2mm, post length=1mm}] (3,-1) -- (3+1,0); 
 \node[draw, thick, circle, minimum size=1.5cm, inner sep=0pt] (vertex) at (3+1+1.5/2,0) {};
 \draw[decorate, decoration={snake, amplitude=.4mm, segment length=2mm, post length=1mm}] (3+1+1.5,0) -- (3+1+1.5+1.1,0)
  node at (6.9,0) {$+$};
  \draw[decorate, decoration={snake, amplitude=.4mm, segment length=2mm, post length=1mm}] (7.2,0) -- (7.2+1,0); 
  \draw[decorate, decoration={snake, amplitude=.4mm, segment length=2mm, post length=1mm}] (7.2,1) -- (7.2+1,0); 
 \draw[decorate, decoration={snake, amplitude=.4mm, segment length=2mm, post length=1mm}] (7.2,-1) -- (7.2+1,0); 
 \node[draw, thick, circle, minimum size=1.5cm, inner sep=0pt] (vertex) at (7.2+1+1.5/2,0) {};
    \end{tikzpicture}
\end{equation}

We then construct the transverse vertices using \eqref{eq: Vtras}. The transverse vertices are then organized into the following structures
\be
 \tilde{V}_{3,\mu\nu}&= \fft{H_{12}^2}{4}  \left( P^2 \eta _{\mu \nu} -P_\mu P_\nu \right) \, C^{(0)}(P^2)+\fft{H_{12}^2}{4} \left(\frac{P^2  \eta _{\mu \nu }-P_\mu P_\nu}{d-1}+\left(p_{1\mu }-p_{2\mu }\right) \left(p_{1\nu }-p_{2\nu }\right)\right)\,C^{(2A)}(P^2)\nn\\
&\quad +\frac{1}{2}H_{12} \Big(\fft{H_{12}}{2(d-1)P^2}\big((d-4)P_\mu P_\nu + 3 P^2 \eta_{\mu\nu}\big)+\fft{P^2}{2}\epsilon_{1(\mu}\epsilon_{2\nu)}+\epsilon_1\cdot\epsilon_2\, p_{1(\mu}p_{2\nu)}\nn\\
&\quad -p_2\cdot\epsilon_1 p_{1(\mu}\epsilon_{2\nu)}-p_1\cdot\epsilon_2 p_{2(\mu}\epsilon_{1\nu)}\Big) C^{(2B)}(P^2)\,.
\ee
where $P=p_1+p_2$ and $a_{(\mu}b_{\nu)}=a_{\mu}b_\nu+a_\nu b_\mu$. 

We present our result for scalar loops, while leaving other species to the ancillary file. For scalar loops, we find
\begin{subequations}
\begin{align}
 C_0^{(0)}(p^2)  &= \frac{(d-4) (d-2) \left(-p^2 \left(d \left(4 \xi -1\right)-4 \xi +2\right)-4 m_{0}^2\right)}{2 (d-1)^2 d p^6} I_{\text{tad}}\left(m_{0}\right) \nonumber \\
& \quad + \frac{(d-4) \left(4 (5 d-6) m_{0}^2-(d-2)^2 p^2\right)\left(-p^2 \left(d \left(4 \xi -1\right)-4 \xi +2\right)-4 m_{0}^2\right)}{4 (d-2) (d-1)^2 d p^6} I_{\text{bub}}\left(m_{0},-p^2\right) \nonumber \\
& \quad -\frac{16 m_{0}^4 \left(4 m_{0}^2-p^2 \left(-4 d \xi +4 \xi +d-2\right)\right)}{(d-2) (d-1) d p^6} I_{\text{tri}}\left(m_{0},-p^2\right), \\
C_{0}^{(2A)}(p^2)  &= \frac{(d-2)\left(12 \left(3 d^2+10 d+2\right) m_{0}^2+(d-4) p^2\right)}{(d-1) d (d+1) (d+2) p^6} I_{\text{tad}}\left(m_{0}\right) \nonumber \\
& \quad -\frac{\left(8 d \left(d^2-12 d+32\right) p^2 m_{0}^2-48 \left(d^3+12 d^2-16 d-12\right) m_{0}^4+(d-4) (d-2)^2 p^4\right)}{2 d \left(d^4-5 d^2+4\right) p^6} I_{\text{bub}}\left(m_{0},-p^2\right) \nonumber \\
& \quad -\frac{16 m_{0}^4 \left(12 m_{0}^2-(d+2) p^2\right) }{d \left(d^2-4\right) p^6}I_{\text{tri}}\left(m_{0},-p^2\right), \\
 C_{0}^{(2B)}(p^2)  &= \frac{2 (d-2)\left(4 \left(d^2+6 d+2\right) m_{0}^2+d p^2\right)}{(d-1) d (d+1) (d+2) p^4} I_{\text{tad}}\left(m_{0}\right) \nonumber \\
& \quad +\frac{\left(8 \left(d^3-6 d^2+6 d+4\right) p^2 m_{0}^2+16 \left(d^3-12 d^2+8 d+12\right) m_{0}^4+(d-2)^2 d p^4\right)}{d \left(d^4-5 d^2+4\right) p^4} I_{\text{bub}}\left(m_{0},-p^2\right) \nonumber \\
& \quad + \frac{128 m_{0}^6 }{d \left(d^2-4\right) p^4}I_{\text{tri}}\left(m_{0},-p^2\right)\,.
\end{align}
\end{subequations}

\subsection{Results in $d=4$ and the conformal limit}
\label{ssec: d4 amps}

Let's first apply our building blocks to compute the massless nonperturbative amplitudes in $d=4$. In $d=4$, a massive $2$-form is dual to a vector, and similarly a $3$-form is dual to a scalar. Therefore, it is only necessary to consider standard low-spin matters. To keep the discussion short, we only present external $+--+$ helicity configuration, where we get for the $f$ function defined in \eqref{eq:Mdef}
\be
 f(s,u) = \frac{8\pi G}{stu} + \sum_{\mathcal{N}=0,1,4,6,8} n_{j} f^{\rm 1-loop,GR}_j (s,u)
+\frac{3}{16}\frac{t^2 C^{(0)}(-t)^2}{2t+\Sigma_0(-t)}
+\frac{(6su-t^2)C^{(2A)}(-t)^2}{24(t-\Sigma_2(-t))}\,.\label{eq: f in 4D}
\ee
We refer to $\mathcal{N}=0$ as the scalar case. The structures in vertices $C$ should be understood as those with summation over species. Note that $C^{(2B)}$ does not contribute for external momenta and polarizations in $d=4$, since it is a Gauss-Bonnet type structure.

In the massless limit, the master integrals are simple with the dimensional regularization $d=4-2\epsilon$ of UV divergence
\be
& I_{\rm tad}(0)=0\,,\quad I_{\rm bub}(0,s)=\fft{1}{(4\pi)^2}\Big(\fft{1}{\epsilon}+2+\log\Big(\fft{\mubar^2}{-s}\Big)\Big)\,,\quad \lim_{m\rightarrow 0}I_{\rm tri}(m,s)=-\fft{\log^2\big(\fft{m^2}{-s}\big)}{32 \pi^2 s}\,,\nn\\
& \lim_{m\rightarrow 0} I_{\rm box}(m,s,t)=\fft{1}{(4\pi)^2 st}\Big(2\log\Big(\fft{m^2}{-s}\Big)\log\Big(\fft{m^2}{-t}\Big)-\pi^2\Big)\,.
\ee
It is important to note that there are IR divergences in the triangles and boxes, where the mass is naturally served as the IR regulator.  To have well-defined amplitudes, we have to appropriately perform the renormalization to absorb all the UV $1/\epsilon$ divergence. We simply choose a convenient scheme, which redefines the couplings $(c_{R^2},\delta c,c_{\rm GB})$ to remove all the divergences and absorb all the constant terms in $(\Sigma^{(0)}, \Sigma^{(2)}, C^{(2B)})$.
It is straightforward to explicitly write down the self-energy and the effective vertices
\be
& \Sigma^{(0)}(p^2)=-\fft{8\pi G p^4}{6}\Big(144 c_{R^2}^{\rm ren}+ \fft{1}{16\pi^2}\log\Big(\fft{\mubar^2}{p^2}\Big)\big((1-6\xi)^2 n_0+n_1-16n_{\fft{3}{2}}+83n_2\big)\Big)\,,\nn\\
& \Sigma^{(2)}(p^2)=-16\pi G p^4 \big(\delta c^{\rm ren}+ c \log\Big(\fft{\mubar^2}{p^2}\Big)\big)\,,\nn\\
& C^{(0)}(p^2)=-\frac{8\pi G \left(1080 n_2 \log ^2\left(\frac{m^2}{p^2}\right)+(90 \xi -19) n_0+33 n_1-6575 n_2-14 n_{\frac{1}{2}}-28 n_{\frac{3}{2}}\right)}{4320 \pi^2  p^2}\,,\nn\\
& C^{(2A)}(p^2)=\frac{8\pi G \left(n_0+3 n_1+5 n_2-4 n_{\frac{1}{2}}-8 n_{\frac{3}{2}}\right)}{2880 \pi^2  p^2}\,,\nn\\
& C^{(2B)}(p^2)=8\pi G\times 8 c_{\rm GB}^{\rm ren}-\fft{8\pi G}{8\pi^2} n_2 \log^2\Big(\fft{m^2}{p^2}\Big)-8\pi G \times 4 (a-c) \log\Big(\fft{\mubar^2}{p^2}\Big)\,,\label{eq: Sigma and C in 4D massless}
\ee
where $a$ and $c$ are
generalization of central charges in $d=4$ CFT
\be \label{ca massless}
c=\frac{n_0+13 n_1+95 n_2+6 n_{\frac{1}{2}}-148 n_{\frac{3}{2}}}{1920 \pi ^2}\,,\quad a=\frac{n_0+63 n_1-115 n_2+11 n_{\frac{1}{2}}-218 n_{\frac{3}{2}}}{5760 \pi ^2}\,.
\ee
These reduce to usual central charges of conformal field theory when taking $n_{3/2}=n_2\equiv 0$ and then specifying the massless modes to be
\be
n_0=n_S-n_V\,,\quad n_1=n_V\,,\quad n_{\fft{1}{2}}=\fft{n_{\rm Weyl}}{2}\,,
\ee
which gives, in agreement with  \cite{Duff:1977ay,Christensen:1978gi}:
\be
c=\fft{n_S+12n_V+3n_{\rm Weyl}}{1920\pi^2}\,,\quad a=\fft{2n_S+124n_V+11n_{\rm Weyl}}{11520\pi^2}\,.
\ee
This is of course not a coincidence, since, as we discussed in section \ref{sec: general}, the self-energy and effective vertices are simply stress-tensor correlators
in the non-gravitational theory.
Conformal stress-tensor two-point functions are usually parametrized as
\be
\fft{\Sigma^{(2)}(p^2)}{8\pi G}\big|_{\rm CFT}=2 p^4 c \log\Big(\fft{\mubar^2}{p^2}\Big)\,.
\ee
Additionally, in CFT we need to have $\Sigma^{(0)}=0$, which is indeed achieved using the conformal coupling
$\xi = 1/6$ for each scalar (and using that a massless gauge field is the $m\to 0$ limit of a massive vector
plus $-1$ scalar with $\xi=0$).

When including gravitinos and gravitons in the loop,
the physical significance of $c$ and $a$ in \eqref{ca massless} becomes less clear.  The gravitino contribution to $c$ is even negative!  This negativity of the $TT$ two-point function does not violate unitarity, however, because in a theory with a gravitino one also needs to have dynamical gravity, and stress tensor correlators are no longer gauge invariant observables.\footnote{
One might have thought that the graviton self-energy must have a definite sign by unitarity because it can be measured from the $S$-matrix of scalars that interact only gravitationally.  However, it seems that in any situation in which a gravitino loop can contribute to the self-energy, there must also exist other diagrams where the gravitino couples directly to the external legs, so no physical process  measures \emph{only} the self-energy.
} 
The matter contents for which $c=0$ (or $a=0$) seem to be rather random ($a$ and $c$ do not vanish with maximal supergravity).

We conclude that the graviton self-energy and effective vertex are not sensible physical observables, and that the $n_{3/2}$ and $n_2$ contributions to them carry limited significance.

Let's end by recording the one-loop amplitudes. They  take significantly simpler forms when organized using the supersymmetry decomposition \eqref{susy decomp}. 
At the integrand level, we have verified that our results agree with \cite{Bern:2021ppb} for all helicity configurations.
As another sanity check, we reproduce the $\mathcal{N}=8$ multiplets results in $d=4$ for any mass \cite{Green:1982sw,Bern:1998ug,Bern:2021ppb,Bern:2022yes}:
\be
f^{1-{\rm loop}}_{\mathcal{N}=8}(s,u)= (8\pi G)^2 \big(I_{\rm box}(m,s,t)+I_{\rm box}(m,s,u)+I_{\rm box}(m,t,u)\big)\,,
\ee
while $h$ and $g$ are zero.  To give a taste of the expressions, we show the massless limit in $d=4$:
\be
f^{1-{\rm loop}}_{\mathcal{N}=0}(s,u)&=\fft{(8\pi G)^2}{16\pi^2} \Big(-\frac{\pi ^2 s^3 u^3}{2 (s+u)^8}-\frac{s^3 u^3 \log ^2\left(\frac{u}{s}\right)}{2 (s+u)^8}+\frac{2 s^4+23 s^3 u+222 s^2 u^2+23 s u^3+2 u^4}{360 (s+u)^6}\nn\\
&\quad+\frac{(s-u) \left(s^4+9 s^3 u+46 s^2 u^2+9 s u^3+u^4\right) \log \left(\frac{u}{s}\right)}{60 (s+u)^7}\Big)\,,\nn\\
f^{1-{\rm loop}}_{\mathcal{N}=1}(s,u)&=\fft{(8\pi G)^2}{16\pi^2} \Big(  \frac{-\pi ^2 s^2 u^2}{(s+u)^6}+\frac{s^2+14 s u+u^2}{12 (s+u)^4}-\frac{s^2 u^2 \log ^2\left(\frac{u}{s}\right)}{(s+u)^6}+\frac{(s-u) \left(s^2+8 s u+u^2\right) \log \left(\frac{u}{s}\right)}{6 (s+u)^5}     \Big)\,,\nn\\
f^{1-{\rm loop}}_{\mathcal{N}=4}(s,u)&=\fft{(8\pi G)^2}{16\pi^2} \Big(\frac{1}{2 (s+u)^2}-\frac{\pi ^2 s u}{2 (s+u)^4}-\frac{s u \log ^2\left(\frac{u}{s}\right)}{2 (s+u)^4}+\frac{(s-u) \log \left(\frac{u}{s}\right)}{2 (s+u)^3} \Big)\,,\nn\\
f^{1-{\rm loop}}_{\mathcal{N}=6}(s,u)&=-\fft{(8\pi G)^2}{16\pi^2} \Big(\frac{\pi ^2}{(s+u)^2}+\frac{\log ^2\left(\frac{u}{s}\right)}{(s+u)^2} \Big)\,,\nn\\
f^{1-{\rm loop}}_{\mathcal{N}=8}(s,u)&=\fft{(8\pi G)^2}{64\pi^2}\frac{s \log \left(-\frac{m^2}{t}\right) \log \left(-\frac{m^2}{u}\right)+\log \left(-\frac{m^2}{s}\right) \left(u \log \left(-\frac{m^2}{t}\right)+t \log \left(-\frac{m^2}{u}\right)\right)}{s t u}\,.
\ee
It is worth noting that all contributions are finite in the massless limit except from that of the graviton loop,
for which we have kept a small $m^2$ to provide an infrared regulator in the last line.

\subsection{Higher dimensions}

At the integrand level, we verify that our scalar loop and vector loop match with \cite{Bern:2022yes} up to finite number of contact terms, representing the higher-dimensional operators added to \eqref{Sgrav ders}.
For example, in $d\leq 6$, we can remove all $1/\epsilon$ poles in the on-shell amplitude by adding a multiple of the (divergent) tadpole integral to various operators with up to six derivatives. For example, for scalar loop, we obtain the following low-lying counterterms
\be
& c_{\rm GB}=c_{\rm GB}^{\rm ren} + n_0 \fft{(d-2)}{720 m^2} I_{\rm tad}(m) \,,\nn\\
& \alpha_4^\prime=\alpha_4^{\prime\, \rm ren} -n_0 \fft{(d-4)(d-2)}{3780 m^4} I_{\rm tad}(m) \,,\quad \alpha_4=\alpha_4^{\rm ren}-n_0 \fft{(d-4)(d-2)}{5040 m^4} I_{\rm tad}(m) \,,
\ee
where $\alpha_4$ and $\alpha_4^\prime$ are coefficients of two six-derivative operators, following the convention of \cite{Caron-Huot:2022jli}. It is then clear that there is no one-loop UV divergence in $d=4$. 

As we mentioned in section \ref{ssec: actions}, in higher dimensions, two-form and three-form fields are unavoidable parts of maximal supergravity. The three-form loop in particular would be technically challenging. However, we avoided its explicit calculation by turning supersymmetry around and imposing that the maximal supergravity amplitudes takes the form
\be \label{eq: N8 1-loop}
{\cal M}_{{\cal N}=8}^{1-{\rm loop}} =(8\pi G)^2 {\cal T}(\epsilon_i,p_i)\Big( I_{\rm box}(m,s,t) + I_{\rm box}(m,s,u)+I_{\rm box}(m,t,u)\Big)\,,
\ee
where the tensor structures ${\cal T}(\epsilon_i,p_i)$ reads
\be
{\cal T}(\epsilon_i,p_i)=\big(H_{14}H_{23}+H_{13}H_{24}+H_{12}H_{34}+2(X_{1234}+X_{1243}+X_{1324})\big)^2\,\label{eq: susy structure}
\ee
in terms of the tensor basis \eqref{graviton basis}.
It is worth noting that the massive maximal supergravity amplitudes do not have, for example, the $\mathcal{G}$ polarization in that basis. Nevertheless, all fields from scalar to massive spin-$2$ nontrivially contribute to this tensor structure. Furthermore, since a massive 3-form is duality-equivalent to lower forms in $d\leq 6$, we can construct the duality-subtracted combination in \eqref{susy decomp} which is only nonvanishing when external momenta and polarizations lie in a 7-dimensional subspace, ie. it must be proportional to ${\cal G}$. Using this property we find that \eqref{eq: N8 1-loop} is nontrivially satisfied for all polarizations provided that
\be
{\cal M}_{\bar{(1,1,1)}}^{1-{\rm loop}}=(8\pi G)^2\, \mathcal{G}\,
\Big(\fft{4((7-d)s t+16m^2 u)}{(d-3)u}I_{\rm box}(m,s,t) + \fft{8 (d-7)s^2}{(d-3)t u} I_{\rm tri}(m,s) + {\rm cyc}\Big)\,.
\ee
In an ancillary file, we record the coefficients of all the $29$ tensor structures in \eqref{graviton basis} for all the $1$-loop amplitudes in \eqref{susy decomp} in $d$ dimensions.

\section{Sum rules and the species bound}
\label{sec: sum rules}

A generic problem with species-type bounds is that the very idea of ``counting species'' breaks down near the cutoff,
where any field theory calculation necessarily becomes inapplicable. In particular, one-loop formulas cannot be trusted near the cutoff.
The bounds in this section will be subject to this caveat, which we will partly circumvent in section \ref{sec: QFT breaking}.

\subsection{Resummed propagators do not make sense} \label{ssec: resummed props}

Before discussing the four-graviton amplitudes, let us discuss the resummed graviton propagator including 1-loop self-energies.
It seems that for sufficiently high energies, $GNs\gtrsim 1$,  the propagator develops unphysical poles and it becomes difficult to make sense of it.
As discussed in section \ref{subsec: ingredients}, one might \emph{in principle}
have considered a power-counting scheme where the 1-loop self-energy remains accurate at these energies, so the issue is not merely a power-counting one.

To streamline the discussion, we focus on $d=4$ and massless loops; including masses (or a power-law distribution of masses) does not qualitatively change the conclusions.
For the transverse part of the resummed propagator, we have according to \eqref{eq: Sigma and C in 4D massless}:
\be
P_{2}=
\fft{1}{s+ 16\pi G c s^2 \big(\log[-\mubar^2/s]+\delta c^{\rm ren}/c\big)}\,.
\ee
In quantum field theory, the central charge $c>0$ is strictly positive, however
note that according to \eqref{ca massless}, the gravitino contribution is negative and so in a general gravitational context $c$ could in principle have either sign.  If $c=0$, then the analysis simplifies and the propagator develops a pole at $s=-1/(16\pi G \delta c^{\rm ren})$, where its residue has the wrong sign, ie. it is a ghost-like excitation incompatible with unitarity.
This is a generic feature of ${\rm Riem}^2$ modifications to gravity, which
implies that $|\delta c^{\rm ren}|$ should be small enough that the pole is outside the expected regime of validity of the EFT.

The problem becomes more severe when $c\neq 0$.  Then, depending on the sign of $c$ and choice of $\delta c^{\rm ren}$, we find either a pole at spacelike $s<0$ (problematic!), a pole at complex $s$ (problematic!) or a pole at positive $s$ but wrong-sign residue (problematic!).  Because of the nontrivial $s$-dependence of the self-energy, in particular the $\log s$ term, it is not possible to choose $\delta c^{\rm ren}$ to push this problem parametrically outside $|s|\sim 1/(Gc)$. In other words, it seems that it never makes sense to use the resummed propagator $P_2$: as soon as the renormalized self-energy parametrically becomes important, pathologies appear.

In the absence of gravitinos or gravitons running in loops, such that $c$ is a physical gauge-invariant quantity, this observation (as also proposed in \cite{Dvali:2007wp}) would probably lead to a satisfactory definition of the universal cutoff $\cutoff$. Namely, the position of the first complex or wrong-sign pole determines a scale at which something qualitatively new must be added to the calculation.
However, as discussed at the top of this section, this definition would still only be a parametric one, since it is associated with a breakdown of the approximation that enters its calculation.

It is amusing to put in numerical values. The central charge of the Standard Model including only observed fields is approximately that of 283 scalars, giving $c\approx 0.015$ in the standard normalization (see \eqref{ca massless}), and a scale
$1/\sqrt{16\pi G c}=1.4\times 10^{19}\,{\rm GeV}$.
It seems that, despite the somewhat large effective number of Standard Model fields, the scale at which the 1-loop graviton propagator develops pathologies is not much lower than the naive Planck scale, essentially due to loop factors.

In the presence of virtual gravitinos or gravitons, focusing on $P_2$ seems much less meaningful.
For one, as noted in section \ref{ssec: d4 amps},
$c$ is then not gauge invariant.  In fact it is not even sign definite: gravitinos contribute negatively in the covariant gauge we used (see \eqref{eq: Sigma and C in 4D massless}), and $c=0$ occurs for a strange matter content unrelated to supersymmetry.
This makes it difficult to take seriously the scale $1/(Gc)$.  A related observation is that massive gravitinos and gravitons will necessarily have other interactions, leading to more diagrams than the
self-energy chains shown in Fig.~\ref{fig: Gull and Vfull}.  Self-energy resummation, while reasonable for matter contributions, does not seem to be a reasonable model for gravity at higher loops.

It is amusing to note that the contribution from $P_2$ to the four-graviton amplitude vanishes with any amount of supersymmetry. Indeed, recall that in $d=4$, $C^{(2B)}=0$ identically, and the other vertex according to \eqref{eq: Sigma and C in 4D massless} gives a contribution:
\begin{equation} \label{susy V2A vanishes}
    {\cal M}\supset C^{(2A)} P_2 C^{(2A)}, \quad
    C^{(2A)} \propto {\rm Tr}[(-1)^F]\,.
\end{equation}
This cancellation holds even for massive fields.
(The spectrum does not need to be exactly supersymmetric, only in an average sense.)
This may suggest that with supersymmetry the pathologies of $P_2$ are not necessarily physical, however there may exist other processes (ie. matter-matter scattering) where the same cancellation does not occur and thus the significance of \eqref{susy V2A vanishes} is unclear to us.

Similar pathologies appear in the trace part of the graviton propagator, $P_0$.  Its contribution to graviton scattering takes the form \eqref{eq: f in 4D}
\begin{equation}
{\cal M}\supset C^{(0)}P_0C^{(0)},
\end{equation}
with $C^{(0)}$ in $d=4$ given in \eqref{eq: Sigma and C in 4D massless}. Again, the positions of its pathological singularities seem to be unphysical combinations, and there may also be situations where $C^{(0)}$ vanishes.

We conclude that the (renormalized) graviton self-energy may be a reasonable probe of the scale $\cutoff$ in situations with only standard matter runs in loops but no virtual gravitinos nor gravitons (ie. no Kaluza-Klein modes), but in general it is not.  But even in the former case, it does not make physical sense to use the resummed propagator at energies where the self-energy dominates.

\subsection{Review of gravitational sum rule formalism}
\label{subsec: sum rule}

In the non-gravitational context (amplitudes without a graviton pole), general methods to constrain EFTs using the forward limit of dispersive sum rules have been discussed in \cite{Adams:2006sv,deRham:2017avq,Caron-Huot:2020cmc, Bellazzini:2020cot, Tolley:2020gtv,Arkani-Hamed:2020blm}.  Here, we review the extension to the gravitational context, following closely \cite{Caron-Huot:2021rmr,Caron-Huot:2022ugt} (see also \cite{Tokuda:2020mlf,Alberte:2020jsk,Caron-Huot:2024tsk} for other discussions and  \cite{Henriksson:2022oeu,Hong:2023zgm,Albert:2024yap} for other applications).

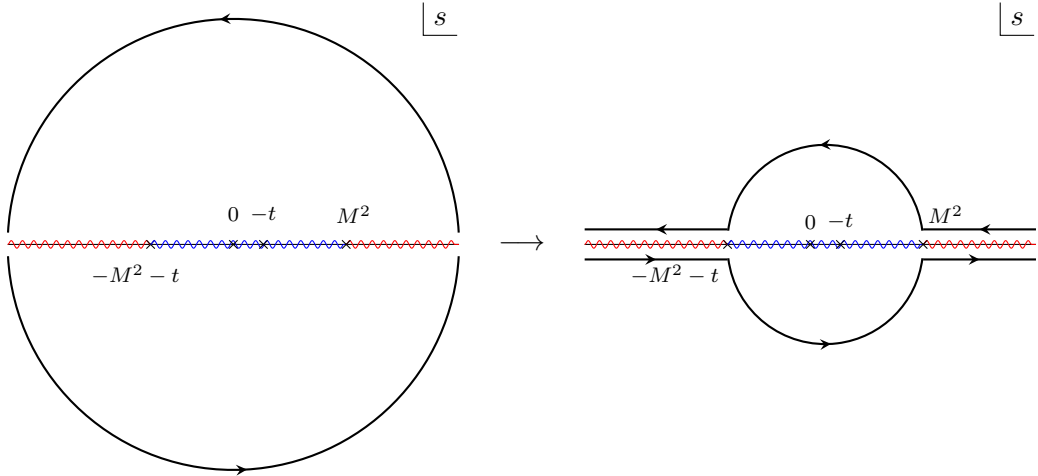
\begin{figure}[t]
\centering 
\begin{tikzpicture}[decoration={markings, 
    mark= at position 0.52 with {\arrow{stealth}}}]
	\draw (-3,0) -- (3,0);
    \draw[thick, rotate=3, postaction={decorate}] (-3,0) arc (-180:-6:3) ;
    \draw[thick, rotate=3, postaction={decorate}] (3,0) arc (0:174:3) ;
    \draw[red, decorate, decoration={snake=zigzag,segment length=1.5mm, amplitude=0.5mm}]       (-3,0) -- (-1.1,0);
    \fill (-1.1,0) node[red, cross=2pt] {};
    \draw[blue, decorate, decoration={snake=zigzag,segment length=1.5mm, amplitude=0.5mm}]       (-1.1,0) -- (1.5,0);
	\draw[red, decorate, decoration={snake=zigzag,segment length=1.5mm, amplitude=0.5mm}]       (1.5,0) -- (3,0);
    \fill (0,0) node[cross=2pt] {};
	\draw (0.4,0) node[yshift=0.4cm] {\scriptsize $-t$};
    \fill (0.4,0) node[cross=2pt] {};
    \fill (1.5,0) node[cross=2pt] {};
	\node (a) at (2.75,3) {$s$};
	\draw (a.north west) -- (a.south west) -- (a.south east);
	\draw (0,0) node[yshift=0.4cm] {\scriptsize $0$};
	\draw (1.5,0) node[xshift=0.1cm,yshift=0.4cm] {\scriptsize $M^2$};
	\draw (-1.5,0) node[yshift=-0.4cm,xshift=0.2cm] {\scriptsize $-M^2 -t$};
	\end{tikzpicture}
    \raisebox{86pt}{$\quad\longrightarrow\quad$}
\begin{tikzpicture}
	\draw (-3,0) -- (3,0);
    \draw[white, thick, rotate=-3, postaction={decorate}] (3,0) arc (0:-174:3) ;
    \draw[white, thick, rotate=3, postaction={decorate}] (3,0) arc (0:174:3) ;
    \draw[thick, rotate=7, postaction={decorate}, decoration={markings, 
        mark= at position 0.52 with {\arrow{stealth}}}] (-1.105,-0.05) arc (-180:-14:1.3) ;
    \draw[thick, rotate=7, postaction={decorate}, decoration={markings, 
        mark= at position 0.52 with {\arrow{stealth}}}] (1.5,0) arc (0:166:1.3) ;
	\draw[thick, thick, postaction={decorate}, decoration={markings, 
            mark= at position 0.5 with {\arrow{stealth}}}] (1.5,-0.2) --(3,-0.2) ;
	\draw[thick, thick, postaction={decorate}, decoration={markings, 
            mark= at position 0.5 with {\arrow{stealth}}}] (-3,-0.2) --(-1.1,-0.2) ;
    \draw[thick, postaction={decorate}, decoration={markings, 
            mark= at position 0.5 with {\arrow{stealth}}}] (3,0.2) --(1.5,0.2) ;
	\draw[thick, thick, postaction={decorate}, decoration={markings, 
            mark= at position 0.5 with {\arrow{stealth}}}] (-1.1,0.2) --(-3,0.2) ;
    \draw[red, decorate, decoration={snake=zigzag,segment length=1.5mm, amplitude=0.5mm}]       (-3,0) -- (-1.1,0);
    \fill (-1.1,0) node[cross=2pt] {};
	\draw[red, decorate, decoration={snake=zigzag,segment length=1.5mm, amplitude=0.5mm}]       (1.5,0) -- (3,0);
	 \draw[blue, decorate, decoration={snake=zigzag,segment length=1.5mm, amplitude=0.5mm}]       (-1.1,0) -- (1.5,0);
    \fill (0,0) node[cross=2pt] {};
	\draw (0.4,0) node[yshift=0.3cm] {\scriptsize $-t$};
    \fill (0.4,0) node[cross=2pt] {};
    \fill (1.5,0) node[cross=2pt] {};
	\draw (1.5,0) node[xshift=0.3cm,yshift=0.4cm] {\scriptsize $M^2$};
	\node (a) at (2.75,3) {$s$};
	\draw (a.north west) -- (a.south west) -- (a.south east);
	\draw (0,0) node[yshift=0.3cm] {\scriptsize $0$};
	\draw (-1.5,0) node[yshift=-0.4cm,xshift=-0.3cm] {\scriptsize $-M^2-t$};
	\end{tikzpicture}
\caption{The contour deformation for the sum rules in \eqref{B2B3}. The red branch cut represents the UV branch cut above the species scale $M$, and the blue branch cut represents all possible low-energy branch cuts.}\label{fig: contour}
\end{figure}

In order to benefit from the $\sim t^4$ behavior of the helicity prefactor in \eqref{eq:Mdef}, we consider contour integrals at large energies with $u$ fixed, so that $t\approx -s$ as $s\to\infty$:
\begin{subequations} \label{B2B3} \begin{align}
 B_2(p) &\equiv \oint_{\mathcal{C}_+\cup\, \mathcal{C}_-} \frac{ds}{2\pi i} (s-t) f(s,u=-p^2)\,,\\
 B_3(p) &\equiv -\oint_{\mathcal{C}_+\cup\, \mathcal{C}_-} \frac{ds}{2\pi i} f(s,u=-p^2)\,,
\end{align}\end{subequations}
where $\mathcal{C}_\pm$ are the two halves of the infinite-energy circles shown in Fig.~\ref{fig: contour}. The idea is to exploit the vanishing of \eqref{B2B3}, together with
analyticity in the upper-half $s$-plane, to relate physics below and above the scale $M$, as shown in Fig.~\ref{fig: contour}.
These sum rules are called ``superconvergent'' because we have not introduced any denominator (also known as subtraction terms): this makes them automatically insensitive to all analytic terms at low energies.

Morally, the vanishing of the arcs at infinity is related to the Froissart bound $\lim_{|s|\to\infty} \mathcal{M}/s^2= 0$,
or equivalently $s^2f\to0$ at fixed $u$.  However, the Froissart bound is known to not apply to graviton scattering with fixed exchange momentum in $d=4$,
due to the long-range nature of gravitational interactions \cite{Haring:2022cyf}.
This seems to be a mostly technical problem with a simple physical solution.
Together with the singular nature of the graviton pole as $p\to 0$,
it is the key reason why we consider scattering at small impact parameter rather than of plane waves.

In practice, this is achieved by integrating the above sum rules with focusing wavepackets $\psi_k(p)$.\footnote{These were called ``smearing'' functions in \cite{Caron-Huot:2021rmr} due to their role in momentum space. Here we adopt a terminology
that highlights the focusing role of the wavepacket in impact parameter space.}
It will be important to use momenta in a finite range $0<p<M$.
After integrating against suitable $\psi(p)$, corresponding physically to functions whose Fourier transform vanish sufficiently
fast at large impact parameter, the integrals along  in \eqref{B2B3} vanish as ${\cal C}_\pm$ are taken to infinity, giving the sum rules:
\be\begin{aligned}\label{low equals high}
&\mathcal{F}:=\sum_{k=2,3}
 \int_0^{M} pdp\, \psi_k(p) B_k(p)=0
\\ &\Rightarrow\quad -\sum_{k=2,3}\int_0^{M} pdp\,\psi_k(p) B_k(p)\Big|_{\rm low}
= \sum_{k=2,3}\int_0^{M} pdp\,\psi_k(p) B_k(p)\Big|_{\rm high}\,.
\end{aligned}\ee
The low-energy contribution accounts for arcs going from $t=M^2$ to $s=M^2$:
\begin{subequations}\begin{align}
 -B_2(p)\Big|_{\rm low} &=
 \sum_{\pm}\int_{M^2}^{p^2-M^2} \frac{ds}{2\pi i} (p^2-2s) f(s,-p^2) = \frac{8\pi G}{p^2}+\mbox{loops}\,,
\label{B2low}\\
 -B_3(p)\Big|_{\rm low} &=
 \sum_{\pm}\int_{M^2}^{p^2-M^2} \frac{ds}{2\pi i} f(s,-p^2) = 0 + \mbox{loops}\,,
\end{align}\end{subequations}
where $\sum_{\pm}$ refers to the two halves of the small arc shown in the RHS of Fig \ref{fig: contour}. The high-energy contribution can be parametrized by its partial wave decomposition:
\begin{subequations}\begin{align}
 B_2(p)\Big|_{\rm high} &= 16\int_{M^2}^\infty \frac{ds}{s^4}\,(2s-p^2)
 \left[\sum_{J\geq 0,\,\rm even} |\bar{c}_J^{++}(s)|^2 P_J\left(1-\tfrac{2p^2}{s}\right)
 +\sum_{J\geq 4} |\bar{c}_J^{+-}(s)|^2 \tilde{d}_{4,4}^J \left(1-\tfrac{2p^2}{s}\right)\right]\,,
\label{B2high}
\\
 B_3(p)\Big|_{\rm high} &= 16\int_{M^2}^\infty \frac{ds}{s^4}
 \left[\sum_{J\geq0,\,\rm even} |\bar{c}_J^{++}(s)|^2 P_J(1-\frac{2p^2}{s})
 -\sum_{J\geq 4} |\bar{c}_J^{+-}(s)|^2 \tilde{d}_{4,4}^J(1-\frac{2p^2}{s})\right]\,,
\end{align}\end{subequations}
where $\tilde{d}_{a,b}$ is related to Wigner $D$-functions \eqref{Wigner d} by slipping off the ``helicity factors'' 
\be
d_{h,h'}^J = \left(\frac{1+x}{2}\right)^{\frac{h+ h'}{2}}  \left(\frac{1-x}{2}\right)^{\frac{h-h'}{2}}  \tilde{d}_{h,h'}^J\,.
\ee
$P_J(x)$ is the general dimensional Legendre polynomials
\be
P_J(x)=\,{}_2F_1(-J, J{+}d{-}3, (d {-}2)/2, (1 - x)/2)\,.\label{eq: PJ}
\ee
The $|\bar{c}_J|^2=|c_J|^2(2J+1)$ are unknown positive quantities which represent probabilities to scatter at high energies.

\subsection{A first view on the high-spin onset scale: fixed impact parameters}
\label{ssec:dispersive bounds}

We will now use these sum rules to demonstrate the phenomenon of high-spin onset, as a refined version of Fig. \ref{fig: compare b scalar} shown in the introduction,
focusing here on $d=4$.

The basic idea is to choose a wavefunction $\psi_2$ that corresponds to transforming \eqref{B2low} and \eqref{B2high} to impact parameter space.  A subtlety is that we have to be careful to use only transverse momenta $|p|\leq M$ in the Fourier transform.  We thus consider:
\be
B_2(b)= \int_0^{M}dp \,(1-p)^2 p J_0(bp) B_2(p)\,. \label{eq: sum rule b toy}
\ee
The factor $(1-p)^2$ smoothens the cutoff.
It has another, less obvious, property: the Fourier-conjugate function
$\psi(b)=\int_0^M dp (1-p)^2 p J_0(p b)$ is positive.
This means physically that \eqref{eq: sum rule b toy} could also be written as a convolution $d^2b$ of $\widehat{B}_2(b)$ with a positive smearing function with spread of order $M^{-1}$. This ensures that
the action of \eqref{eq: sum rule b toy} on states with sufficiently large $s\gg M^2$ in \eqref{B2high} is automatically positive.  (Positivity at all $s>M^2$ will require slightly more complicated wavefunctions discussed in the next section,
which will also include $B_3$.)

The family of sum rules \eqref{eq: sum rule b toy}, labelled by $b$, allow us to explain the main phenomenom.
Evaluating $B_2(b)$ for low-energy Einstein gravity using \eqref{B2low}, we find
\be
B_2(b)\Big|_{\rm grav}=-\frac{1}{8} b^2 \, _2F_3\left(1,1,2,2,2,-\frac{b^2}{4}\right)+(\pi  \pmb{H}_1(b)-2) J_0(b)+\frac{(1-\pi  b \pmb{H}_0(b)) J_1(b)}{b}-\log m_{\rm IR}\,,\label{eq: low grav b}
\ee
where $\pmb{H}$ is Struve function of the second kind.
The logarithmic divergence is the usual infrared divergence of the Shapiro time delay in $d=4$ and would be absent if considering analogous sum rules in higher dimensions.

\begin{figure}[t]
\centering\includegraphics[width=0.9\textwidth]{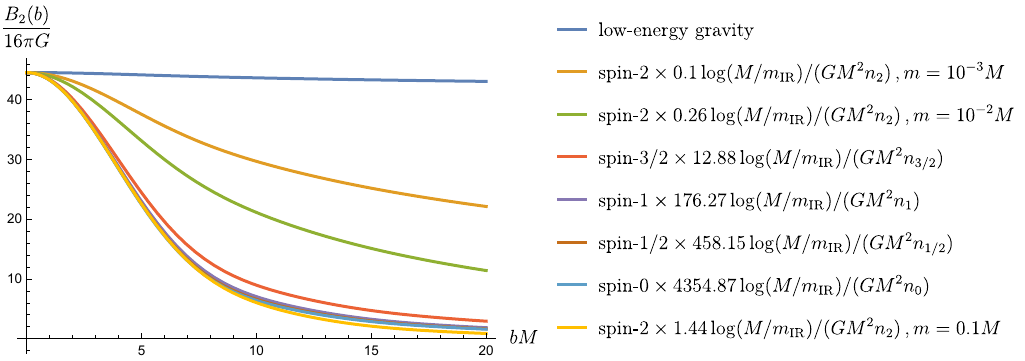}
\caption{Illustration of the (non-optimal) sum rules \eqref{eq: sum rule b toy}, which for each $b$ measures Newton's constant $G$, together with the 1-loop contributions from $N$ light fields of various types. The total of shown 1-loop contributions from $s<M^2$, and unknown positive contributions from energies $s>M^2$,
must add up to the low-energy gravity prediction.
For interpretation of the numbers in the legends, see the main text. The 1-loop contributions all have a similarly rapid decaying shape, and therefore, there is no way for their sum to saturate the low-energy prediction.  We used $m_{\rm IR}=10^{-20}M$ for illustration.}\label{fig: B2 impact 4d}
\end{figure}

The arc integrals \eqref{B2low} also receive nontrivial contributions at one-loop from matter fields with mass $m<M/2$.  Their calculation is technically much more involved.  A shortcut is to deform the contour in Fig.~ \ref{fig: contour} to wrap around the discontinuities, namely,
\be
B_2(p)= \int \fft{ds}{\pi} (2s-p^2) {\rm Im}\, f(s,-p^2)\,.
\ee
The discontinuities of master integrals with fixed $u=-p^2$ are detailed in Appendix \ref{app: branch}; see \eqref{eq: fix u Disc} for more details. We then numerically perform the integral of \eqref{eq: sum rule b toy}. The main result are the shapes of the impact parameter sum rules which are displayed in Fig.~\ref{fig: B2 impact 4d} for different light fields, and contrasted with the graviton pole contribution in \eqref{eq: low grav b}.

As already sketched in the introduction, we can now clearly see that there has to exist bounds on $N(M)$ so that the matter loops never exceed the maximum ``budget'' allowed by Newton's constant at low energies, which is first saturated at the smallest impact parameter we can access with the relevant energy, $b\lesssim M^{-1}$. We make such ``species bound'' manifest in the legends of Fig \ref{fig: B2 impact 4d}. For example, one can read $n_0\leq 4354.87 \log(M/m_{\rm IR})/(G M^2)$. 
The different numbers in the legends are thus suggestive of the relative contributions from different types of matter fields to the species scale, ie. a massive spin-1 field counts as much as about 25 scalars (despite having just three polarization modes).
Note however that the sum rules \eqref{eq: sum rule b toy}
are rather simple-minded and are neither optimal nor fully rigorous (because they are not positive for all $s>M^2$, as mentioned above.)  They will be improved below.

The main message we would like to highlight here
is that all the computable loops have the \emph{wrong shape}.  They decay at large $b$ like $\sim b^{-2}$ for light fields and $\sim e^{-2bm}$ for massive fields and there is no way to take any positive linear combinations of them to obtain the flat total $\sim \log(b)$ required by low-energy gravity.
This means that there necessarily exists states with $m\geq M$ whose contribution at large spin, $bM\sim 2JM/m \gg 1$, is significantly larger than any loop effect.  This is the phenomenon of \emph{high-spin onset}.

\subsection{A dispersive bound on the number of light species} \label{ssec: dispersive species}

We can now describe a strategy for finding species-type bounds: we search for wavefunctions $\psi_2(p)$, $\psi_3(p)$ that make the right-hand-side of \eqref{low equals high} \emph{positive} for all $s>M^2$, and such that the contribution of each matter loop on the left-hand-side is \emph{negative} for any choice of mass $m<M/2$.

The bounds here are rigorously valid for any cutoff $M$ such that we can use the one-loop approximation below the cutoff.
Of course, as mentioned in introduction, all that one can really bound in this way is the number of fields parametrically lighter than the universal field theory cutoff $\cutoff$, ie. modes for which one can trust the one-loop approximation.  For this reason, we didn't try too hard to optimize the bounds.

Another slightly unsatisfactory feature in $d=4$ must be mentioned. Positivity at very large $m$ requires that $\psi_2(0)\neq 0$: this is simply the statement that the Fourier transform of a positive function of impact parameter cannot vanish at zero momentum.
On the other hand, this behavior is incompatible with convergence of the sum rules on the graviton pole, which require $\psi(0)=0$.
The resulting bounds thus necessarily depend logarithmically on an infrared cutoff, schematically:
\be
  8\pi G \log \frac{M^2}{m_{\rm IR}^2}  - \frac{(8\pi G)^2}{M^2} \sum_i c_i n_i \geq 0\,,
\ee
which gives species-like bounds up to an infrared logarithm $c_in_i < \frac{\Mpl^2}{M^2} \log\frac{M^2}{m_{\rm IR}^2}$,
where the $c_i$ are $\mathcal{O}(1)$ constants to be determined.
This is still much stronger than the quadratically divergent bounds one would obtain from expanding around the forward limit.
In the context of AdS/CFT, it has been shown that the bounds uplift to rigorous CFT sum rules
with $m_{\rm IR}^{-1}\approx R_{\rm AdS}$ \cite{Caron-Huot:2021enk}.  This suggests that
$m_{\rm IR}^{-1}$ can be interpreted as the distance from which the scattered particles are sent in---we see no reason why this size would have to be parametrically larger than $M$, and so we will treat $\log \frac{M^2}{m_{\rm IR}^2}$ as an unknown large but $\mathcal{O}(1)$ constant.

\begin{figure}[t]
\centering\includegraphics[width=0.9\textwidth]{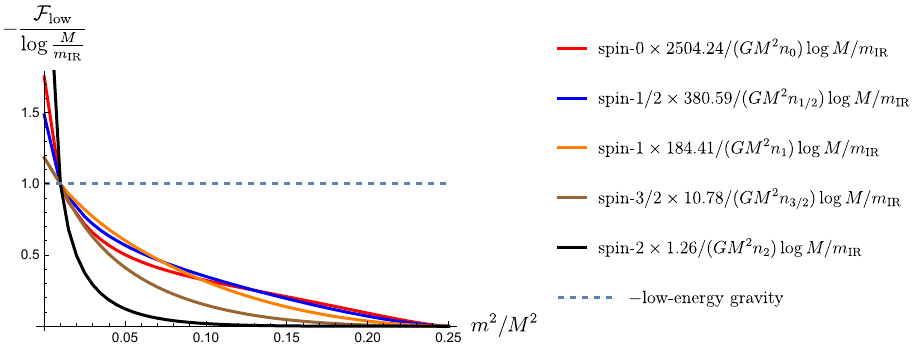}
\caption{Sum rules at low-energy measured by the wavefunctions \eqref{eq: wavefunction} for all species with $j\leq 2$ in $d=4$. The sign definiteness of these curves ensures the existence of a bound on the number of each respective species.
The legends show the normalizations factor which align the curves at $m=M/10$.
}\label{fig: species}
\end{figure}

Using the numerical techniques detailed in \cite{Caron-Huot:2022ugt}, we then find that the following pair of wavefunctions
\be
&\psi_2(p)=(1-p)^5 \big(0.5+2.3 p + 6 p^2-13.4 p^3+45.6 p^4\big)\,,\nn\\
& \psi_3(p)=(1-p)^5\big(0.7+3.6 p+9.4 p^2 + 28.8 p^3\big)\label{eq: wavefunction}
\ee
gives a species bound for the number of massless particles:
\be
n_0+5.6 n_{1/2}+9.2 n_1 +157 n_{3/2} 
+ 491.7 n_2\log\frac{M}{m_2}
< \big(1429.6 \log\fft{M}{m_{\rm IR}}-1735.6\big) \fft{1}{G M^2}\,.
\ee
Note as mentioned previously that the contribution from spin-2 fields is singular in the massless limit in $d=4$.  If we take all the particles to have mass $m=M/10$, we get instead the bound
\be n_0+6.6 n_{1/2}+13.6 n_1 +232.4 n_{3/2} 
+ 1983.9 n_2
< \big(2504.2 \log\fft{M}{m_{\rm IR}}-3040.3\big) \fft{1}{G M^2}\,.
\ee
In general, the relative contributions of fields with various mass
is shown in Fig.~\ref{fig: species}.
The relative contributions of different fields are somewhat different than in Fig.~\ref{fig: B2 impact 4d}, although the general hierarchies are preserved, ie. a massive spin-two field has the same weight as a very large number of scalar fields.

\subsection{Comments on Wilson coefficients from modes with $\Lambda< m < \cutoff$}
\label{sec: above the cutoff}

Here we briefly discuss the low-energy perspective on ``non-maximal'' EFTs whose cutoff $\Lambda$ is below
the universal cutoff $\cutoff$, if one assumes that low-energy Wilson coefficients are dominated by integrating out calculable matter fields.  This setup was considered for example in \cite{Bern:2021ppb} and \cite{Caron-Huot:2022ugt}
and gives bounds on the number of calculable modes which can exist above the scale $\Lambda$\footnote{See also \cite{AccettulliHuber:2020oou,Alviani:2024sxx} for the generalization to one-loop effects from nonminimal couplings.}. Note that this line of logic is distinct from that considered above.

The causality constraints on Wilson coefficients controlling modifications to gravity have been numerically established in $d = 4$ in \cite{Caron-Huot:2022ugt} and generalized to higher dimensions in \cite{Caron-Huot:2022jli}.  Here we contrast these bounds with the Wilson coefficients arising in calculable models.

\begin{table}[t]
\centering
\begin{tabular}{c|cc}
spin & $\alpha_2$ & $\alpha_4$ \\\hline
$0$ & $-\frac{G N m^6}{34560\pi^4}$ & $\frac{G N m^4}{322560 \pi^4}$ \\[1mm]
$\frac{1}{2}$ & $-\frac{7G N d_F m^6}{276480\pi^4}$ & $-\frac{G N d_F m^4}{322560 \pi^4}$ \\[1mm]
$1$ & $\frac{G N m^6}{5760 \pi ^4}$ & $\frac{G N m^4}{35840 \pi ^4}$ \\[2mm]
$\frac{3}{2}$ & $\frac{23 G N d_F m^6}{34560 \pi ^4}$ & $-\frac{G N d_F m^4}{40320 \pi ^4}$ \\[1mm]
$2$ & $-\frac{149 G N m^6}{34560 \pi ^4}$ & $\frac{11 G N m^4}{80640 \pi ^4}$ \\[1mm]
$(1,1)$ & $-\frac{7 G N m^6}{11520 \pi ^4}$ & $\frac{G N m^4}{8960 \pi ^4}$ \\[1mm]
$(1,1,1)$ & $-\frac{73 G N m^6}{11520 \pi ^4}$ & $\frac{G N m^4}{3840 \pi ^4}$\\[1mm]
Sharipro-Virasoro & $0$ & $0$ \\[1mm]
heterotic string & $\fft{2}{M^2}$ & $0$ \\[1mm]
bosonic string & $\fft{4}{M^2}$ & $\fft{4}{M^4}$
\end{tabular}
\caption{Wilson coefficients $(\alpha_2,\alpha_4)$ from different UV models.}
\label{tab: Wilson coefficients}
\end{table}

\begin{figure}[t]
\centering\includegraphics[width=0.7\textwidth]{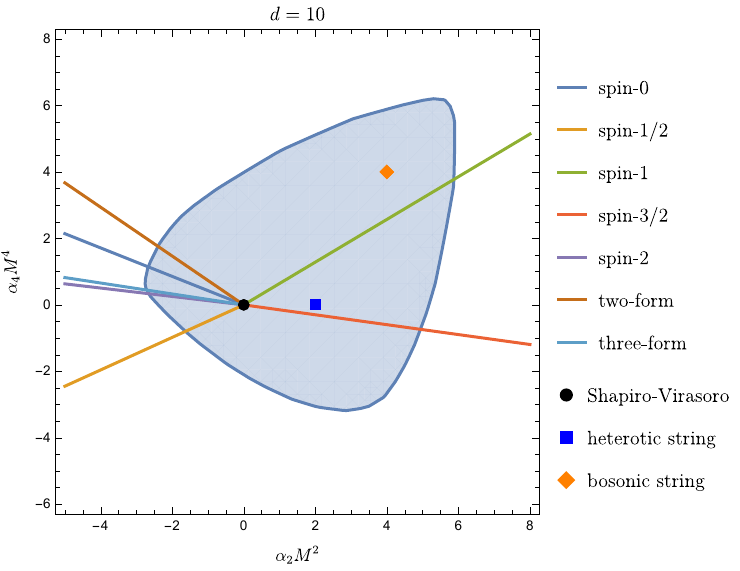}
\caption{Allowed region for three-graviton couplings $(\alpha_2,\alpha_4)$,  from \cite{Caron-Huot:2022jli}. The lines display the loop amplitudes, and the dots display the string amplitudes. The Shapiro-Virasoro amplitude lies at the origin due to maximal supersymmetry.}\label{fig: Wilson coeff 10d}
\end{figure}

We restrict to $d=10$ here and only look at the Wilson coefficients $(\alpha_2,\alpha_4)$ that control higher-derivative corrections to the three-graviton vertex, whose causality bounds were established in \cite{Caron-Huot:2022jli} following earlier ideas of \cite{Camanho:2014apa}.
We follow the conventions of \cite{Caron-Huot:2022jli} so that the gravitational EFT schematically expands as
\be
S=\fft{1}{16\pi G}\int d^{10}x \sqrt{-g} \Big(R + \fft{\alpha_2}{4} C^2 + \fft{\alpha_4}{12}\big(3C_{\mu\nu\rho\sigma}C^{\rho\sigma}\,_{\alpha\beta}C^{\alpha\beta\mu\nu}-4 C_{\mu\nu\rho\sigma}C^{\nu\alpha\sigma\beta}C_{\alpha}\,^\mu\,_\beta\,^\rho\big)+\cdots\Big).
\ee
We computed the heavy mass expansion of the loop amplitudes computed in section \ref{sec: amplitudes} to obtain $(\alpha_2,\alpha_4)$ by matching to the low-energy gravitational EFT amplitudes \cite{Caron-Huot:2022jli}. For string theory, we simply expand the superstring, as well as the heterotic string and bosonic string amplitudes (reviewed in Appendix \ref{app: string amp}) to obtain $(\alpha_2,\alpha_4)$. The resulting Wilson coefficients are shown in Table \ref{tab: Wilson coefficients}. A sanity check is that any amount of supersymmetry ensures $\alpha_4=0$. We then compare the specific models in Table \ref{tab: Wilson coefficients} with the bounds provided by \cite{Caron-Huot:2022jli}, which are shown in Fig.~\ref{fig: Wilson coeff 10d}.

Loop amplitudes are labeled by colored lines in Fig. \ref{fig: Wilson coeff 10d}, as they give rise to Wilson coefficients scaling with the number of species. Different species correspond to different directions.
Note that cancellations between different directions are possible, as indeed required since $\alpha_2$ and $\alpha_4$ vanish for supersymmetric spectra.
As already noted by \cite{Caron-Huot:2022ugt,Caron-Huot:2022jli}, the lines corresponding to loop amplitudes extend outside of the allowed region when $N$ becomes too large. This suggests a naive version of the species bound, however, which turns out to yield huge numbers in $d=10$, around $GNM^{8} \lesssim 10^{7}$.
However, the interpretation of these numbers is fundamentally unclear since Wilson coefficients like $\alpha_2$ and $\alpha_4$ can always be modified by fine-tuning bare couplings. It is thus hard to see how robust bounds on the number of species could be obtained from low-energy Wilson coefficients.

The dispersive species bound derived in the preceding subsection are distinct since they bound calculable modes below the cutoff.  We would thus expect significantly different numbers in this case. The extension of subsection \ref{ssec: dispersive species} to $d>4$ should be technically feasible and we leave it to future work. In $d>4$ the bounds will be infrared finite.
Since we expect field theory methods to break down near the universal cutoff $\cutoff$, however, these bounds will necessarily only apply to modes parametrically lighter than the cutoff.  To define the cutoff more sharply, we turn to a different strategy in the next section.

\section{The high-spin onset scale $\hs$ and gravity's UV completion}
\label{sec: QFT breaking}

We now explore the candidate sharp definition of the QFT breaking scale proposed in introduction: $\cutoff=\hs$, the scale at which the high-spin content of graviton scattering amplitudes ``onsets''.

The strategy is inspired by the fixed impact parameters dispersive sum rules in Fig.~\ref{fig: B2 impact 4d},
but we now work spin-by-spin. Specifically, we seek to measure the contributions a given spin to a sum rule for Newton's constant. For any SO($d-1$) representation $\rho$, we consider the following moment:
\be
 \mathcal{G}_\rho(s^\ast)=\fft{d-2}{\pi}\int_{0}^{s^\ast} \fft{ds}{s^{\fft{d}{2}}} {\rm Im}\, a_\rho(s)\,.\label{eq: moments}
\ee
We will show that computable contributions $\mathcal{G}$ from matter loops decay extraordinary fast with spin.
This is related the ``low-spin-dominance'' phenomenon observed in \cite{Bern:2021ppb}. However, we will also stress that other UV scenarios which unitarize the graviton scattering amplitude, such as the Virasoro-Shapiro amplitude in string theory, the eikonal approximation at large impact parameter, or strongly interacting physics at the Planck scale,
are all qualitatively different. While they are still all low-spin dominated in the sense of \cite{Bern:2021ppb}, they display a numerically richer high-spin content.

The high-spin richness of an amplitude can be quantified by comparing the moments $\mathcal{G}_\rho$
for adjacent spins:
\be
 \frac{\mathcal{G}_{\rho+2}(s^\ast)}{\mathcal{G}_{\rho}(s^\ast)}\,, \label{hs criterion}
\ee
where here and below $\rho+2$ denotes the representation whose Young Tableau is obtained from that of $\rho$
by adding two boxes to the first row.  In general we refer to length of the first row of $\rho$ as the ``spin'' $J_\rho$ of this representation.

The main reason why we use integrated spectral densities in \eqref{hs criterion} is to smooth out possible resonances: point-like comparison of $a_{\rho+2}(s)$ and $a_\rho(s)$ would assign too much significance to narrow resonances that have small overall coupling.  The reason for the particular weight in \eqref{eq: moments} is that it endows the moments ${\cal G}$ with
the units of Newton's constant $G$. (We use dimensionless partial waves normalized so that $|1+i a_\rho|\leq 1$ by unitarity, see subsection \ref{subsec: partial wave}.)

Finally, let us comment on the normalization of \eqref{eq: moments}, which is chosen such that in the eikonal model, where the amplitude is simply the exponential of the tree-level partial wave in general relativity (which has a simple energy scaling: $\delta^{\rm GR}_\rho = s^{\fft{d-2}{2}}\widehat{\delta}^{\rm GR}_\rho$), the moment recovers the eikonal phase:
\be \label{G eik}
a_\rho^{\rm eik} = i \left(1-\exp\left(i s^{\fft{d-2}{2}}\widehat{\delta}_\rho\right)\right)\quad
\Rightarrow\quad \lim_{s^\ast\to\infty} {\cal G}^{\rm eik}_\rho(s^\ast) = \widehat{\delta}_\rho^{\rm GR}\,.
\ee
The eikonal approximation is expected to be reliable at large enough impact parameters (see discussion in \cite{Giddings:2009gj,Haring:2022cyf}).
Thus we expect that ${\cal G}_\rho(s^\ast)\to \widehat{\delta}^{\rm GR}_\rho$ for large spin and sufficiently large $s^\ast$,
for any representation $\rho$ which can appear in graviton-graviton scattering, 
where at large spin (see below):
\be
 \widehat{\delta}^{\rm GR}_\rho\to \frac{8\pi G}{J_\rho^{d-4}}
 2^{1-d} \pi^{1-\frac{d}{2}} \Gamma \left(\tfrac{d-4}{2}\right) \qquad \mbox{(large spin)}\. \label{delta GR}
\ee
Although it is not clear whether the limit \eqref{G eik} must be saturated for arbitrary ${\cal G}$ that correspond to ultraviolet-complete gravity amplitudes, the GR phase shift $\widehat{\delta}_\rho^{\rm GR}$
provides a useful baseline for interpreting the value of ${\cal G}_\rho(s^\ast)$.

\subsection{Loops versus UV completions: $d=10$ with maximal supersymmetry}
\label{ssec: sugra}

We start with a warm-up with maximally supersymmetric partial waves in $d=10$.
In the supersymmetric theory, the graviton lives in the same supermultiplet as a scalar, therefore sharing the universal polarizations with the Einstein tree-level amplitude
\be
\mathcal{M}^{\rm susy} = \mathcal{T}(\epsilon_i,p_i) \bar{\mathcal{M}}\label{eq: susy amp}
\ee
with $\mathcal{T}$ in \eqref{eq: susy structure}.
The technical simplification of this setup lies in our ability to directly decompose  $\bar{\mathcal{M}}$ into scalar partial waves and read off the supersymmetric spectral density using the scalar partial wave (see \cite{Arkani-Hamed:2022gsa} for further discussion). 
Susy partial waves have a single label, $J$:
\be \label{susy partial waves}
\bar{\mathcal{M}}(s,t) =  s^{-4-\frac{d-4}{2}}\sum_{J\,\rm even}
a_J(s)n_J P_J(1+2t/s)
\ee
where $n_J=\frac{(4\pi)^{d/2}(2J+d-3)}{\pi\Gamma(\frac{d-2}{2})} (J{+}1)_{d-4}$
and $P_J(x)$ is given in \eqref{eq: PJ}. 
For the low-energy supergravity amplitude $\bar{\mathcal{M}}=\frac{8\pi G}{stu}$, this gives the known
partial wave $a_J^{\rm sugra}= s^{\frac{d-2}{2}}\widehat{\delta}_J^{\rm sugra}$ with
\be
\widehat{\delta}^{\rm sugra}_J=8\pi G \frac{2^{1-d} \pi ^{1-\frac{d}{2}} \Gamma \left(\frac{d-4}{2}\right) \Gamma (J+1)}{\Gamma (d+J-3)},
\ee
which refines \eqref{delta GR} into an expression valid also for finite spins.

\begin{figure}[t]
\centering\includegraphics[width=0.9\textwidth]{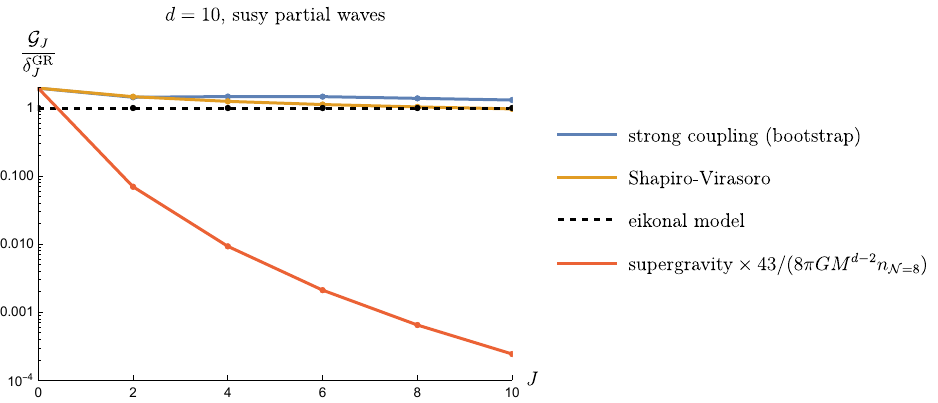}
\caption{Comparison of moments of for ten-dimensional supersymmetric partial waves for string amplitudes, strongly coupled amplitudes, the eikonal model, and loops of $n_{\mathcal{N}=8}$ light supermultiplets.
In all cases the moments are integrated to $s^\ast=\infty$, but the limit is saturated rather
rapidly above the respective scales ($s^\ast\sim M_s^2$ or $M_{\rm pl}^2$).}\label{fig: d10 susy compare}
\end{figure}

The first UV scenario we consider is weakly coupled string theory, where in the (sub-Planckian) energy range of interest $\bar{\mathcal{M}}$ is given by the Shapiro-Virasoro amplitude
\be
\bar{\mathcal{M}}^{\rm string} = \frac{8\pi G}{stu} \fft{\Gamma\big(1- \fft{\alpha^\prime s}{4}\big)\Gamma\big(1-\fft{\alpha^\prime  t}{4}\big)\Gamma\big(1- \fft{\alpha^\prime u}{4}\big)}{\Gamma\big(1+\fft{\alpha^\prime  s}{4}\big)\Gamma\big(1+ \fft{\alpha^\prime t}{4}\big)\Gamma\big(1+ \fft{\alpha^\prime u}{4}\big)}\,.\label{eq: superstring}
\ee
Setting $\alpha^\prime=4/M_s^2$, the imaginary part of the amplitude is given as
\be
{\rm Im}\,\bar{\mathcal{M}}^{\rm string} = \pi \sum_{n=1}^{\infty} \big(-{\rm Res}_{s=n M_s^2}\, \bar{\mathcal{M}}^{\rm string} \big) \delta\big(s-n M_s^2\big)\,.
\ee

\begin{figure}[t]
\centering\includegraphics[width=0.9\textwidth]{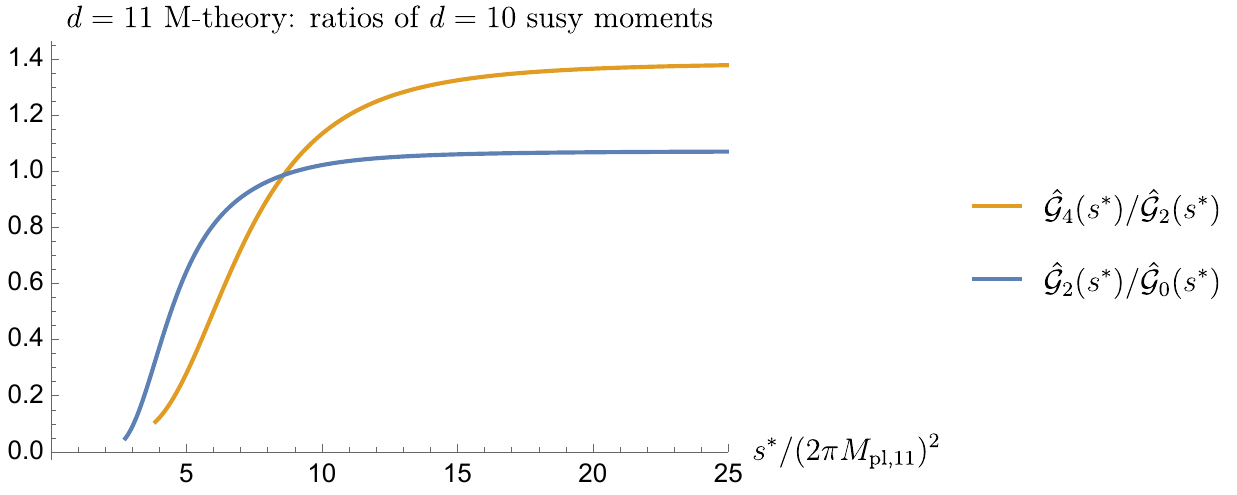}
\caption{
Adjacent-spin ratios of $d=10$ partial wave moments integrated up to $s^\ast$ in the scenario of $M$-theory compactified on a large circle, computed by dimensionally reducing the strongly-coupled $d=11$ supergravity amplitude bootstrapped in \cite{Guerrieri:2022sod}.
High-spin onset is seen to happen near
the 11-dimensional Planck scale. We truncated at low $s^\ast$ where the ratio becomes noisy.}\label{fig: integrated 11d10d}
\end{figure}

The second UV scenario we consider is strong coupling, where an amplitude which \emph{minimizes}
the first higher-derivative correction ($\sim {\rm Riem}^4$) to supergravity has been obtained using
the nonperturbative S-matrix bootstrap in \cite{Guerrieri:2021ivu,Guerrieri:2022sod}.\footnote{We thank the authors of \cite{Guerrieri:2022sod} for sending us the
partial wave data which we used to make figures \ref{fig: d10 susy compare} and \ref{fig: integrated 11d10d}.}
The idea there is that minimizing the departure from GR is a way of making the cutoff as high as possible, and indeed the resulting extremal amplitude only displays structure at the Planck scale.

Both of these UV scenarios are contrasted in Figure \ref{fig: d10 susy compare} with the computable contributions of $N$ supermultiplets at low energies.
In the figure we chose $N$ so that $\mathcal{G}_{J=0}$ is roughly the same for all theories.

It is evident from the figure that the high-spin content of candidate UV completions is very similar to that of the eikonal model,  to which they asymptote at large spins, while supermultiplet loops decay much more rapidly.
We stress that in absolute units, all partial waves do decay rapidly with spin:
${\cal G}_J^{\rm eik}\sim J^{-6}$ and ${\cal G}^{\rm loops}_J\sim J^{-12}$ in $d=10$.
The essential point is that loops decay comparatively faster with spin than UV completions, which seem flat in Figure \ref{fig: d10 susy compare} because we divide by $\widehat{\delta}^{\rm eik}_J\sim J^{-6}$.

To avoid confusion, we also stress that, due to the nature  of the supersymmetric partial wave decomposition in \eqref{susy partial waves}, $J=0$ in that formula accounts physically for spin-4 states. Thus, in the present subsection, ``higher-spin'' means $J\geq 0$, whereas in the rest of the paper where we do not employ supersymmetry, ``higher-spin'' means $J\geq 4$.

\begin{figure}[t]
\centering\includegraphics[width=0.9\textwidth]{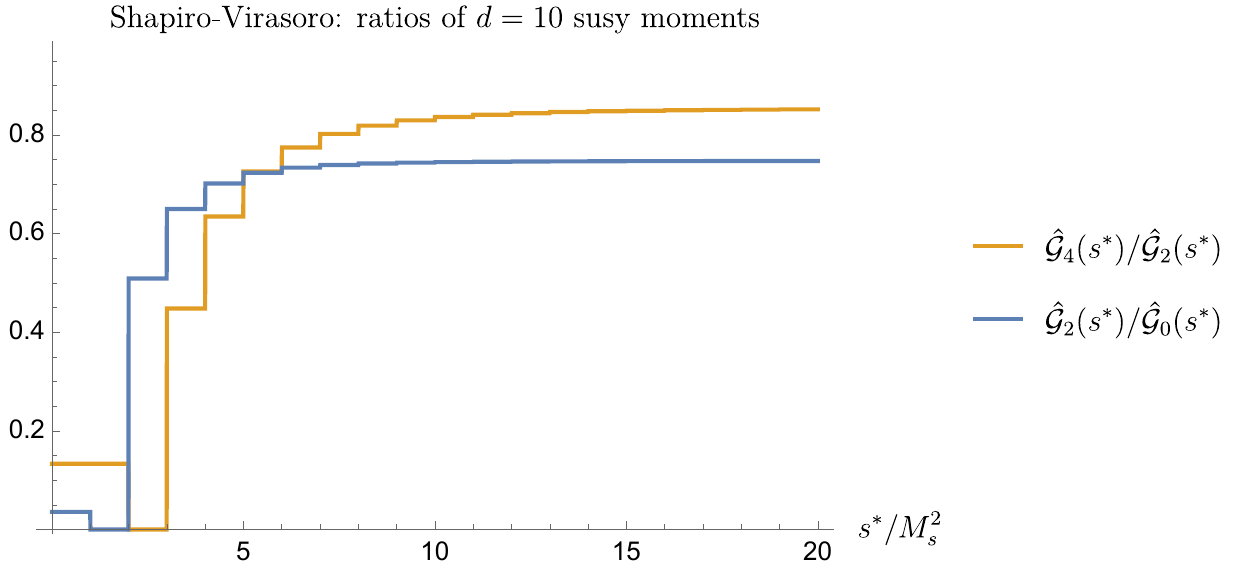}
\caption{Adjacent-spin ratios of $d=10$ partial wave moments integrated up to $s^\ast$ for the Veneziano-Shapiro amplitude. High-spin onset happens rapidly after the second mass level.}\label{fig: integrated 10d VS}
\end{figure}

As a quantitative illustration of the adjacent-spin ratios \eqref{hs criterion}, we consider two models in $d=10$. In Figure \ref{fig: integrated 10d VS} we display the moments for the supersymmetric partial waves \eqref{susy partial waves} extracted
from the Shapiro-Virasoro amplitude.
At energies below the string scale we use the one-loop formula
${\cal G}_J\propto (\widehat{\delta}_J^{\rm sugra})^2$,
which yields small but nonzero ratios.  Note that in the plot we divided by the eikonal model to get a more meaningful normalization:
\be
\hat{\cal G}_J(s)\equiv {\cal G}_J(s)/\widehat{\delta}^{\rm sugra}_J.
\ee
The figure shows a sequence of plateaus corresponding to each mass level, which quickly rise to an $\mathcal{O}(1)$ value. This shows that $\hs \approx M_s$ in string theory.  More precisely, defining for example the scale of high-spin onset for susy partial waves by
$\frac{\hat{\cal G}_2(\hs)}
{\hat{\cal G}_0(\hs)}\approx 0.5$,
one gets either
$\hs= M_s\sqrt{2}$ or $\hs= M_s\sqrt{3}$
depending on the precise criterion.

We also consider the dimensional reduction of $d=11$ $M$-theory on a large circle of circumference
$R_{11}M_{\rm pl,11}\gg 1$.  Then we have a 10-dimensional theory with a large number of light Kaluza-Klein modes
(corresponding to the light D0 branes of strongly coupled IIA superstring theory).
In this scenario, at energies $R_{11}^{-1}\ll \sqrt{s}\ll M_{\rm pl,11}$ we recover the amplitude of a 11-dimensional theory with a single massless graviton supermultiplet running in the loop,
for which ${\rm Im}\,a_{J,11}\propto (\widehat{\delta}^{{\rm GR}}_{J,11})^2$.  Reducing
to $d=10$, we get integrated spectral densities
that are proportional to $(s^\ast)^{9/2}$ in this regime such that adjacent-spin ratios are $s^\ast$-independent, equal to small constants which we find numerically to be
\be
\frac{\hat{\cal G}_2(s^\ast)}
{\hat{\cal G}_0(s^\ast)} \approx 0.026\,, \quad
\frac{\hat{\cal G}_4(s^\ast)}
{\hat{\cal G}_2(s^\ast)}\approx 0.104\,, \quad (\mbox{for } R_{11}^{-2}\ll s^\ast\ll M^2_{\rm pl,11}).
\label{M theory 10d}
\ee
When we approach the 11-dimensional Planck scale, which represents the scale $\cutoff$ of the 10-dimensional theory, we expect the
ratio to increase sharply until it saturates to a value close to unity as suggested by the flatness of the upper curves in Fig.~\ref{fig: d10 susy compare}.

Of course, our knowledge of strongly interacting $M$-theory amplitudes is limited, but we find it very satisfying that exactly such a transition is visible in the bootstrapped amplitudes of strongly coupled $d=11$ supergravity which were
generously provided to us by the authors of
 \cite{Guerrieri:2022sod}.
We dimensionally reduced their $d=11$ amplitude and integrated over $s$ according to \eqref{eq: moments} with $d=10$.  The result for the ratios in \eqref{M theory 10d} at higher energies is displayed in Fig.~\ref{fig: integrated 11d10d}.
Defining the scale of high-spin onset for example by 
$\frac{\hat{\cal G}_2(\hs)}
{\hat{\cal G}_0(\hs)}=0.5$ gives $\Lambda_o\approx 4\pi M_{\rm pl,11}$ where
$M_{\rm pl,11}=(8\pi G^{(11)})^{-9}$ is the 11-dimensional Planck scale.\footnote{At low $s^\ast$ the numerical results of \cite{Guerrieri:2022sod} seem consistent with \eqref{M theory 10d} but show significant oscillations.  It is not clear to us whether these are significant, given the small size of the amplitudes in this region.}
This demonstrate that, in a decompactification scenario,  high-spin onset happens at the higher-dimensional Planck scale.

These two examples support the identification of $\hs$ with the universal field theory cutoff $\cutoff$: in both stringy and decompactification limits, high-spin onset happens at the expected scales, $M_s$ and $M_{\rm pl,d+k}$.

\subsection{Loops versus UV completions in $d=10$ without supersymmetry}

We can perform a similar partial wave analysis for non-supersymmetric matter loops,
by applying the general partial wave decomposition reviewed in section \ref{subsec: partial wave} to
the matter loops computed in section \ref{sec: amplitudes}.  The main technical
complication is that partial waves exist in many SO$(d{-}1)$ Young Tableaux.  Using orthogonality of partial waves,
we extracted the corresponding moments ${\cal G}_\rho$
by integrating the (imaginary part of the) calculated amplitudes against the partial waves given in the ancillary files of \cite{Caron-Huot:2022jli}. For reference,  we provide the low-lying partial wave data of tree-level graviton scattering $\hat{\delta}^{\rm GR}$ in Table \ref{tab: GR phase}.

\begin{table}[t]
\centering
\begin{tabular}{c|c|c|c|c|c}
$\rho$ & $(8\pi G)^{-1}\hat{\delta}_\rho^{\rm GR}$ & $\rho$ & $(8\pi G)^{-1}\hat{\delta}_\rho^{\rm GR}$ & $\rho$ & $(8\pi G)^{-1}\hat{\delta}_\rho^{\rm GR}$ \\\hline
$(0)$ & $0.31$ &  (2,2)& (8.77,0.63) & (3,3,1)  & 1.4 \\
$(2)$ & $(3.92,0.47)$ & (3,2) & 2.8 & (5,3,1) & 1.2 \\
$(4)$ & $(141.85,2.19,0.63)$ & (4,2) & (6.57,2.6,0.79)& (4,3) & 1 \\
$(6)$ & $(14.9,1.76,0.74)$ & (5,2) & 1.9 & (5,3) & 2 \\
(1,1,1)& 1.2 & (6,2) & (3.31,1.74,0.87) & (6,3) & 1 \\
(3,1,1)& (21.94,1.02) & (2,2,2) & 3.2 & (4,4) & 1 \\
(5,1,1) & (4.57,1.03) & (4,2,2) & 1.55 & (6,4)& 1 \\
(2,1) & 0.6 & (6,2,2) & 1.27  \\
(3,1) & 1.35 & (3,2,1)& 1.8  \\
(4,1) & (5.76,0.74) & (4,2,1) & 2.91  \\
(5,1) & (11.07,1.23) & (5,2,1) & 1.4  \\
(6,1) & (3.04,0.83) & (6,2,1) & 1.89  \\
\end{tabular}
\caption{Eigenvalues of the GR phase shift $\hat{\delta}_\rho^{\rm GR}$ for various low-lying partial wave representations in $d=10$.}
\label{tab: GR phase}
\end{table}
 
Relaxing supersymmetry also allows us to consider graviton scattering for other string amplitudes,
such as the heterotic string and bosonic string, reviewed in appendix \ref{app: string amp}.
(The bosonic string amplitude has an unphysical tachyon pole, but its residue does not contribute to higher-spin partial waves and so we ignored this issue.)

\begin{figure}[t]
\centering\includegraphics[width=0.8\textwidth]{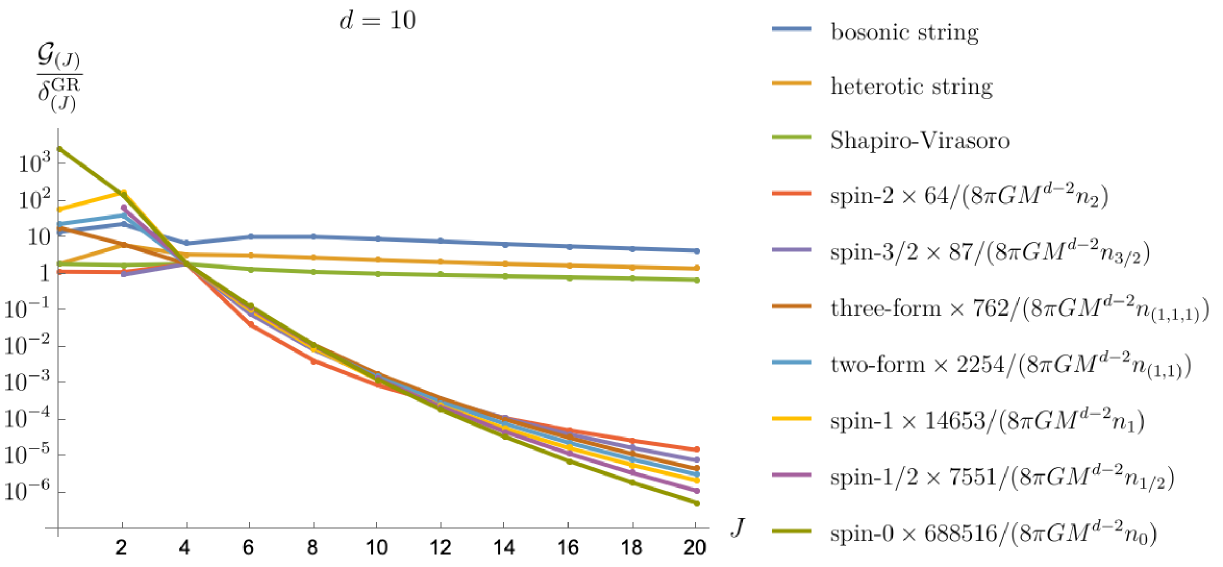}
\caption{Comparison of partial wave moments in the traceless symmetric representation $\rho=(J)$
for various string amplitude and computable massless matter loops.
Matter loops were normalized to match the eikonal model at spin $J=4$.
Treating ${\cal G}_{(4)}/\delta^{\rm GR}_{(4)}\lesssim O(1)$ as an overall budget, we see that (roughly)
the maximum number of spin-2 species which can exist below any scale $M$ is $n_2\lesssim 64/(8\pi G M^{d-2})$,
whereas the allowed number of scalars $n_0$ is larger by a factor $688516/64$.}\label{fig: d10 sym J}
\end{figure}

The resulting moments are shown for traceless symmetric representations $(J)$ for all matter fields in Fig \ref{fig: d10 sym J}.
Again we observe a qualitative distinction between all string models of the UV, whose moments are relatively flat after dividing by the GR partial waves $\delta^{\rm GR}$ (where we take the trace of the phase shift matrix), while all loop scenarios decay much more rapidly.

We also briefly explored the partial waves for the Veneziano amplitude for 10D supergluons.  Although it is hard to do an apple-to-apple comparison since the external states are different, we observed a rapid decay with $J$ similar to matter. We conclude that the flat behavior of the Shapiro-Virasoro data in Fig.~\ref{fig: d10 sym J} is specific to string models \emph{which unitarize gravity}.

\begin{figure}[t]
\centering\includegraphics[width=0.8\textwidth]{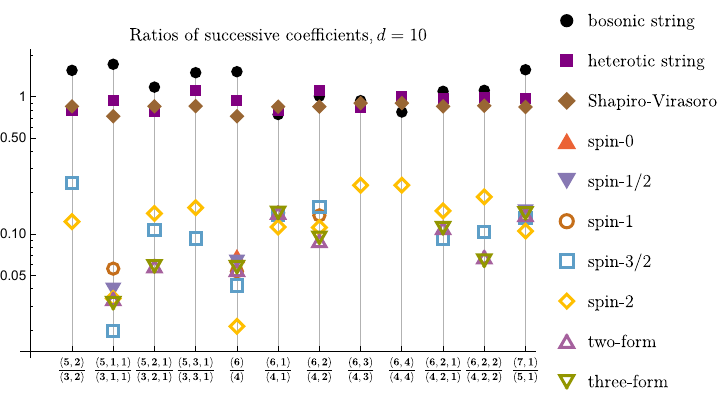}
\caption{Ratios $R_\rho$ of successive moments for various representations $\rho$ and amplitudes.
A clear gap is visible between computable loops and UV completions, for all representations.
}\label{fig: 10ratios}
\end{figure}

We also find that the traceless symmetric representation $(J)$ is not special in any way. Partial waves in all representations display similar behavior.
In  Fig.~\ref{fig: 10ratios} we show the ratios
\be \label{R ratio}
R_\rho(s^\ast)=\frac{{\cal G}_{\rho}(s^\ast)/\widehat{\delta}^{\rm GR}_{\rho}}
{{\cal G}_{\rho-2}(s^\ast)/\widehat{\delta}^{\rm GR}_{\rho-2}}
\ee
for various amplitudes for a variety of representations.
For matter loops, we consider $s^\ast\gg M$ in which case the ratios essentially do not depend on $s^\ast$.
We can easily observe that for sufficiently high spins, these ratios are always significantly larger (by a factor 4 or more) for strings than for loops.\footnote{We found a single exception: the ratio $R_{5,1}$ is larger for spin-2 and three-form fields than for the Shapiro-Virasoro amplitude.  For this Tableau shape, we simply take $R_{7,1}$ instead to be the first high-spin case.}
The gap between these values gives estimates for the constants $C_\rho$ which can be used to define the high-spin onset scale $\hs$ in \eqref{def cutoff}.

\subsection{$d=5$ and $d=4$ with mass distributions}

We now repeat the same exercise for other dimensions. As prototypes, we focus on $d=5$ and $d=4$. In lower dimensions, taking the massless approximations for towers of spin-$2$ and spin-$3/2$ particles may be misleading because their contributions to the sum rules may not decay fast enough with increasing $J$. This then shows a fake high spin prevalence, as we neglect significant contributions from massive particles. The worst situation occurs in four dimensions, where taking the massless limit of spin-$2$ amplitudes simply yields the infrared divergence. To deal with this subtlety, we have to sum over all possible masses with certain distributions $\rho(m)$
\be
\int_{m=0}^{m=\fft{M}{2}} d\rho(m) = n\,.
\ee
For simplicity, we only consider a mass distribution corresponding to a single extra dimension,
\be
\rho(m)= n \fft{2m}{M}\,.
\ee
To simplify the analysis without losing generality and validity, whenever the mass distribution is needed, we actually count the contribution to $s^\ast$ where we estimate the dispersive moments \eqref{eq: moments}.

\begin{figure}[t]
\centering\includegraphics[width=0.9\textwidth]{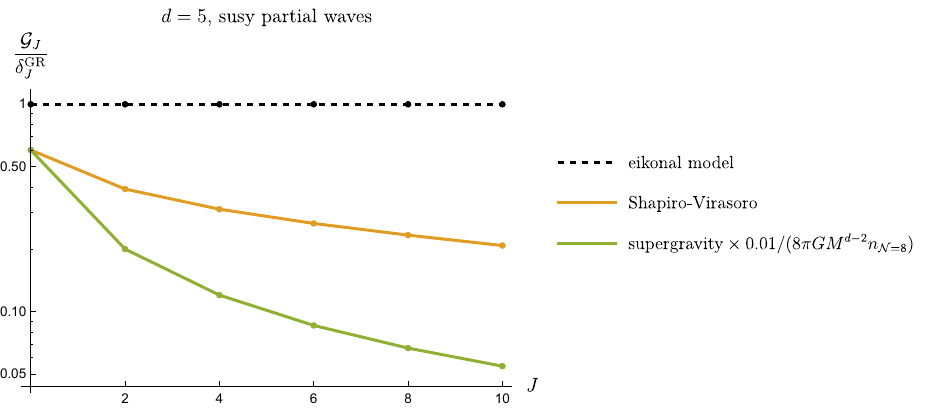}
\caption{Five-dimensional comparison of supersymmetric partial wave coefficients between the string amplitude, supergravity and the eikonal model.}\label{fig: d5 susy compare}
\end{figure}

It is also instructive to start with the supersymmetric partial wave in $d=5$, as shown in Fig \ref{fig: d5 susy compare}. As expected, supergravity exhibits much faster decay behavior at large spin, even without the mass distribution.

\begin{figure}[t]
\centering\includegraphics[width=0.9\textwidth]{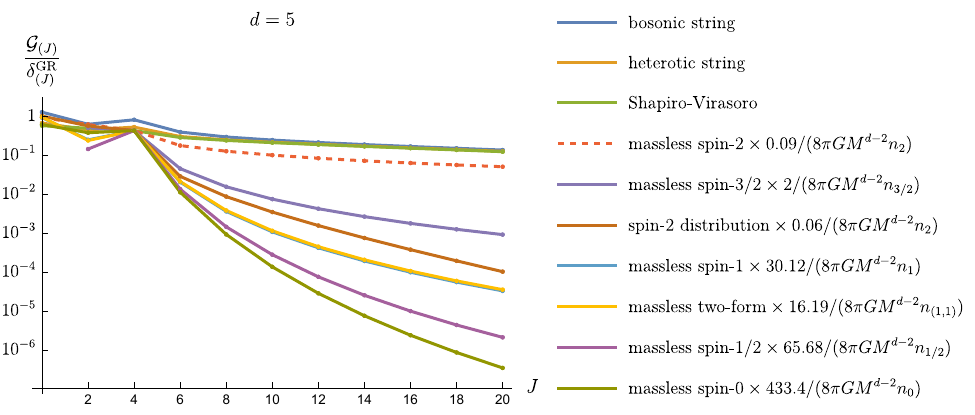}
\caption{Five-dimensional comparison of partial wave coefficients in the traceless symmetric representation (even spin) between the string amplitudes and matter loops. The red dashed line is for the massless limit of spin-$2$ species, which is not physical.}\label{fig: d5 sym J}
\end{figure}

\begin{figure}[t]
\centering\includegraphics[width=0.8\textwidth]{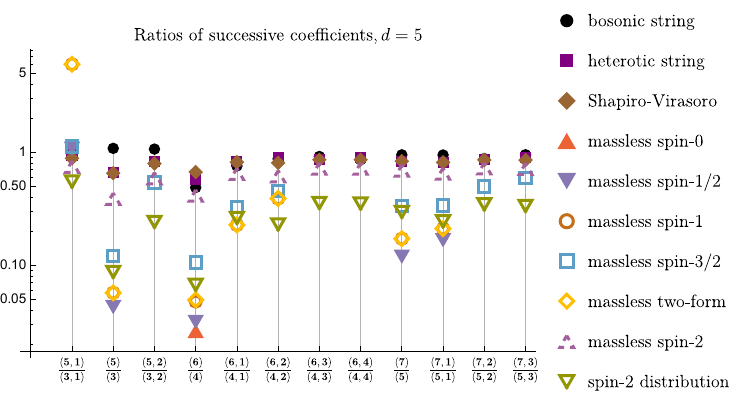}
\caption{Ratios of successive coefficients of partial waves in five dimensions for different matter fields, in comparison with the string amplitudes.
While the scenario of a large number of massless spin-2 fields (dashed triangle) gives ratios close to those of strings, the more physically relevant scenario of a distribution of massive spin-2 fields yields smaller ratios.
}\label{fig: 5ratios}
\end{figure}

\begin{figure}[htbp]
    \centering
    \begin{subfigure}[b]{0.9\textwidth}
        \centering
        \includegraphics[width=\textwidth]{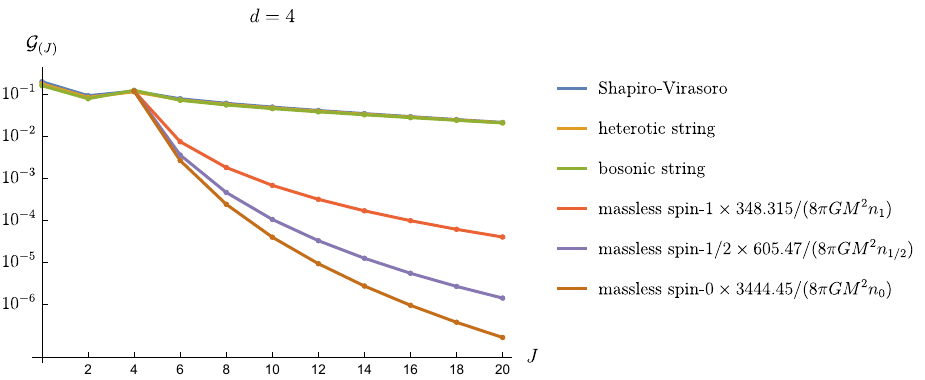} 
        \label{fig: low spin}
        \caption{}
    \end{subfigure}
    \hfill
    \begin{subfigure}[a]{0.95\textwidth}
        \centering
        \includegraphics[width=\textwidth]{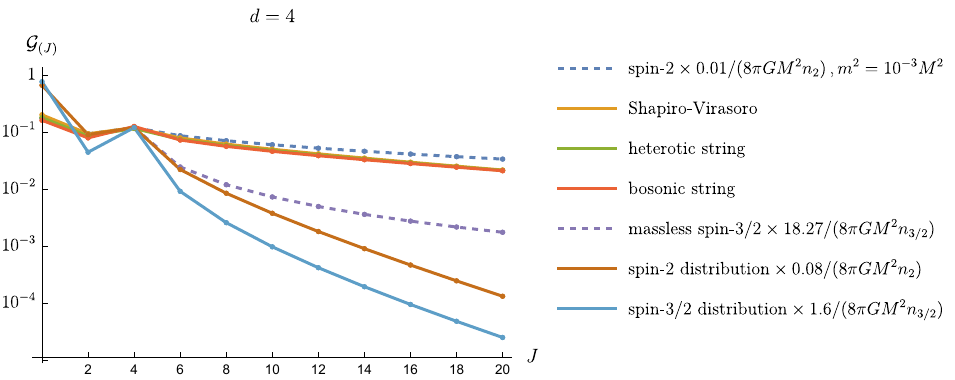} 
        \label{fig: gravtino and graviton}
        \caption{}
    \end{subfigure}
    \caption{Four-dimensional comparison of partial wave coefficients in the traceless symmetric representation (even spin) between the string amplitudes and matter loops. The top plot shows ordinary matter while the second plot shows gravitino and graviton loops, the dashed lines labelling unphysical cases: large numbers of nearly massless spin-$3/2$ and spin-$2$ species (as opposed to power-law mass distributions of these fields).}
    \label{fig: 4devenJ}
\end{figure}

Next, we display the traceless symmetric representation. In $d=5$, there are both even spin and odd spin; we only present the even spin in Fig. \ref{fig: d5 sym J}. The behavior for odd $J$ follows a similar pattern. We observe that all string amplitudes behave similarly flat by increasing the spin, demonstrating significant high-spin prevalence. In contrast, the matter loops are significantly dominated by the low-spin states. It is worth noting that we use the dashed line to display the massless limit of spin-$2$ particles to emphasize that their dispersive moments are not physical without being weighted by the mass distribution. More generally, we demonstrate the ratios $R_\rho$ (with $J=3,5,7)$ of all amplitudes in all representation, see Fig \ref{fig: 5ratios}. The pattern for matter loops shows that although there could exist anomalously large ratios for certain states, there always exists representations with highly suppressed ratios. On the other hand, all ratios are around $\mathcal{O}(1)$ for string amplitudes!

We can repeat this analysis in $d=4$. Here, it is worth noting that the phase shift of Einstein gravity contains an IR divergence in $d=4$. For this reason, we do not normalize the moments by the Einstein phase shift. We display the even spin traceless symmetric representation in Fig. \ref{fig: 4devenJ}, separating it into two subfigures for the $j<3/2$ (\ref{fig: 4devenJ}a) and $j=3/2, 2$ cases ((\ref{fig: 4devenJ}b)). In subfigure (\ref{fig: 4devenJ}b), we also contrast the massless limit (with a numerical IR regulator for spin-2) with an actual distribution of masses. In these plots we added in quadrature the couplings to different external graviton helicity.

\subsection{Partial wave moments in AdS/CFT}

Our considerations readily extent to theories of gravity in AdS$_{d}$ space, where by the
AdS/CFT duality the natural observables are correlation functions in the boundary conformal field theory (CFT$_{d-1}$).
The four-graviton scattering amplitude, in particular, becomes the correlation function of four stress tensors.

The analog of the partial waves $a_\rho(s)$ in \eqref{eq: moments} are the three-point couplings squared
$f_{TT{\cal O}_{\Delta,\rho}}^2$ which appear in the conformal block decomposition of the $\< TTTT\>$ correlator. More precisely, for proper normalization, we should divide by the analog of the disconnected part of the S-matrix, which is the OPE data of the generalized-free-fields correlator (GFF) (a product of factorized $\<TT\>$ two-point functions).
Furthermore, taking the imaginary part of the amplitude corresponds to its so-called ``double-discontinuity'' which multiplies the data by a factor $2\sin^2(\frac{\pi}{2}(\ldots))$ \cite{Caron-Huot:2017vep}.
Thus, altogether  (see \cite{Paulos:2016fap,Caron-Huot:2021enk,vanRees:2023fcf,vanRees:2022zmr,Chang:2023szz} for more details): 
\begin{equation}
 R_{\rm AdS}^2\int ds\, {\rm Im}\,a_\rho(s) \,\Longleftrightarrow\,
 4\sum_{\Delta} 2\sin^2\big(\tfrac{\pi}{2}(\Delta-J-2\Delta_T)\big)
 \frac{(f_{TT{\cal O}_{\Delta,\rho}})^2}{
 (f_{TT{\cal O}_{\Delta,\rho}}^{\rm GFF})^2}\Delta, \label{CFT sum 1}
\end{equation}
where the sum is over the scaling dimensions of operators in the ${\rm SO}(d{-}1)$ representation $\rho$, and the denominator is regarded as a smooth function of $\Delta$.\footnote{When couplings with multiple tensor structures exist, we assume in \eqref{CFT sum 1} that a basis is used in which
$(f_{TT{\cal O}_{\Delta,\rho}}^{\rm GFF})^2$ is diagonal at large $\Delta$, as constructed in \cite{Li:2021snj} and earlier in AdS$_4$/CFT$_3$ in \cite{Caron-Huot:2021kjy}.}
(There is an extra factor of 4 on the right because of the state spacing for GFF, $\int d(\Delta^2)\to 4\sum_\Delta$.)
Thus, up to inverse powers of $R_{\rm AdS}$ (or equivalently, of $\Delta$), the moments \eqref{eq: moments} can be written in ${\rm CFT}_{d-1}$ as
\begin{equation} \label{CFT sum}
    {\cal G}^{\rm CFT}_\rho(\Delta^*) \equiv
    8\frac{d-2}{\pi}
\sum_{\Delta} \sin^2\big(\tfrac{\pi}{2}(\Delta-J-2\Delta_T)\big)
 \frac{(f_{TT{\cal O}_{\Delta,\rho}})^2}{
 (f_{TT{\cal O}_{\Delta,\rho}}^{\rm GFF})^2}\Delta^{1-d}\,.
\end{equation}

A famous conjecture is that any CFT with a large-$N$ expansion and a large gap of higher-spin single-trace operators should be holographic, in the sense of being described by a local effective theory of gravity down to lengths $R_{\rm AdS}/\Delta_{\rm gap}$ in the bulk \cite{Heemskerk:2009pn}.
Essentially, the assumptions mean that the higher-spin contribution to sums like \eqref{CFT sum} receive negligible contributions from states with $\Delta-J<\Delta_{\rm gap}$. 

Important aspects of this conjecture are now well established.
For example, it has been shown that contributions from states with twist $\tau>\Delta_{\rm gap}$ to certain CFT sum rules admit a $1/\Delta_{\rm gap}$ expansion which makes these sum rules identical to S-matrix dispersion relations \cite{Caron-Huot:2021enk} (see \cite{Chang:2023szz} for technical extension to spinning operators),
enabling to uplift S-matrix dispersive sum rules to CFT sum rules.  It is then shown in \cite{Caron-Huot:2022jli} that dispersion relations for four-graviton S-matrices with corresponding spectral assumptions (namely ${\rm Im}\, a_\rho(s)=0$ for $J>2$ and $s<4M_{\rm gap}^2$) imply that the amplitude is equal to Einstein's prediction plus corrections suppressed by inverse powers of $M_{\rm gap}$.
Together, these show that $\<TTTT\>$ in holographic theories can only differ from the prediction of  Einstein's theory in the bulk by $1/\Delta_{\rm gap}$ corrections.

We expect that the partial wave moments in \eqref{CFT sum} will enable to  strengthen this result,
by allowing for small but nonzero contributions from states below $\Delta_{\rm gap}$, ie. loops in the bulk theory.  In particular, it should now be possible to make do without the ``large-$N$ expansion'' assumption and deal directly with CFTs at finite $N$, ie. ${\rm SU}(N_c)$ gauge theory with $N_c=100$.

It would be interesting to evaluate the
adjacent-spin-ratios in various CFTs that are known to not be dual to local gravity,  ie. the 3D critical Ising model, and to verify that high-spin onset then happens at low $\Delta$.
In general, we expect a bound of the form $\Delta_{\rm o}< C\,(c_T)^{1/(d-2)}$ on the scale of high-spin onset in any ${\rm CFT}_{d-1}$. Other S-matrix conjectures stated below should also have direct CFT analogs. 

\section{Conclusion}
\label{sec: conclusion}

In this paper we analyzed graviton-graviton scattering amplitudes at moderate energies, below or near the Planck scale.
We considered calculable one-loop effects at low energies as well as  examples of ultraviolet completions,
namely weakly coupled strings or strongly coupled Planck-scale physics.

Our main motivation stems from Kramers-Kronig-type dispersion relations which relate Newton's constant at low energies to positive scattering probabilities at other energies. Thus, Newton's constant sets an overall ``budget'' for graviton scattering at arbitrary energies. It is natural to ask how this budget can be distributed among known and unknown ingredients, and what this tells us about gravitational physics at various length scales.

Our main technical results are explicit formulas for the one-loop contributions of various fields of spin $j\leq 2$
to graviton-graviton scattering, and their partial wave decompositions.
This was then used to quantify, in various ways, how much of the graviton scattering ``budget'' can be carried by field-theoretic contributions. Our main conclusions are two-fold.

First, in section \ref{sec: sum rules}, we obtain, for any cutoff $M$ in $d=4$,
a species-like bound of the form $G M^2 N\lesssim \mathcal{O}(\log)$ on the number $N$ of species below the scale $M$ for which the one-loop approximation is reliable.
This gives a relatively sharp notion of the species bound.
In cases where all the species are ordinary matter fields of spin $\leq 1$, somewhat stronger estimates can be obtained by considering the graviton propagator, as discussed in section \ref{ssec: resummed props}.

Second, figures \ref{fig: d10 sym J}-\ref{fig: 4devenJ} reveal
a clear numerical distinction between the high-spin content of any field-theoretic contribution, and that of ultraviolet completions of gravity, in various space-time dimensions.
By their nature, field theory correlators exhibit either exponential or strong power-law decay. This motivates the conjecture that the scale of \emph{high-spin onset} of the graviton amplitude represents a universal cutoff, such that at lengths shorter than $1/\cutoff$ spacetime, cannot be described using local fields.

The high-spin onset scale $\hs$ has several appealing properties.
As demonstrated in section \ref{ssec: sugra}, for weakly coupled string theory, the onset scale coincides with the mass of spinning string scales on the graviton trajectory.  In scenarios where a large extra dimension decompactifies, the onset scale automatically picks up the Planck mass of the higher-dimensional theory, even though the graviton-graviton amplitude of the compactified theory is still far from saturating unitarity.

It is also worth noting that $\hs$ is expected to be insensitive to spinning bound states held together by ordinary field-theoretic interactions, since all states enter weighted by their couplings to two gravitons. Thus, for example, near the positronium threshold $\sqrt{s}\approx 2m_e$,
the graviton cross-section is still low-spin dominated  despite the existence of many high-spin bound states, simply because the graviton wavelength is much shorter than their size. Furthermore, when integrating to $s^\ast\gtrsim 4m_e^2$, one can evaluate \eqref{eq: moments} perturbatively using constituent fields and avoid any discussion of bound states altogether.\footnote{One may object that perturbation theory breaks down when $\sqrt{s}/(2m_e)=1+O(\alpha^2)$, necessitating Coulomb resummation. However, this near-threshold region can be avoided by deforming the contour in \eqref{eq: moments} away from the real axis.}
Similarly, $\hs$ is insensitive to the existence of many spinning mesons bound by the QCD string, since stress tensor correlators at distances $\ll \Lambda_{\rm QCD}^{-1}$ can be computed using approximately free quark and gluon fields, rather than mesons.  The QCD string is very different from a fundamental string because it admits a local stress tensor.

Many questions remain.

A main corollary of the conjecture $\cutoff=\hs$ is that, at distances shorter than $1/\hs$, \emph{all} scattering amplitudes must be profoundly modified, not only those of gravitons. Can this be shown directly?
Experimentally, this would immediately establish that $\hs> \mathcal{O}({\rm TeV})$.

Proving the converse would also be interesting: is effective field theory guaranteed to work at lengths larger than $1/\hs$, possibly in a higher dimensional sense?  Tentatively, this means that as far as low-multiplicity scattering amplitudes of matter and gravitons are concerned, one has an ordinary non-gravitational QFT possibly strongly interacting with itself but weakly coupled to gravity, together with a list of spin-3/2 and spin-2 fields which interact weakly among themselves and with the QFT. Irrelevant interactions should be bounded by inverse powers of $\hs$.

Technical refinements on the moments \eqref{eq: moments} could be useful. Are there universal lower and upper bounds on the ``budget'' ${\cal G}_\rho(\infty)$ for each representation $\rho$, in any UV-complete graviton S-matrix,
\be
 C^{\rm min}_\rho \leq {\cal G}_\rho(\infty)/{\widehat{\delta}^{\rm GR}_\rho} \leq C^{\rm max}_\rho\,?
\ee
Furthermore, does there exist ``thresholds'', such that 
if ${\cal G}_\rho(s)/{\widehat{\delta}^{\rm GR}_\rho}>C^{\rm thres}_\rho$ in \emph{some} representation $\rho$, then
onset becomes unavoidable in all representations?
What are best choices for the constants $C_\rho$ in \eqref{def cutoff}, besides that they should separate the scenarios in figure \ref{fig: 10ratios}?

Could the onset condition be phrased in a technically simpler way that would not require computing partial waves?  Can the condition be somehow related to arguments involving the entropy of small black holes?

One can also ask if matter-matter interactions can significantly modify the shape of the 1-loop partial waves in ie. figure \ref{fig: 4devenJ}? This question could be answered by studying stress-tensor four-point correlators in non-gravitational QFTs (for spins $J_\rho>2$ and to leading order in gravity-matter interactions, graviton self-interactions do not contribute to ${\rm Im\,}a_\rho$).
For three-point stress tensor correlators in CFT, it is known  that indeed an arbitrary unitary theory lies within the convex hull of free theories \cite{Hofman:2008ar,Zhiboedov:2013opa}.

Are there constraints on the matter content below $\hs$?
The 1-loop contributions from different fields contribute very differently to different partial waves.
For example, scalar loops only contribute to symmetric traceless tensor representations $(J)$ whereas only loops of spin-two particles contribute to $(J,3)$ and $(J,4)$ representations, as can be seen in figure \ref{fig: 10ratios}.  One could also refine the partial waves by looking at different tensor structures,
ie. in $d=4$ one has separate couplings to $++$ and $+-$ helicity states.  It might be, for example, that saturating the budget in some structures but not in others is incompatible with crossing symmetry at high energies.  This could rule out scenarios where there is a large number of ordinary matter fields but no graviton KK modes, or perhaps even scenarios where the spectrum is not supersymmetric on the average, providing a concrete step towards the emergent string conjecture mentioned in introduction.

Can the identification $\hs=\cutoff$ lead to progress on conjectures involving the rate of change $\partial_\phi \log\Lambda$ of the cutoff along moduli space, such as the ``pattern'' of \cite{Castellano:2023stg,Castellano:2023jjt}?
At the very least, it suggests a novel definition of $\partial_\phi \log\Lambda$ using three-point coupling between a light scalar field and two higher-spin states at the onset scale. Indeed, such couplings determine how the onset mass is influenced by the scalar (in much the same way that Yukawa couplings set fermion masses in the Standard Model).

\acknowledgments We thank Aidan Herderschee, Juan Maldacena, Brian McPeak, Sebastian Mizera, Julio Parra-Martinez, David Simmons-Duffin, Piotr Tourkine for useful discussions and comments. 
The work of S.C.H. is supported by the National Science and Engineering Council of Canada (NSERC), the Canada Research Chair program, reference number CRC-2022-00421, and the Simons Collaboration on the Nonperturbative Bootstrap. The work of Y.Z.L is supported by the US National Science Foundation under Grant No. PHY- 2209997.

\appendix 

\section{Ghosts for Rarita-Schwinger and two-form fields}
\label{app: ghost}

In this appendix, we review the derivation of ghost actions for spin-$3/2$ Rarita-Schwinger particle and two-form.

\paragraph{Rarita-Schwinger} The ghost of Rarita-Schwinger can be derived by following the standard FP procedure. A simple way is to consider the gauge-fixing conditions for $\Psi$ and $\bar{\Psi}$ separately
\be
\gamma^\mu \slashed{D}\bar{\Psi}_\mu = \bar{\omega}\,,\quad \gamma^\mu \Psi_\mu=\omega\,,
\ee
where we treat $\omega$ and $\bar{\omega}$ independently. Then we can follow the FP trick to insert the Dirac delta functions in the path integral
\be
Z[g]=\int D\Psi D\bar{\Psi} \, \delta\big(\gamma\cdot\Psi -\omega\big) \delta\big(\gamma\cdot \slashed{D} \Psi -\bar{\omega}\big) {\rm det}[\slashed{D}]^{-1} {\rm det}[\slashed{D}\slashed{D}]^{-1} e^{-S[\Psi,\bar{\Psi},g]}\,,
\ee
where the determinants come from the infinitesimal gauge transformation
\be
\delta\Psi_\mu = \partial_\mu \epsilon\,.
\ee
Because the partition function does not depend on $\omega$ and $\bar{\omega}$, we can shift the partition function by
\be
{\rm const}=\int D\omega D\bar{\omega} e^{-\fft{1}{\xi}\int \bar{\omega}\omega}\,.
\ee
Integrating out $\omega, \bar{\omega}$ yields
\be
Z[g]=\int D\Psi D\bar{\Psi} {\rm det}[\slashed{D}]^{-3} e^{-S[\Psi,\bar{\Psi},g] + S_{\rm gf}}\,,\label{eq: Z 3/2}
\ee 
where the gauge-fixing term is
\be
S_{\rm gf}\sim \fft{1}{\xi}\int d^dx \sqrt{g} \,\bar{\Psi}_\mu \gamma^{\mu}\gamma^\nu\gamma^\rho D_\nu \Psi_\rho\,.
\ee
We choose $\xi=4/(1-d)$ to remove additional $1/p^2$ pole in the propagator. The partition function \eqref{eq: Z 3/2} contains $ {\rm det}[\slashed{D}]^{-3}$, suggesting that the FP ghosts of Rarita-Schwinger contributes like $-3$ of fermions.

\paragraph{Two-form} The theory with massless two-form is known as the reducible gauge theory \cite{Siegel:1980jj,Batalin:1983ggl}. A subtlety arises from the fact that its FP ghosts are also gauge fields, therefore requesting the ghosts for ghosts. For $2$-form, this fact is manifest by the gauge transformation
\be
\delta B_{\mu\nu} = \nabla_{[\mu} \theta_{\nu]}\,.
\ee
However, the naive application of the standard FP quantization does not produce the correct results. The essential reason is that the gauge-fixing functions are constrained. We consider the gauge-fixing function by inserting a Dirac delta function
\be
\delta\Big(\nabla^\mu B_{\mu\nu} -\omega_\nu\Big)\,.
\ee
However, different from ordinary gauge fields, this gauge-fixing function introduces nontrivial field $\omega$, which is constrained by
\be
\nabla^\nu \omega_\nu=\nabla^\mu \nabla^\nu B_{\mu\nu}\equiv 0\,.
\ee
The constraints on the gauge-fixing function completely spoil the FP quantization procedure. There are several procedures are developed to systematically deal with this reducible gauge theory \cite{Schwarz:1978cn,Schwarz:1979ae,Obukhov:1982dt,Buchbinder:1988tj}, the famous one is Batalin-Vilkovisky formalism as the generalization of BRST  \cite{Batalin:1983ggl}. 

Here, we focus on $2$-form, which is simple enough to avoid those intricate quantization procedures, and we simply review a slight generalization of FP procedure in \cite{Siegel:1980jj}. We follow the same routine to make the delta function manifest
\be
Z[g]=\int DB\, \delta\big(\nabla^\mu B_{\mu\nu}-\omega_\nu\big) {\rm det}[\nabla\cdot d] e^{-S[B,g]}\,,
\ee
where $d$ is the exterior derivative. As we mentioned, now inserting $e^{-\omega^2}$ does not give the averaging due to $\nabla\cdot \omega=0$, namely
\be
\int D\omega e^{-\fft{1}{\xi}\int \omega^2} \delta(\nabla\cdot \omega) \neq {\rm const}\,.
\ee
To rescue this, we now disentangle the constraint on $\omega$ by introducing its own ``gauge-fixing'' function $\delta(\nabla \cdot\omega-\omega^\prime)$. Then one obvious constant averaging is
\be
\int D\omega D\omega^\prime\,\delta(\nabla\cdot \omega-\omega^\prime) e^{-\fft{1}{\xi}\int \big(\omega^2+(\omega^\prime)^2+\omega^\prime \fft{1}{\nabla^2}\omega^\prime\Big)}  {\rm det}[\nabla]={\rm const}\,.
\ee
To prove this, we can integrate $\omega$, this integration cancels ${\rm det}[\nabla]$ and sets $\omega_\mu=1/\nabla_\mu \omega^\prime$ to eliminate $\omega^\prime 1/\nabla^2 \omega^\prime$ in the exponent. The remaining is the trivial Gaussian $\omega^\prime$. To proceed, we can integrate $\omega^\prime$ first and then integrate $\omega$
\be
Z[g]=\int DB\, \,{\rm det}[\nabla\cdot d]{\rm det}[\nabla] e^{-S[B,g]+S_{gf}}\,,
\ee
where the gauge-fixing term is
\be
S_{gf}\sim \fft{1}{\xi}\int d^dx\sqrt{g} \nabla^\mu B_{\mu\nu} \nabla^\rho B_{\rho}\,^\nu\,.
\ee
We can now count the ghosts. First, we have ${\rm det}[\nabla\cdot d]$, this term gives the ghosts as $-2$ spin-$1$ gauge fields. Since this ghost is itself a gauge field, it induces further ghost behaving like $+4$ scalars. Besides, we have ${\rm det}[\nabla]$, contributing $-1$ scalars. Therefore, we have ghosts for ghosts contributing as $+3$ real scalars. To keep the terminology light, we may say that the ghosts for $2$-form contribute like $-2$ vectors and $-1$ real scalar. This is consistent with the degrees of freedom for massless $2$-form. 

\section{Discontinuities of 1-loop master integrals in $d$ dimensions}
\label{app: branch}

In this appendix, we provide the necessary formulas for taking the discontinuity of master integrals. 

In general dimensions, we can obtain the imaginary part of all master integrals in the physical $s$-channel kinematics $s>4m^2, t<0$. This is sufficient for computing the spectral density of one-loop amplitudes in section \ref{sec: QFT breaking}.
\be
& {\rm Im}\, I_{\rm bub}(s,m)= \frac{\pi  \Gamma \left(\frac{d-2}{2}\right)}{(4\pi)^{\fft{d}{2}}\Gamma (d-2)}s^{\frac{d-4}{2}} \beta (s,m)^{d-3}\,,\nn\\
& {\rm Im}\, I_{\rm tri}(s,m)=  \frac{\pi  \Gamma \left(\frac{d-2}{2}\right)}{(4\pi)^{\fft{d}{2}}\Gamma (d-2)}\frac{2^{d-3} m^{d-4} \beta (s,m)^{d-3}}{s}\, _2F_1\left(\frac{d-3}{2},\frac{d-2}{2},\frac{d-1}{2},\beta (s,m)^2\right)\,,\nn\\
& {\rm Im}\, I_{\rm box}(s,t,m)= -\frac{\pi  \Gamma \left(\frac{d-2}{2}\right)}{(4\pi)^{\fft{d}{2}}\Gamma (d-2)}\frac{2^{d-2} m^{d-4} \beta (s)^{d-3}}{s t \beta(s,t,m)^2}\times \nn\\
& \Big((d-3) \left(1-\beta (s)^2\right)^{2-\frac{d}{2}} \left(1-\frac{1}{\beta (s,t)^2}\right)^{2-\frac{d}{2}} F_1\left(\frac{1}{2},\frac{4-d}{2},1,\frac{3}{2},\frac{1}{\beta (s,t)^2},\frac{\beta (s)^2}{\beta (s,t)^2}\right)\nn\\
& -(d-4) \, _2F_1\left(\frac{d-3}{2},\frac{d-2}{2},\frac{d-1}{2},\beta (s)^2\right) \, _2F_1\left(\frac{d-3}{2},1,\frac{3}{2},\frac{1}{\beta (s,t)^2}\right)\Big)\,.\label{eq: s-channel Im}
\ee
For ${\rm Im}\, I_{\rm box}(s,u,m)$, we simply repace $t\rightarrow u$. All others discontinuities are identically zero.

In section \ref{sec: sum rules}, we apply the dispersive sum rules to constrain the number of species in $d=4$, where we have to evaluate arc integrals for the low-energy contribution to sum rules, coming from massive fields below the cutoff. The arc integral can be performed by wrapping around the discontinuities but with fixed-$u$ or fixed-$t$. Let's fix $u=-p^2<0$. In addition to the normal threshold captured by the preceding formula, the discontinuities now also include anomalous thresholds (especially for $I_{\rm box}(s,t,m)$).
Specializing here to $d=4$, a careful analysis yields in the $s$ plane:
\be
{\rm Disc}\, I_{\rm bub}(s,m)&=2i\pi \,\beta(s,m) \theta(s-4m^2)\,,\nn\\
{\rm Disc}\, I_{\rm bub}(t,m)&=-2i\pi \,\beta(p^2-s,m) \theta(p^2-s-4m^2)\,,\nn\\
{\rm Disc}\, I_{\rm tri}(s,m)&= \frac{2i\pi \theta(s-4 m^2)}{s} \log \left(\frac{1+\beta (s,m)}{1-\beta (s,m)}\right) \,,\nn\\
{\rm Disc}\, I_{\rm tri}(t,m)&=- \frac{2i\pi  \theta(p^2-s-4 m^2)}{s} \log \left(\frac{1+\beta (p^2-s,m)}{1-\beta (p^2-s,m)}\right)\,,\nn\\
{\rm Disc}\, I_{\rm box}(s,u,m)&=\frac{4\pi i \theta(s-4 m^2)}{p^2 s \beta \left(s,-p^2,m\right)} \log \left(\frac{\beta \left(s,-p^2,m\right)+\beta (s,m)}{\beta \left(s,-p^2,m\right)-\beta (s,m)}\right)  \,,\nn\\
{\rm Disc}\, I_{\rm box}(t,u,m)&=-\frac{4\pi i  \theta(-4 m^2+p^2-s)}{p^2 \left(p^2-s\right) \beta \left(p^2-s,-p^2,m\right)} \log \left(\frac{\beta \left(p^2-s,m\right)+\beta \left(p^2-s,-p^2,m\right)}{\beta \left(p^2-s,-p^2,m\right)-\beta \left(p^2-s,m\right)}\right) \,,\nn\\
{\rm Disc}\, I_{\rm box}(s,t,m)&=4\pi i \Big\{\,\frac{\theta(4 m^2-s) \theta(-4 m^2+p^2-s)  \log \left(\frac{\beta \left(p^2-s,m\right)+\beta \left(s,p^2-s,m\right)}{\beta \left(s,p^2-s,m\right)-\beta \left(p^2-s,m\right)}\right)}{s \left(p^2-s\right) \beta \left(s,p^2-s,m\right)} \nn\\
& +\frac{\theta(s-4 m^2) \theta(4 m^2-p^2+s) \log \left(\frac{\beta \left(s,p^2-s,m\right)+\beta (s,m)}{\beta \left(s,p^2-s,m\right)-\beta (s,m)}\right)}{s \left(s-p^2\right) \beta \left(s,p^2-s,m\right)}\nn\\
&-\frac{\theta(s-4 m^2) \theta(-4 m^2+p^2-s) \left(\log \left(\frac{\beta \left(s,p^2-s,m\right)+\beta (s,m)}{\beta (s,m)-\beta \left(s,p^2-s,m\right)}\right)-\log \left(\frac{\beta \left(p^2-s,m\right)+\beta \left(s,p^2-s,m\right)}{\beta \left(p^2-s,m\right)-\beta \left(s,p^2-s,m\right)}\right)\right)}{s \left(p^2-s\right) \beta \left(s,p^2-s,m\right)}\Big\}\,,\label{eq: fix u Disc}
\ee
while others are zero.

\section{Tree-level string amplitudes}
\label{app: string amp}

In this appendix, we collect the tree-level closed string amplitudes with arbitrary polarizations so that we can use them in higher dimensions. We take the open string amplitudes from \cite{Kawai:1985xq} and perform the double copy to explicitly write down the closed string amplitudes.

As we have already mentioned in the main text, due to the maximal supersymmetry, the Shapiro-Virasoro (superstring) amplitude is proportional to the tree-level Einstein gravity. We then have \eqref{eq: susy amp} and \eqref{eq: superstring}, which we copy here but setting $\alpha^\prime =4$ to keep the expression light
\be
\mathcal{M}^{SV}=-8\pi G  \fft{\Gamma\big(- s\big)\Gamma\big(-t\big)\Gamma\big(-u\big)}{\Gamma\big(1+s\big)\Gamma\big(1+t\big)\Gamma\big(1+ u\big)} \mathcal{T}(\epsilon_i,p_i) \.,
\ee
where $\mathcal{T}(\epsilon_i,p_i)$ is explicitly given in \eqref{eq: susy structure}. As expected, this tensor structure is the square of the tensor structure of supersymmetric open string amplitude $ \mathcal{T}=(\mathcal{K}^{{\rm susy}})^2$ where \cite{Kawai:1985xq}
\be
\mathcal{K}^{\rm susy}(\epsilon_i,p_i)=H_{14}H_{23}+H_{13}H_{24}+H_{12}H_{34}+2(X_{1234}+X_{1243}+X_{1324})\,,
\ee

The tensor structure of the bosonic open string is
\be
& \mathcal{K}^{\rm bos}(\epsilon_i,p_i)=(1+s)(1+u)(s u -t-1) H_{14}H_{23}+(1+t)(1+s)(s t -u-1) H_{13}H_{24}\nn\\
& +(1+t)(1+u)(t u-s-1) H_{12}H_{34}+ 2(1+s)(1+t)\big((1-2u) X_{1234}+(1-2t)X_{1243}+(1-2s)X_{1324}\big)\nn\\
&+4 S(1+s)(1+t)(1+u)
\ee
Using the KLT relation \cite{Kawai:1985xq}, the bosonic and heterotic closed string amplitudes are given by
\be
& \mathcal{M}^{\rm bos}=-8\pi G \fft{(\mathcal{K}^{{\rm bos}})^2}{(1+s)^2(1+t)^2(1+u)^2} \fft{\Gamma\big(- s\big)\Gamma\big(-t\big)\Gamma\big(-u\big)}{\Gamma\big(1+s\big)\Gamma\big(1+t\big)\Gamma\big(1+ u\big)} \,,\nn\\
&  \mathcal{M}^{\rm het}=-8\pi G \fft{\mathcal{K}^{{\rm bos}}\mathcal{K}^{\rm susy}}{(1+s)(1+t)(1+u)} \fft{\Gamma\big(- s\big)\Gamma\big(-t\big)\Gamma\big(-u\big)}{\Gamma\big(1+s\big)\Gamma\big(1+t\big)\Gamma\big(1+ u\big)}\,.
\ee

\bibliographystyle{JHEP}
\bibliography{refs}

\providecommand{\href}[2]{#2}\begingroup\raggedright\begin{thebibliography}{10}

\bibitem{Donoghue:1994dn}
J.~F. Donoghue, \emph{{General relativity as an effective field theory: The
  leading quantum corrections}},
  \href{http://dx.doi.org/10.1103/PhysRevD.50.3874}{\emph{Phys. Rev. D} {\bf
  50} (1994) 3874--3888}, [\href{https://arxiv.org/abs/gr-qc/9405057}{{\tt
  gr-qc/9405057}}].

\bibitem{Donoghue:2022eay}
J.~F. Donoghue, \emph{{Quantum General Relativity and Effective Field Theory}}.
\newblock 2023,
  \href{http://dx.doi.org/10.1007/978-981-19-3079-9\_1-1}{10.1007/978-981-19-3079-9\_1-1}.

\bibitem{Ooguri:2006in}
H.~Ooguri and C.~Vafa, \emph{{On the Geometry of the String Landscape and the
  Swampland}},
  \href{http://dx.doi.org/10.1016/j.nuclphysb.2006.10.033}{\emph{Nucl. Phys. B}
  {\bf 766} (2007) 21--33}, [\href{https://arxiv.org/abs/hep-th/0605264}{{\tt
  hep-th/0605264}}].

\bibitem{Brennan:2017rbf}
T.~D. Brennan, F.~Carta and C.~Vafa, \emph{{The String Landscape, the
  Swampland, and the Missing Corner}},
  \href{http://dx.doi.org/10.22323/1.305.0015}{\emph{PoS} {\bf TASI2017} (2017)
  015}, [\href{https://arxiv.org/abs/1711.00864}{{\tt 1711.00864}}].

\bibitem{Palti:2019pca}
E.~Palti, \emph{{The Swampland: Introduction and Review}},
  \href{http://dx.doi.org/10.1002/prop.201900037}{\emph{Fortsch. Phys.} {\bf
  67} (2019) 1900037}, [\href{https://arxiv.org/abs/1903.06239}{{\tt
  1903.06239}}].

\bibitem{vanBeest:2021lhn}
M.~van Beest, J.~Calder\'on-Infante, D.~Mirfendereski and I.~Valenzuela,
  \emph{{Lectures on the Swampland Program in String Compactifications}},
  \href{https://arxiv.org/abs/2102.01111}{{\tt 2102.01111}}.

\bibitem{Arkani-Hamed:2006emk}
N.~Arkani-Hamed, L.~Motl, A.~Nicolis and C.~Vafa, \emph{{The String landscape,
  black holes and gravity as the weakest force}},
  \href{http://dx.doi.org/10.1088/1126-6708/2007/06/060}{\emph{JHEP} {\bf 06}
  (2007) 060}, [\href{https://arxiv.org/abs/hep-th/0601001}{{\tt
  hep-th/0601001}}].

\bibitem{Harlow:2022ich}
D.~Harlow, B.~Heidenreich, M.~Reece and T.~Rudelius, \emph{{Weak gravity
  conjecture}},
  \href{http://dx.doi.org/10.1103/RevModPhys.95.035003}{\emph{Rev. Mod. Phys.}
  {\bf 95} (2023) 035003}, [\href{https://arxiv.org/abs/2201.08380}{{\tt
  2201.08380}}].

\bibitem{Dvali:2000xg}
G.~R. Dvali and G.~Gabadadze, \emph{{Gravity on a brane in infinite volume
  extra space}},
  \href{http://dx.doi.org/10.1103/PhysRevD.63.065007}{\emph{Phys. Rev. D} {\bf
  63} (2001) 065007}, [\href{https://arxiv.org/abs/hep-th/0008054}{{\tt
  hep-th/0008054}}].

\bibitem{Veneziano:2001ah}
G.~Veneziano, \emph{{Large N bounds on, and compositeness limit of, gauge and
  gravitational interactions}},
  \href{http://dx.doi.org/10.1088/1126-6708/2002/06/051}{\emph{JHEP} {\bf 06}
  (2002) 051}, [\href{https://arxiv.org/abs/hep-th/0110129}{{\tt
  hep-th/0110129}}].

\bibitem{Dvali:2007hz}
G.~Dvali, \emph{{Black Holes and Large N Species Solution to the Hierarchy
  Problem}}, \href{http://dx.doi.org/10.1002/prop.201000009}{\emph{Fortsch.
  Phys.} {\bf 58} (2010) 528--536},
  [\href{https://arxiv.org/abs/0706.2050}{{\tt 0706.2050}}].

\bibitem{Dvali:2007wp}
G.~Dvali and M.~Redi, \emph{{Black Hole Bound on the Number of Species and
  Quantum Gravity at LHC}},
  \href{http://dx.doi.org/10.1103/PhysRevD.77.045027}{\emph{Phys. Rev. D} {\bf
  77} (2008) 045027}, [\href{https://arxiv.org/abs/0710.4344}{{\tt
  0710.4344}}].

\bibitem{Brustein:2009ex}
R.~Brustein, G.~Dvali and G.~Veneziano, \emph{{A Bound on the effective
  gravitational coupling from semiclassical black holes}},
  \href{http://dx.doi.org/10.1088/1126-6708/2009/10/085}{\emph{JHEP} {\bf 10}
  (2009) 085}, [\href{https://arxiv.org/abs/0907.5516}{{\tt 0907.5516}}].

\bibitem{vandeHeisteeg:2022btw}
D.~van~de Heisteeg, C.~Vafa, M.~Wiesner and D.~H. Wu, \emph{{Moduli-dependent
  Species Scale}},  \href{https://arxiv.org/abs/2212.06841}{{\tt 2212.06841}}.

\bibitem{Cribiori:2022nke}
N.~Cribiori, D.~L\"ust and G.~Staudt, \emph{{Black hole entropy and
  moduli-dependent species scale}},
  \href{http://dx.doi.org/10.1016/j.physletb.2023.138113}{\emph{Phys. Lett. B}
  {\bf 844} (2023) 138113}, [\href{https://arxiv.org/abs/2212.10286}{{\tt
  2212.10286}}].

\bibitem{vandeHeisteeg:2023dlw}
D.~van~de Heisteeg, C.~Vafa, M.~Wiesner and D.~H. Wu, \emph{{Species scale in
  diverse dimensions}},
  \href{http://dx.doi.org/10.1007/JHEP05(2024)112}{\emph{JHEP} {\bf 05} (2024)
  112}, [\href{https://arxiv.org/abs/2310.07213}{{\tt 2310.07213}}].

\bibitem{Cribiori:2023ffn}
N.~Cribiori, D.~Lust and C.~Montella, \emph{{Species entropy and
  thermodynamics}},
  \href{http://dx.doi.org/10.1007/JHEP10(2023)059}{\emph{JHEP} {\bf 10} (2023)
  059}, [\href{https://arxiv.org/abs/2305.10489}{{\tt 2305.10489}}].

\bibitem{Basile:2024dqq}
I.~Basile, N.~Cribiori, D.~Lust and C.~Montella, \emph{{Minimal black holes and
  species thermodynamics}},
  \href{http://dx.doi.org/10.1007/JHEP06(2024)127}{\emph{JHEP} {\bf 06} (2024)
  127}, [\href{https://arxiv.org/abs/2401.06851}{{\tt 2401.06851}}].

\bibitem{Lee:2019wij}
S.-J. Lee, W.~Lerche and T.~Weigand, \emph{{Emergent strings from infinite
  distance limits}},
  \href{http://dx.doi.org/10.1007/JHEP02(2022)190}{\emph{JHEP} {\bf 02} (2022)
  190}, [\href{https://arxiv.org/abs/1910.01135}{{\tt 1910.01135}}].

\bibitem{Castellano:2022bvr}
A.~Castellano, A.~Herr\'aez and L.~E. Ib\'a\~nez, \emph{{The emergence proposal
  in quantum gravity and the species scale}},
  \href{http://dx.doi.org/10.1007/JHEP06(2023)047}{\emph{JHEP} {\bf 06} (2023)
  047}, [\href{https://arxiv.org/abs/2212.03908}{{\tt 2212.03908}}].

\bibitem{Blumenhagen:2023yws}
R.~Blumenhagen, A.~Gligovic and A.~Paraskevopoulou, \emph{{The emergence
  proposal and the emergent string}},
  \href{http://dx.doi.org/10.1007/JHEP10(2023)145}{\emph{JHEP} {\bf 10} (2023)
  145}, [\href{https://arxiv.org/abs/2305.10490}{{\tt 2305.10490}}].

\bibitem{Basile:2023blg}
I.~Basile, D.~L\"ust and C.~Montella, \emph{{Shedding black hole light on the
  emergent string conjecture}},
  \href{http://dx.doi.org/10.1007/JHEP07(2024)208}{\emph{JHEP} {\bf 07} (2024)
  208}, [\href{https://arxiv.org/abs/2311.12113}{{\tt 2311.12113}}].

\bibitem{Bedroya:2024ubj}
A.~Bedroya, R.~K. Mishra, Wiesner and Max, \emph{{Density of States, Black
  Holes and the Emergent String Conjecture}},
  \href{https://arxiv.org/abs/2405.00083}{{\tt 2405.00083}}.

\bibitem{Etheredge:2022opl}
M.~Etheredge, B.~Heidenreich, S.~Kaya, Y.~Qiu and T.~Rudelius,
  \emph{{Sharpening the Distance Conjecture in diverse dimensions}},
  \href{http://dx.doi.org/10.1007/JHEP12(2022)114}{\emph{JHEP} {\bf 12} (2022)
  114}, [\href{https://arxiv.org/abs/2206.04063}{{\tt 2206.04063}}].

\bibitem{Castellano:2023stg}
A.~Castellano, I.~Ruiz and I.~Valenzuela, \emph{{Universal Pattern in Quantum
  Gravity at Infinite Distance}},
  \href{http://dx.doi.org/10.1103/PhysRevLett.132.181601}{\emph{Phys. Rev.
  Lett.} {\bf 132} (2024) 181601},
  [\href{https://arxiv.org/abs/2311.01501}{{\tt 2311.01501}}].

\bibitem{Castellano:2023jjt}
A.~Castellano, I.~Ruiz and I.~Valenzuela, \emph{{Stringy evidence for a
  universal pattern at infinite distance}},
  \href{http://dx.doi.org/10.1007/JHEP06(2024)037}{\emph{JHEP} {\bf 06} (2024)
  037}, [\href{https://arxiv.org/abs/2311.01536}{{\tt 2311.01536}}].

\bibitem{Dvali:2010vm}
G.~Dvali and C.~Gomez, \emph{{Species and Strings}},
  \href{https://arxiv.org/abs/1004.3744}{{\tt 1004.3744}}.

\bibitem{Caron-Huot:2021rmr}
S.~Caron-Huot, D.~Mazac, L.~Rastelli and D.~Simmons-Duffin, \emph{{Sharp
  Boundaries for the Swampland}},
  \href{http://dx.doi.org/10.1007/jhep07(2021)110}{\emph{JHEP} {\bf 07} (2021)
  110}, [\href{https://arxiv.org/abs/2102.08951}{{\tt 2102.08951}}].

\bibitem{Adams:2006sv}
A.~Adams, N.~Arkani-Hamed, S.~Dubovsky, A.~Nicolis and R.~Rattazzi,
  \emph{{Causality, analyticity and an IR obstruction to UV completion}},
  \href{http://dx.doi.org/10.1088/1126-6708/2006/10/014}{\emph{JHEP} {\bf 10}
  (2006) 014}, [\href{https://arxiv.org/abs/hep-th/0602178}{{\tt
  hep-th/0602178}}].

\bibitem{Camanho:2014apa}
X.~O. Camanho, J.~D. Edelstein, J.~Maldacena and A.~Zhiboedov, \emph{{Causality
  Constraints on Corrections to the Graviton Three-Point Coupling}},
  \href{http://dx.doi.org/10.1007/JHEP02(2016)020}{\emph{JHEP} {\bf 02} (2016)
  020}, [\href{https://arxiv.org/abs/1407.5597}{{\tt 1407.5597}}].

\bibitem{Bern:2021ppb}
Z.~Bern, D.~Kosmopoulos and A.~Zhiboedov, \emph{{Gravitational effective field
  theory islands, low-spin dominance, and the four-graviton amplitude}},
  \href{http://dx.doi.org/10.1088/1751-8121/ac0e51}{\emph{J. Phys. A} {\bf 54}
  (2021) 344002}, [\href{https://arxiv.org/abs/2103.12728}{{\tt 2103.12728}}].

\bibitem{Heemskerk:2009pn}
I.~Heemskerk, J.~Penedones, J.~Polchinski and J.~Sully, \emph{{Holography from
  Conformal Field Theory}},
  \href{http://dx.doi.org/10.1088/1126-6708/2009/10/079}{\emph{JHEP} {\bf 10}
  (2009) 079}, [\href{https://arxiv.org/abs/0907.0151}{{\tt 0907.0151}}].

\bibitem{Giddings:2009gj}
S.~B. Giddings and R.~A. Porto, \emph{{The Gravitational S-matrix}},
  \href{http://dx.doi.org/10.1103/PhysRevD.81.025002}{\emph{Phys. Rev. D} {\bf
  81} (2010) 025002}, [\href{https://arxiv.org/abs/0908.0004}{{\tt
  0908.0004}}].

\bibitem{Caron-Huot:2021enk}
S.~Caron-Huot, D.~Mazac, L.~Rastelli and D.~Simmons-Duffin, \emph{{AdS bulk
  locality from sharp CFT bounds}},
  \href{http://dx.doi.org/10.1007/JHEP11(2021)164}{\emph{JHEP} {\bf 11} (2021)
  164}, [\href{https://arxiv.org/abs/2106.10274}{{\tt 2106.10274}}].

\bibitem{Elvang:2013cua}
H.~Elvang and Y.-t. Huang, \emph{{Scattering Amplitudes}},
  \href{https://arxiv.org/abs/1308.1697}{{\tt 1308.1697}}.

\bibitem{Chowdhury:2019kaq}
S.~D. Chowdhury, A.~Gadde, T.~Gopalka, I.~Halder, L.~Janagal and S.~Minwalla,
  \emph{{Classifying and constraining local four photon and four graviton
  S-matrices}}, \href{http://dx.doi.org/10.1007/JHEP02(2020)114}{\emph{JHEP}
  {\bf 02} (2020) 114}, [\href{https://arxiv.org/abs/1910.14392}{{\tt
  1910.14392}}].

\bibitem{Caron-Huot:2022jli}
S.~Caron-Huot, Y.-Z. Li, J.~Parra-Martinez and D.~Simmons-Duffin,
  \emph{{Graviton partial waves and causality in higher dimensions}},
  \href{http://dx.doi.org/10.1103/PhysRevD.108.026007}{\emph{Phys. Rev. D} {\bf
  108} (2023) 026007}, [\href{https://arxiv.org/abs/2205.01495}{{\tt
  2205.01495}}].

\bibitem{Duff:1977ay}
M.~J. Duff, \emph{{Observations on Conformal Anomalies}},
  \href{http://dx.doi.org/10.1016/0550-3213(77)90410-2}{\emph{Nucl. Phys. B}
  {\bf 125} (1977) 334--348}.

\bibitem{Duff:1993wm}
M.~J. Duff, \emph{{Twenty years of the Weyl anomaly}},
  \href{http://dx.doi.org/10.1088/0264-9381/11/6/004}{\emph{Class. Quant.
  Grav.} {\bf 11} (1994) 1387--1404},
  [\href{https://arxiv.org/abs/hep-th/9308075}{{\tt hep-th/9308075}}].

\bibitem{Scherk:1978ta}
J.~Scherk and J.~H. Schwarz, \emph{{Spontaneous Breaking of Supersymmetry
  Through Dimensional Reduction}},
  \href{http://dx.doi.org/10.1016/0370-2693(79)90425-8}{\emph{Phys. Lett. B}
  {\bf 82} (1979) 60--64}.

\bibitem{Scherk:1979zr}
J.~Scherk and J.~H. Schwarz, \emph{{How to Get Masses from Extra Dimensions}},
  \href{http://dx.doi.org/10.1016/0550-3213(79)90592-3}{\emph{Nucl. Phys. B}
  {\bf 153} (1979) 61--88}.

\bibitem{Siegel:1980jj}
W.~Siegel, \emph{{Hidden Ghosts}},
  \href{http://dx.doi.org/10.1016/0370-2693(80)90119-7}{\emph{Phys. Lett. B}
  {\bf 93} (1980) 170--172}.

\bibitem{Batalin:1983ggl}
I.~A. Batalin and G.~A. Vilkovisky, \emph{{Quantization of Gauge Theories with
  Linearly Dependent Generators}},
  \href{http://dx.doi.org/10.1103/PhysRevD.28.2567}{\emph{Phys. Rev. D} {\bf
  28} (1983) 2567--2582}.

\bibitem{Kimura:1980aw}
T.~Kimura, \emph{{Quantum Theory of Antisymmetric Higher Rank Tensor Gauge
  Field in Higher Dimensional Space-time}},
  \href{http://dx.doi.org/10.1143/PTP.65.338}{\emph{Prog. Theor. Phys.} {\bf
  65} (1981) 338}.

\bibitem{Dunbar:1994bn}
D.~C. Dunbar and P.~S. Norridge, \emph{{Calculation of graviton scattering
  amplitudes using string based methods}},
  \href{http://dx.doi.org/10.1016/0550-3213(94)00385-R}{\emph{Nucl. Phys. B}
  {\bf 433} (1995) 181--208}, [\href{https://arxiv.org/abs/hep-th/9408014}{{\tt
  hep-th/9408014}}].

\bibitem{Bern:1995db}
Z.~Bern and A.~G. Morgan, \emph{{Massive loop amplitudes from unitarity}},
  \href{http://dx.doi.org/10.1016/0550-3213(96)00078-8}{\emph{Nucl. Phys. B}
  {\bf 467} (1996) 479--509}, [\href{https://arxiv.org/abs/hep-ph/9511336}{{\tt
  hep-ph/9511336}}].

\bibitem{Bern:1993tz}
Z.~Bern and A.~G. Morgan, \emph{{Supersymmetry relations between contributions
  to one loop gauge boson amplitudes}},
  \href{http://dx.doi.org/10.1103/PhysRevD.49.6155}{\emph{Phys. Rev. D} {\bf
  49} (1994) 6155--6163}, [\href{https://arxiv.org/abs/hep-ph/9312218}{{\tt
  hep-ph/9312218}}].

\bibitem{Green:1982sw}
M.~B. Green, J.~H. Schwarz and L.~Brink, \emph{{N=4 Yang-Mills and N=8
  Supergravity as Limits of String Theories}},
  \href{http://dx.doi.org/10.1016/0550-3213(82)90336-4}{\emph{Nucl. Phys. B}
  {\bf 198} (1982) 474--492}.

\bibitem{Bern:1998ug}
Z.~Bern, L.~J. Dixon, D.~C. Dunbar, M.~Perelstein and J.~S. Rozowsky, \emph{{On
  the relationship between Yang-Mills theory and gravity and its implication
  for ultraviolet divergences}},
  \href{http://dx.doi.org/10.1016/S0550-3213(98)00420-9}{\emph{Nucl. Phys. B}
  {\bf 530} (1998) 401--456}, [\href{https://arxiv.org/abs/hep-th/9802162}{{\tt
  hep-th/9802162}}].

\bibitem{Buric:2023ykg}
I.~Buric, F.~Russo and A.~Vichi, \emph{{Spinning partial waves for scattering
  amplitudes in d dimensions}},
  \href{http://dx.doi.org/10.1007/JHEP10(2023)090}{\emph{JHEP} {\bf 10} (2023)
  090}, [\href{https://arxiv.org/abs/2305.18523}{{\tt 2305.18523}}].

\bibitem{Caron-Huot:2014lda}
S.~Caron-Huot and J.~M. Henn, \emph{{Iterative structure of finite loop
  integrals}}, \href{http://dx.doi.org/10.1007/JHEP06(2014)114}{\emph{JHEP}
  {\bf 06} (2014) 114}, [\href{https://arxiv.org/abs/1404.2922}{{\tt
  1404.2922}}].

\bibitem{Passarino:1978jh}
G.~Passarino and M.~J.~G. Veltman, \emph{{One Loop Corrections for e+ e-
  Annihilation Into mu+ mu- in the Weinberg Model}},
  \href{http://dx.doi.org/10.1016/0550-3213(79)90234-7}{\emph{Nucl. Phys. B}
  {\bf 160} (1979) 151--207}.

\bibitem{Chetyrkin:1981qh}
K.~G. Chetyrkin and F.~V. Tkachov, \emph{{Integration by parts: The algorithm
  to calculate $\beta$-functions in 4 loops}},
  \href{http://dx.doi.org/10.1016/0550-3213(81)90199-1}{\emph{Nucl. Phys. B}
  {\bf 192} (1981) 159--204}.

\bibitem{Latosh:2024lhl}
B.~Latosh, \emph{{FeynGrav 3.0}},  \href{https://arxiv.org/abs/2406.14872}{{\tt
  2406.14872}}.

\bibitem{Capper:1973pv}
D.~M. Capper, G.~Leibbrandt and M.~Ramon~Medrano, \emph{{Calculation of the
  graviton selfenergy using dimensional regularization}},
  \href{http://dx.doi.org/10.1103/PhysRevD.8.4320}{\emph{Phys. Rev. D} {\bf 8}
  (1973) 4320--4331}.

\bibitem{Capper:1973mv}
D.~M. Capper and M.~J. Duff, \emph{{THE ONE LOOP NEUTRINO CONTRIBUTION TO THE
  GRAVITON PROPAGATOR}},
  \href{http://dx.doi.org/10.1016/0550-3213(74)90582-3}{\emph{Nucl. Phys. B}
  {\bf 82} (1974) 147--154}.

\bibitem{Capper:1973bk}
D.~M. Capper, \emph{{ON QUANTUM CORRECTIONS TO THE GRAVITON PROPAGATOR}},
  \href{http://dx.doi.org/10.1007/BF02735608}{\emph{Nuovo Cim. A} {\bf 25}
  (1975) 29}.

\bibitem{Capper:1974ed}
D.~M. Capper, M.~J. Duff and L.~Halpern, \emph{{Photon corrections to the
  graviton propagator}},
  \href{http://dx.doi.org/10.1103/PhysRevD.10.461}{\emph{Phys. Rev. D} {\bf 10}
  (1974) 461--467}.

\bibitem{DeMeyer:1974ed}
H.~E. De~Meyer, \emph{{Massive spin 1/2 contribution to the graviton
  propagator}}, \href{http://dx.doi.org/10.1007/BF02822259}{\emph{Lett. Nuovo
  Cim.} {\bf 11S2} (1974) 498--502}.

\bibitem{tHooft:1974toh}
G.~'t~Hooft and M.~J.~G. Veltman, \emph{{One loop divergencies in the theory of
  gravitation}}, {\emph{Ann. Inst. H. Poincare A Phys. Theor.} {\bf 20} (1974)
  69--94}.

\bibitem{Burns:2014bva}
D.~Burns and A.~Pilaftsis, \emph{{Matter Quantum Corrections to the Graviton
  Self-Energy and the Newtonian Potential}},
  \href{http://dx.doi.org/10.1103/PhysRevD.91.064047}{\emph{Phys. Rev. D} {\bf
  91} (2015) 064047}, [\href{https://arxiv.org/abs/1412.6021}{{\tt
  1412.6021}}].

\bibitem{Christensen:1978gi}
S.~M. Christensen and M.~J. Duff, \emph{{Axial and Conformal Anomalies for
  Arbitrary Spin in Gravity and Supergravity}},
  \href{http://dx.doi.org/10.1016/0370-2693(78)90857-2}{\emph{Phys. Lett. B}
  {\bf 76} (1978) 571}.

\bibitem{Bern:2022yes}
Z.~Bern, E.~Herrmann, D.~Kosmopoulos and R.~Roiban, \emph{{Effective Field
  Theory islands from perturbative and nonperturbative four-graviton
  amplitudes}}, \href{http://dx.doi.org/10.1007/JHEP01(2023)113}{\emph{JHEP}
  {\bf 01} (2023) 113}, [\href{https://arxiv.org/abs/2205.01655}{{\tt
  2205.01655}}].

\bibitem{deRham:2017avq}
C.~de~Rham, S.~Melville, A.~J. Tolley and S.-Y. Zhou, \emph{{Positivity bounds
  for scalar field theories}},
  \href{http://dx.doi.org/10.1103/PhysRevD.96.081702}{\emph{Phys. Rev. D} {\bf
  96} (2017) 081702}, [\href{https://arxiv.org/abs/1702.06134}{{\tt
  1702.06134}}].

\bibitem{Caron-Huot:2020cmc}
S.~Caron-Huot and V.~Van~Duong, \emph{{Extremal Effective Field Theories}},
  \href{https://arxiv.org/abs/2011.02957}{{\tt 2011.02957}}.

\bibitem{Bellazzini:2020cot}
B.~Bellazzini, J.~Elias~Mir\'o, R.~Rattazzi, M.~Riembau and F.~Riva,
  \emph{{Positive Moments for Scattering Amplitudes}},
  \href{https://arxiv.org/abs/2011.00037}{{\tt 2011.00037}}.

\bibitem{Tolley:2020gtv}
A.~J. Tolley, Z.-Y. Wang and S.-Y. Zhou, \emph{{New positivity bounds from full
  crossing symmetry}},  \href{https://arxiv.org/abs/2011.02400}{{\tt
  2011.02400}}.

\bibitem{Arkani-Hamed:2020blm}
N.~Arkani-Hamed, T.-C. Huang and Y.-T. Huang, \emph{{The EFT-Hedron}},
  \href{https://arxiv.org/abs/2012.15849}{{\tt 2012.15849}}.

\bibitem{Caron-Huot:2022ugt}
S.~Caron-Huot, Y.-Z. Li, J.~Parra-Martinez and D.~Simmons-Duffin,
  \emph{{Causality constraints on corrections to Einstein gravity}},
  \href{http://dx.doi.org/10.1007/JHEP05(2023)122}{\emph{JHEP} {\bf 05} (2023)
  122}, [\href{https://arxiv.org/abs/2201.06602}{{\tt 2201.06602}}].

\bibitem{Tokuda:2020mlf}
J.~Tokuda, K.~Aoki and S.~Hirano, \emph{{Gravitational positivity bounds}},
  \href{http://dx.doi.org/10.1007/JHEP11(2020)054}{\emph{JHEP} {\bf 11} (2020)
  054}, [\href{https://arxiv.org/abs/2007.15009}{{\tt 2007.15009}}].

\bibitem{Alberte:2020jsk}
L.~Alberte, C.~de~Rham, S.~Jaitly and A.~J. Tolley, \emph{{Positivity Bounds
  and the Massless Spin-2 Pole}},
  \href{http://dx.doi.org/10.1103/PhysRevD.102.125023}{\emph{Phys. Rev. D} {\bf
  102} (2020) 125023}, [\href{https://arxiv.org/abs/2007.12667}{{\tt
  2007.12667}}].

\bibitem{Caron-Huot:2024tsk}
S.~Caron-Huot and J.~Tokuda, \emph{{String loops and gravitational positivity
  bounds: imprint of light particles at high energies}},
  \href{https://arxiv.org/abs/2406.07606}{{\tt 2406.07606}}.

\bibitem{Henriksson:2022oeu}
J.~Henriksson, B.~McPeak, F.~Russo and A.~Vichi, \emph{{Bounding violations of
  the weak gravity conjecture}},
  \href{http://dx.doi.org/10.1007/JHEP08(2022)184}{\emph{JHEP} {\bf 08} (2022)
  184}, [\href{https://arxiv.org/abs/2203.08164}{{\tt 2203.08164}}].

\bibitem{Hong:2023zgm}
D.-Y. Hong, Z.-H. Wang and S.-Y. Zhou, \emph{{Causality bounds on scalar-tensor
  EFTs}}, \href{http://dx.doi.org/10.1007/JHEP10(2023)135}{\emph{JHEP} {\bf 10}
  (2023) 135}, [\href{https://arxiv.org/abs/2304.01259}{{\tt 2304.01259}}].

\bibitem{Albert:2024yap}
J.~Albert, W.~Knop and L.~Rastelli, \emph{{Where is tree-level string
  theory?}},  \href{https://arxiv.org/abs/2406.12959}{{\tt 2406.12959}}.

\bibitem{Haring:2022cyf}
K.~H\"aring and A.~Zhiboedov, \emph{{Gravitational Regge bounds}},
  \href{http://dx.doi.org/10.21468/SciPostPhys.16.1.034}{\emph{SciPost Phys.}
  {\bf 16} (2024) 034}, [\href{https://arxiv.org/abs/2202.08280}{{\tt
  2202.08280}}].

\bibitem{AccettulliHuber:2020oou}
M.~Accettulli~Huber, A.~Brandhuber, S.~De~Angelis and G.~Travaglini,
  \emph{{Eikonal phase matrix, deflection angle and time delay in effective
  field theories of gravity}},
  \href{http://dx.doi.org/10.1103/PhysRevD.102.046014}{\emph{Phys. Rev. D} {\bf
  102} (2020) 046014}, [\href{https://arxiv.org/abs/2006.02375}{{\tt
  2006.02375}}].

\bibitem{Alviani:2024sxx}
E.~Alviani and A.~Falkowski, \emph{{Matching and positivity beyond minimal
  coupling}},  \href{https://arxiv.org/abs/2408.03439}{{\tt 2408.03439}}.

\bibitem{Arkani-Hamed:2022gsa}
N.~Arkani-Hamed, L.~Eberhardt, Y.-t. Huang and S.~Mizera, \emph{{On unitarity
  of tree-level string amplitudes}},
  \href{http://dx.doi.org/10.1007/JHEP02(2022)197}{\emph{JHEP} {\bf 02} (2022)
  197}, [\href{https://arxiv.org/abs/2201.11575}{{\tt 2201.11575}}].

\bibitem{Guerrieri:2021ivu}
A.~Guerrieri, J.~Penedones and P.~Vieira, \emph{{Where Is String Theory in the
  Space of Scattering Amplitudes?}},
  \href{http://dx.doi.org/10.1103/PhysRevLett.127.081601}{\emph{Phys. Rev.
  Lett.} {\bf 127} (2021) 081601},
  [\href{https://arxiv.org/abs/2102.02847}{{\tt 2102.02847}}].

\bibitem{Guerrieri:2022sod}
A.~Guerrieri, H.~Murali, J.~Penedones and P.~Vieira, \emph{{Where is M-theory
  in the space of scattering amplitudes?}},
  \href{http://dx.doi.org/10.1007/JHEP06(2023)064}{\emph{JHEP} {\bf 06} (2023)
  064}, [\href{https://arxiv.org/abs/2212.00151}{{\tt 2212.00151}}].

\bibitem{Caron-Huot:2017vep}
S.~Caron-Huot, \emph{{Analyticity in Spin in Conformal Theories}},
  \href{http://dx.doi.org/10.1007/JHEP09(2017)078}{\emph{JHEP} {\bf 09} (2017)
  078}, [\href{https://arxiv.org/abs/1703.00278}{{\tt 1703.00278}}].

\bibitem{Paulos:2016fap}
M.~F. Paulos, J.~Penedones, J.~Toledo, B.~C. van Rees and P.~Vieira, \emph{{The
  S-matrix bootstrap. Part I: QFT in AdS}},
  \href{http://dx.doi.org/10.1007/JHEP11(2017)133}{\emph{JHEP} {\bf 11} (2017)
  133}, [\href{https://arxiv.org/abs/1607.06109}{{\tt 1607.06109}}].

\bibitem{vanRees:2023fcf}
B.~C. van Rees and X.~Zhao, \emph{{Flat-space Partial Waves From Conformal OPE
  Densities}},  \href{https://arxiv.org/abs/2312.02273}{{\tt 2312.02273}}.

\bibitem{vanRees:2022zmr}
B.~C. van Rees and X.~Zhao, \emph{{Quantum Field Theory in AdS Space instead of
  Lehmann-Symanzik-Zimmerman Axioms}},
  \href{http://dx.doi.org/10.1103/PhysRevLett.130.191601}{\emph{Phys. Rev.
  Lett.} {\bf 130} (2023) 191601},
  [\href{https://arxiv.org/abs/2210.15683}{{\tt 2210.15683}}].

\bibitem{Chang:2023szz}
C.-H. Chang, Y.~Landau and D.~Simmons-Duffin, \emph{{Spinning dispersive CFT
  sum rules and bulk scattering}},
  \href{https://arxiv.org/abs/2311.04271}{{\tt 2311.04271}}.

\bibitem{Li:2021snj}
Y.-Z. Li, \emph{{Notes on flat-space limit of AdS/CFT}},
  \href{http://dx.doi.org/10.1007/JHEP09(2021)027}{\emph{JHEP} {\bf 09} (2021)
  027}, [\href{https://arxiv.org/abs/2106.04606}{{\tt 2106.04606}}].

\bibitem{Caron-Huot:2021kjy}
S.~Caron-Huot and Y.-Z. Li, \emph{{Helicity basis for three-dimensional
  conformal field theory}},
  \href{http://dx.doi.org/10.1007/JHEP06(2021)041}{\emph{JHEP} {\bf 06} (2021)
  041}, [\href{https://arxiv.org/abs/2102.08160}{{\tt 2102.08160}}].

\bibitem{Hofman:2008ar}
D.~M. Hofman and J.~Maldacena, \emph{{Conformal collider physics: Energy and
  charge correlations}},
  \href{http://dx.doi.org/10.1088/1126-6708/2008/05/012}{\emph{JHEP} {\bf 05}
  (2008) 012}, [\href{https://arxiv.org/abs/0803.1467}{{\tt 0803.1467}}].

\bibitem{Zhiboedov:2013opa}
A.~Zhiboedov, \emph{{On Conformal Field Theories With Extremal a/c Values}},
  \href{http://dx.doi.org/10.1007/JHEP04(2014)038}{\emph{JHEP} {\bf 04} (2014)
  038}, [\href{https://arxiv.org/abs/1304.6075}{{\tt 1304.6075}}].

\bibitem{Schwarz:1978cn}
A.~S. Schwarz, \emph{{The Partition Function of Degenerate Quadratic Functional
  and Ray-Singer Invariants}},
  \href{http://dx.doi.org/10.1007/BF00406412}{\emph{Lett. Math. Phys.} {\bf 2}
  (1978) 247--252}.

\bibitem{Schwarz:1979ae}
A.~S. Schwarz, \emph{{The Partition Function of a Degenerate Functional}},
  \href{http://dx.doi.org/10.1007/BF01223197}{\emph{Commun. Math. Phys.} {\bf
  67} (1979) 1--16}.

\bibitem{Obukhov:1982dt}
Y.~N. Obukhov, \emph{{THE GEOMETRICAL APPROACH TO ANTISYMMETRIC TENSOR FIELD
  THEORY}}, \href{http://dx.doi.org/10.1016/0370-2693(82)90752-3}{\emph{Phys.
  Lett. B} {\bf 109} (1982) 195--199}.

\bibitem{Buchbinder:1988tj}
I.~L. Buchbinder and S.~M. Kuzenko, \emph{{QUANTIZATION OF THE CLASSICALLY
  EQUIVALENT THEORIES IN THE SUPERSPACE OF SIMPLE SUPERGRAVITY AND QUANTUM
  EQUIVALENCE}},
  \href{http://dx.doi.org/10.1016/0550-3213(88)90047-8}{\emph{Nucl. Phys. B}
  {\bf 308} (1988) 162--190}.

\bibitem{Kawai:1985xq}
H.~Kawai, D.~C. Lewellen and S.~H.~H. Tye, \emph{{A Relation Between Tree
  Amplitudes of Closed and Open Strings}},
  \href{http://dx.doi.org/10.1016/0550-3213(86)90362-7}{\emph{Nucl. Phys. B}
  {\bf 269} (1986) 1--23}.

\end{thebibliography}\endgroup

\end{document}